\documentclass[11pt]{article}

\usepackage[utf8]{inputenc} 

\usepackage{geometry}             
\usepackage{longtable}
\usepackage{booktabs}    
\usepackage{paralist}    
\usepackage{verbatim}    
\usepackage{subfig}      

\usepackage[pdfstartview=FitH]{hyperref}
\usepackage{graphicx} 
\usepackage{parskip}
\usepackage{setspace}
\usepackage{xcolor}      

\usepackage{authblk}
\usepackage{amsthm}
\usepackage{amsmath}     
\usepackage{amssymb}     
\usepackage{mathrsfs}    
\usepackage{amscd}       
\usepackage{tikz-cd}     
\usepackage{array}       

\usepackage{fancyhdr} 
\usepackage{thmtools}

\usepackage{amsthm}
\newtheorem{theorem}{Theorem}
\newtheorem{proposition}[theorem]{Proposition}

\newtheorem{definition}[theorem]{Definition}
\newtheorem{observation}[theorem]{Observation}

\newtheorem*{definitionnn}{Definition}

\numberwithin{equation}{section}
\numberwithin{figure}{section}
\numberwithin{table}{section}

\def\tightlist{}

\title{ Marginal Price Optimization \\ \large A new framework for arbitrage and routing in AMM driven markets \\ v1.0}

\author{
   Stefan Loesch \\
   \texttt{\href{mailto:stefan@bancor.network }{ stefan@bancor.network }}    
   \and
   Mark Richardson \\
   \texttt{\href{mailto:mark@bancor.network }{ mark@bancor.network } }
}

\affil{}
\date{ 27 January 2025 }

\begin{document}

\maketitle

\begin{abstract}
   \setstretch{1.0}
   We introduce a new framework for optimal routing and arbitrage in AMM driven
markets. This framework  improves on the original best-practice convex
optimization by restricting the search to the boundary of the optimal space.
We can parameterize this boundary using a set of prices, and a potentially
very high dimensional optimization problem (2 optimization variables per
curve) gets reduced to to a much lower dimensional root finding problem (1
optimization variable per token, irregardless of number of the curves). Our
reformulation is similar to the dual problem of a reformulation of the
original convex problem. We show our reformulation of the problem is
equivalent to the original formulation except in the case of infinitely
concentrated liquidity, where we provide a suitable approximation. Our
formulation performs far better than the original one in terms of speed -- we
obtain an improvement of up to 200x against Clarabel, the new CVXPY default
solver --  and robustness, especially on levered curves. 
 \\ \\ 
\end{abstract}
\pagebreak

\tableofcontents
\pagebreak

\listoffigures
\addcontentsline{toc}{section}{List of Figures}
\pagebreak

\listoftables
\addcontentsline{toc}{section}{List of Tables}
\pagebreak

\listoftheorems[ignoreall, show={theorem,proposition,lemma,conjecture,proposal,definition,observation}]
\addcontentsline{toc}{section}{List of Definitions and Theorems}
\pagebreak

\nocite{loesch22} \nocite{loesch21} \nocite{rich24} \nocite{angeris21}
\nocite{angeris22} \nocite{diamandis23} \nocite{boyd04}
\nocite{diamond16} \nocite{engel21} \nocite{bancor17} \nocite{carbon22}
\nocite{carbon22l} \nocite{uni18} \nocite{uni18hmd} \nocite{uni20}
\nocite{uni21} \nocite{cvxpy} \nocite{numpy} \nocite{bancor}
\nocite{carbon} \nocite{uniswap} \nocite{fastlane} \nocite{fastlanerepo}
\nocite{nelder65}

\section{Introduction}\label{introduction}

\label{sec:intro}

Part of Bancor's product offering is the FastLane Arbitrage bot
\cite{fastlane}. When developing this bot, we have developed our own
algorithms for solving what we call the ``Arbitrage Problem''\footnote{See definition \ref{def:arbproblem}}, loosely defined as
making risk free money out of a given set of AMMs by trading against
them. We started with what we believe was the standard approach at the
time, the convex regular convex optimization approach proposed in
\cite{angeris21}. This approach worked very well on unlevered curves,
but as soon as we applied it to levered curves, we ran into convergence
issues that we could not ultimately solve. The main purpose of the
FastLane bot was to support the Carbon DeFi protocol
\cite{carbon, carbon22} and when looking into the problem in more
detail, we felt that there were structural reasons why the direct convex
optimization approach from \cite{angeris21} would not work well on
levered curves. We therefore developed our own algorithm, called the
``Marginal Price'' algorithm, which solves the same problem, but which
is significantly faster and which has more benign convergence
properties. It is similar to solving the conjugate convex problem
proposed in \cite{diamandis23} which was published around the time we
put the finishing touches on our arbitrage bot, but it goes somewhat
further, and it is more founded in financial than in purely mathematical
principles.

\subsection{Problem statement}\label{problem-statement}

\label{sec:introps}

Before we go into the details we will provide key results from
\cite{angeris21} that help us to better define the problem space. For
clarity of presentation, the references of this section will lead to
later sections in the paper. We do not want to hide the forest behind
the trees, and whilst the definitions and results referenced are
important enough to warrant formal treatment, the concepts are
sufficiently widely understood that we are confident that a reader even
vaguely familiar with the topic will understand terms like ``AMM''
without having to look up the formal definition.

\begin{definition}[Arbitrage and Routing Problems]\label{def:arbproblem}
Given a set of AMMs\footnote{See definition \ref{def:amm}} of a known state, the \textbf{Arbitrage Problem} is the problem of finding a sequence of trades that will result in a risk-free arbitrage profit\footnote{See definition \ref{def:arbfree}}. The \textbf{Routing Problem} is the problem of finding a sequence of trades on those AMMs that will result in the highest output (or lowest input) of a "target token" when all other token input and output quantities are fixed.
\end{definition}

It turns out that arbitrage and routing are closely related:

\begin{proposition}[Arbitrage and Routing]\label{prop:arbrout}
The Arbitrage Problem is a special case of the Routing Problem where all inputs and outputs except for the target token are fixed at zero.
\end{proposition}

\textbf{Proof.} Solving the routing problem where no other tokens go
into or out of the system is the definition of an arbitrage according to
definition \ref{def:arbfree}. \(\blacksquare\)

In other words -- when solving the routing problem, one is generally
also solving the arbitrage problem because arbitrages subsidize the
desired exchange, and a pure abitrage is optimally routing the
\emph{``null''}. In this paper, we mostly focus on arbitrage because of
our product focus, but transposing the results to routing is generally
straightforward.

\subsection{Automated Market Makers}\label{automated-market-makers}

\label{sec:AMM}

In this section, we briefly discuss the concept of \emph{Automated
Market Makers} (``AMMs'') and their \emph{bonding curves}\footnote{See definition \ref{def:amm} for a formal definition of the concept}.
Here we focus solely on \emph{constant product curves}, including their
levered variety, as they represent all features of interest for our
purposes. We note, however, that most results of this paper generalize
to other types of curves as well, although it helps when certain
quantities can be computed analytically, otherwise interdependent
numerical methods can impose additional challenges and performance may
be poor. We also note that, in our experience, all relevant curves can
be approximated by constant product curves, in segments if need be,
which is the approach we take in practice to deal with curves other than
constant product curves.

Formally, an AMM is a smart contract that allows the user to trade two
or more assets. The seminal version first introduced by Bancor
\cite{US12045807B2, bancor17} was a generalized hypersurface
\(S_{C, \mathbf{r}}\) embedded in \(R^n\), defined by the equation

\begin{equation}\label{eq:bancor}\begin{aligned}
S_{C, \mathbf{r}} = 
\left\{ 
\mathbf{x} \,\left|\, \prod_{i=1}^{n} x_i^{r_i} = C \right.
\right\}
\end{aligned}\end{equation}

where \(\mathbf{x} = (x_1, x_2, \ldots, x_n) \in R^{n+}\) is the vector
of token balances and the \(\mathbf{r} \in R^{n+}\) is the associated
vector of \emph{``reserve weights''}. \(C\) is a constant.

Equivalently, using the log token balances
\(\mathbf{z} = \log\mathbf{x}\) where the log function applies on a
per-component basis, one can transform the curved hypersurface
\(S_{C, \mathbf{r}}\) into a hyperplane \(P_{C, \mathbf{r}}\) given by
the equation

\begin{equation}\label{eq:bancorlog}\begin{aligned}
P_{C, \mathbf{r}} = 
\left\{ 
\mathbf{z} \,\left|\,\sum_{i=1}^{n} z_i r_i = \log C \right.
\right\}
\end{aligned}\end{equation}

This is the version implemented by Bancor \cite{bancor17} and later by
Balancer \cite{balancer19}. The most popular pools on Bancor v1 were all
\emph{two-assets / same-weight}. When Uniswap created their first AMM,
they froze this as a design principle \cite{uni18, uni18hmd} and they,
and many subsequent AMMs including Bancor v2.1/3 and Carbon DeFi, relied
on the well known simplified version of the aforementioned constant
product bonding curve. This simplified invariant function is defined by
the equation

\begin{equation}\label{eq:invariant}\begin{aligned}
x\cdot y = k
\end{aligned}\end{equation}

where \(x\) and \(y\) are the quantities of the two assets, and \(k\) is
the pool invariant. What this equation signifies is that -- ignoring
fees -- the AMM in question will engage in any trade (ie exchange \(x\)
for \(y\) or vice versa) that keeps the pool invariant \(k\) constant.
Adding or removing liquidity, including via fees, will of course change
\(k\). We can express \(y\) as a function of \(x\) as \(y(x)=k/x\). It
is easy to see that the marginal price in units of \(y\) per \(x\) at a
given point \((x,y)\) is given by

\begin{equation}\label{eq:price}\begin{aligned}
p_{\mathrm{marg}} \equiv -\frac {dy}{dx} 
= \frac{k}{x^2} 
= \frac{y^2}{k}
= \frac{y}{x} 
\end{aligned}\end{equation}

from which immediately follows that, at the marginal price, the value of
token holdings in \(x\) and \(y\) will always be the same, in units of
\(x\), or \(y\), or any joint numeraire in which one chooses to express
it.

\subsubsection{Levered liquidity}\label{levered-liquidity}

\begin{figure}[htbp]
    \centering
    \includegraphics[width=\textwidth]{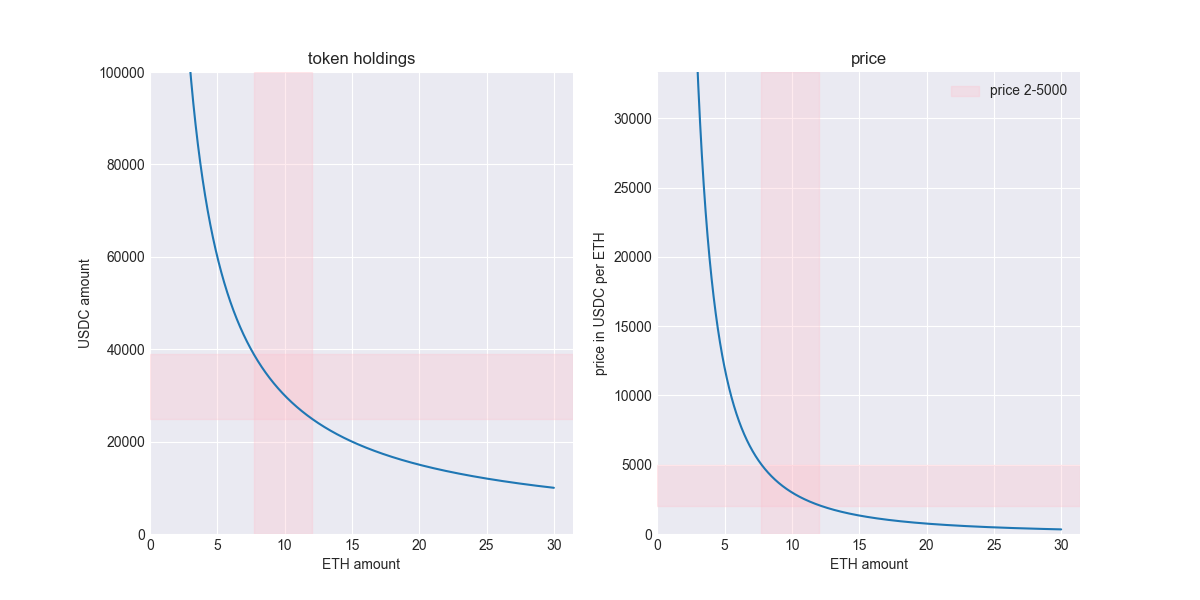}
    \caption[Invariance and price curve for an unlevered AMM]{\textit{Invariance and price curve for an unlevered AMM. }Invariance (aka bonding) curve (left) and associated price curve (right) for an unlevered AMM; the pink area represents an example for a reasonable trading area, and all collateral outside of it is rarely used.}
    \label{fig:cunlev}
\end{figure}

The above curves -- usually referred to as \emph{``unlevered''} curves
for reasons that will become clear in a moment -- trade over the entire
possible price range. This is nice for symmetry and scale invariance
reasons, but it is not particularly efficient in terms of collateral
usage, as most tokens in the AMM are held in reserve for price points
that, in realistic markets, will never be reached. See figure
\ref{fig:cunlev} for an illustration of this.

\begin{figure}[htbp]
    \centering
    \includegraphics[width=\textwidth]{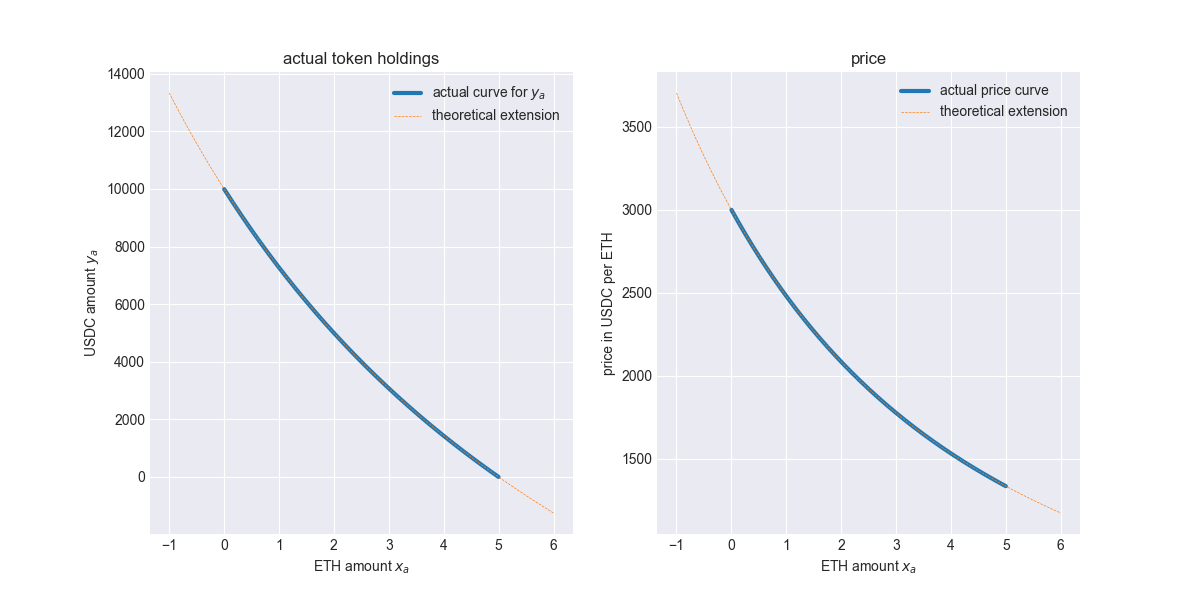}
    \caption[Invariance and price curve for a levered AMM]{\textit{Invariance and price curve for a levered AMM. }Invariance (aka bonding) curve (left) and price curve (right) for a levered AMM; only the area in solid blue corresponds to actual liquidity held by the AMM. The thin curve extending it depicts where the associated unlevered AMM would trade.}
    \label{fig:clev}
\end{figure}

This is where the concept of virtual token balances and
\emph{``amplified''} or \emph{``concentrated''} liquidity curves comes
in, first described by Bancor \cite{US20240119444A1} in 2020, and later
popularized by Uniswap \cite{uni21} in 2021. The idea is simple: we just
posited that a significant proportion of the collateral is allocated to
price points that will never be reached, so we might as well remove it
from the AMM, and if ever the AMM were in a price range where it had to
pay out those removed tokens, it would just halt trading until prices
returned to the range it was prepared to trade in. To formalize this
concept, we introduce virtual token balances\footnote{Some authors
disagree on what exactly is being referred to as \emph{virtual} token
balance, the quantity \(x_v\) or the sum \(x_v+x_a\); herein, we use the
former, but in any case it should always be clear from the context}
\(x_v > 0\) and \(y_v > 0\), and with their help we rewrite the
invariant equation \ref{eq:invariant} as

\begin{equation}\label{eq:invariantl}\begin{aligned}
k = x\cdot y = (x_a+x_v)\cdot(y_a+y_v)
\end{aligned}\end{equation}

where \(x_a \geq 0\) and \(y_a \geq 0\) are the actual token holdings of
the AMM. The trading behavior of the AMM is now slightly modified in
that -- again ignoring fees -- the AMM will accept any trade that (1)
holds \(k\) constant, and importantly (2) maintains \(x_a\) and \(x_b\)
non-negative. What this means in practice is that a levered AMM has two
price boundaries -- one for \(x_a=0\) and one for \(y_a=0\) -- where all
the AMM's collateral is held in the respective other token. An example
for a levered curve is shown in figure \ref{fig:clev} where an amount of
5 ETH or 10,000 USDC (exact mix depends on where we are on the curve) is
deployed over a price range of slightly below 1,500 to 3,000 USDC per
ETH.

This relatively wide range of a levered curve is typical for Bancor's
original concentrated AMM, and its successor, Carbon DeFi. However,
architecture and design decisions made for the Uniswap v3 AMM mandate a
much thinner range: the minimum tick size depends on the fee tier, but
the width of a single Uniswap v3 curve was initially between 10-200
basis points. What is usually referred to as ``the'' Uniswap v3 curve is
actually a collection of independent curves that are located adjacent to
each other, each of which have their own liquidity holdings.

\subsubsection{Directional liquidity and
fees}\label{directional-liquidity-and-fees}

Directional liquidity, also referred to as \emph{``asymmetric
liquidity''}, is a concept that has been formally introduced by Bancor
with its Carbon DeFi product in 2022 \cite{WO2024084480A1, carbon22}. It
means that specific curves only ever trade in one direction, and that if
one wants to offer a two-way market, one has to explicitly specify two
curves for that. Technically, a directional curve in Carbon is
implemented as a levered curve according to equation \ref{eq:invariantl}
with the stipulation that one of \(x_a, y_a\) is always kept at zero. In
other words, whilst in a non-directional AMM, tokens that trade into the
AMM are added to the actual token balances, \(x_a, y_a\) respectively,
to replace those that have been traded out, in a directional AMM the
tokens that are traded in are added to the retained token balances,
\(x_r, y_r\) respectively.

For those retained balances the Carbon AMM design then offers two
choices:

\begin{itemize}
\tightlist
\item
  they are off-curve and inactive, and their only future use is that
  they can be withdrawn by the owner of the curve, or
\item
  they are automatically placed onto another curve that trades into the
  opposite direction, with parameters independent of that of the first
  curve.
\end{itemize}

In a Carbon DeFi context, buy and sell curves are typically
non-overlapping. For example, a curve may buy ETH against USDC between
2,500-2,600 and sell it between 2,900-3,000. However --
\emph{overlapping strategies} are also possible where the parameters are
chosen such that the price of the buy and sell curve move in unison, and
that collateral that moved from the buy to the sell curve is offered
back to the market at only a slightly higher price.

It is an important insight -- and the reason why we wrote ``formally''
introduced above -- that functionally, overlapping strategies are not
different from non-directional curves with fees. For example, a curve
that buys and sells the marginal ETH at 3,000 USDC with a 1\% fee is
functionally equivalent to one curve that buys ETH at 2,970 USDC and
another that sells it at 3,030 USDC, and collateral moves from one curve
to the other upon trading.

Fees will become very important further down this paper because they
actually lead to curve scenarios that are numerically not particularly
well behaved. We will get into the details of this further ahead, but
the key reason is that in the presence of fees there are ``holes'' in
the curve where it does not exist. For example the curve above would not
trade at all for prices between 2,970 and 3,030 USDC, and any algorithm
that performs a local analysis at this price point risks failing.

\subsubsection{Limit orders and infinite
leverage}\label{limit-orders-and-infinite-leverage}

Bancor, via its Carbon DeFi product, also introduced the concept of
limit orders into the AMM space, where the entire liquidity of a curve
is placed at a single price point. Formally, this can be considered as
the limit \(x_\nu, y_\nu \to \infty\) in equation \ref{eq:invariantl}
and in our preprint \cite{rich24}. In practical calculations this of
course does not help us much as we cannot deal numerically with numbers
that are infinite. The Carbon DeFi implementation gets around this by
reparametirizing the problem in terms of \(B=\sqrt{p_{\mathrm{min}}}\)
and \(A=\sqrt{p_{\mathrm{max}}} - \sqrt{p_{\mathrm{min}}}\), where
\(p_{\mathrm{min}}\) and \(p_{\mathrm{max}}\) are the marginal prices at
the boundaries, which allows setting \(A=0\) for limit orders. However,
in our implementation of the arbitrage bot (productized as Bancor's
\emph{``ArbFastLane protocol''}), and throughout this paper, we do not
allow for \(A=0\) as this would lead to numerical instabilities in the
calculations. Instead, we impute a minimum value for \(A\) that is big
enough for numerical stability, and that is then being corrected in a
transaction fine tuning process once arbitrage opportunities have been
identified.

\subsection{Optimization}\label{optimization}

\label{sec:optimization}

Before we dive further into the issue at hand, we want to generally
discuss optimization problems and start building intuition on how they
work, and how they can be solved\footnote{Please also refer to
appendix \ref{app:numerics} for a discussion of the numerical methods we
use}. Generally, an optimization problem has (1) a target function,
and (2) one or more constraints. We then are looking to maximize or
minimize the target function while satisfying the constraints. The
constraints can be equality constraints, inequality constraints, or
both. Equality vs inequality does not usually make a difference as the
solution to a sufficiently well conditioned optimization problem is
found on the boundary of the feasible region. Constraints can be of two
kinds which we call \emph{``(multi-)linear''} or \emph{``explicit''},
respectively. The former consist of \emph{multiple} linear inequalities
and all of them must be satisfied, leading to a non-differentiable
hypersurface where the non-differentiable regions are the boundaries on
which the active constraint switches. What we call explicit constraints
are constraints that can be written in the form \(f(x)=0\) for some
function \(f\). We call an explicit constraint \emph{smooth} if it can
be written using a single smooth constraint function \(f\). Note that
the two forms of constraints are often interchangeable. For example,
every multi-linear constraint can be converted into an explicit
constraint using a piecewise-linear function \(f\), but depending on the
exact use case one or the other may be easier to deal with. Linear
constraints have the advantage that the resulting set will be convex.
Also, in higher dimensions it is usually hard to formulate an explicit
constraint function matching multiple linear constraints.

\begin{figure}[htbp]
    \centering
    \includegraphics[width=\textwidth]{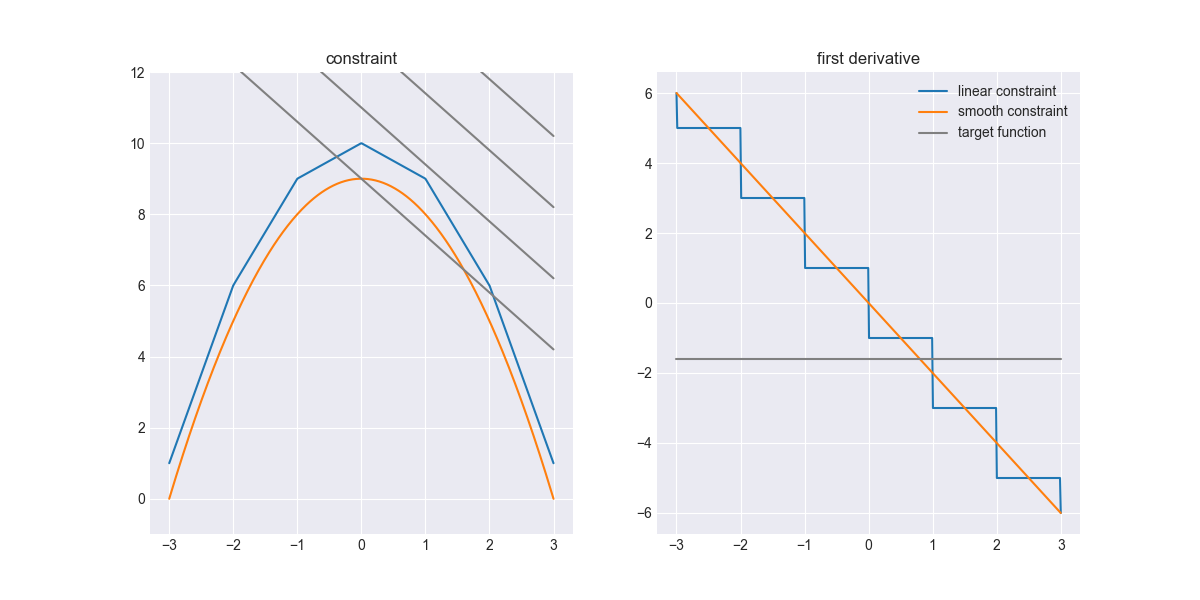}
    \caption[Constraints and target functions]{\textit{Constraints and target functions. }The left panel depicts a smooth constraint (blue) and a piecewise-linear one (orange) as well as the level areas of a linear target function (grey). The right panel shows the respective derivatives, showing that at the optimal point, the derivative of the target function equals the derivative of the constraint.}
    \label{fig:targetfuncs}
\end{figure}

In figure \ref{fig:targetfuncs} on the left hand panel, we have drawn an
optimization problem in one dimension where the blue and orange lines
represent the constraints, linear and smooth respectively, and the grey
lines represent level sets of the target function. We have also drawn
the derivative of the constraint and target functions in the same
figure, on the right hand panel.

In one dimension, this problem is easy to solve: the point where the
target function is optimized in the smooth case is where the level set
is tangent to the constraint. In other words: we look for the point
where the derivative of the target function equals the derivative of the
constraint, which in this case happens slightly below \(x=1\). For the
linear case, we encounter another well known result of optimization
theory which is that the solution to the optimization problem is usually
found on the lowest-dimension boundary, which in this case are the
points where the different constraints meet. In this case the solution
is exactly at \(x=1\). Note that whilst this is not technically a point
where the constraint is tangent to the target function, the derivative
of the constraint flips from above to below the derivative of the target
function. If we would smooth the corners of the constraint function --
say by folding with a \(C^\infty\) kernel -- we would find that the
derivates would again meet very closely to \(x=1\), the exact point
depending on the kernel used.

Ultimately, as we will see in what follows, this problem shows the
essence of what we are aiming to do in this paper, except that the
problem we are solving is higher dimensional, and the functions we
consider are more complex, so that we do have to rely on numerical
methods to find the solution. Specifically, the target function is the
profit made, or rather the outflow in the target token chosen to collect
the profit. The constraints are given by the AMM curves that form the
market, plus the ``self-financing-constraint'' that all token flows must
be accounted for (ie that there is no net token leakage in or out other
than the one explicitly accounted for). In the pure arbitrage case, this
constraint is simply that, on a net basis, all flows in tokens other
than that of the target token are zero. We mention en-passant that for
routing applications one can use other constraints, eg \emph{``there is
a flow of 1,000 USDC into the system''} (or \emph{``1,000 USDC and 3,000
DAI''}), and answering the question \emph{``what is the maximum amount
of ETH that can be extracted and how?''} is the answer to the optimal
routing problem. However, whilst the routing application is interesting
and not particularly hard to execute once the arbitrage application
settled, it is not the focus of this paper.

As shown by Angeris et al in reference \cite{angeris21}, this problem is
a convex optimization problem, and as we will see, we will run into
similar problems that generic convex solvers run into, which is that
smooth problems and linear problems often require different algorithms.
Specifically, smooth problems are often solved using gradient descent
methods, whilst linear problems are often solved using simplex methods,
the latter essentially jumping between the vertices of the feasible
region. The reason why we run into those problems is that we sometimes
have a smooth problem, sometimes a linear problem, and most of the time
a problem that is dominated by one of the two aspects, although it is
hard to predict which one in advance\footnote{In this context see
appendix \ref{app:radome} for the \emph{radome optimization problem}
that provides a geometric example for a problem exposing similar
issues}. Specifically:

\begin{enumerate}
\def\labelenumi{\arabic{enumi}.}
\tightlist
\item
  A market that only consists of unlevered curves is a smooth problem.
\item
  A market that only consists of limit order curves (eg, Carbon curves
  with width zero) is a linear problem
\item
  Levered curves are a mix of the two, depending on their width, or
  rather: how fast the liquidity changes at different points in the
  curve (``ticks'' in Uniswap v3 parlance).
\end{enumerate}

\begin{figure}[htbp]
    \centering
    \includegraphics[width=\textwidth]{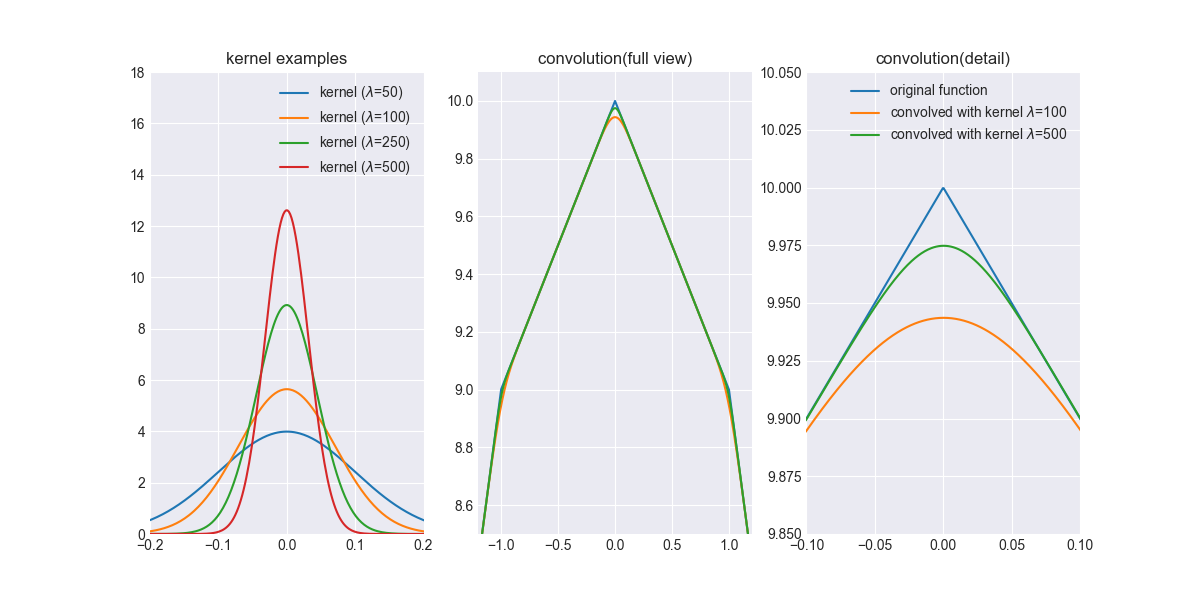}
    \caption[Convolution examples]{\textit{Convolution examples. }The left panel shows a number of Gaussian \(C^\infty\) kernels of different widths. The other panels show a zoomed out (middle) and detailed view of a piecewise linear function (blue) and its convolution with two kernels of different widths (orange and green) which is a \(C^\infty\) function.}
    \label{fig:convolution}
\end{figure}

We can think of the third case above as a ``smoothed'' problem where the
width of the curves corresponds to the smoothing kernel used. We briefly
explain the concept of smoothing kernels in figure
\ref{fig:convolution}. In its left panel, we have drawn a few Gaussian
kernels with different width parameter \(\lambda\). They satisfy the
equation

\begin{equation}\label{eq:kernels}\begin{aligned}
\kappa_\lambda(x) = \sqrt{\frac{\lambda}{\pi}} e^{-\lambda x^2}
\end{aligned}\end{equation}

When we calculate the convolution of the kernel \(\kappa\) with the
target function \(f\) (here, the constraint function), we get

\begin{equation}\label{eq:convolution}\begin{aligned}
f_\kappa(x) = (\kappa * f)(x) = \int \kappa(x-y) f(y) dy
\end{aligned}\end{equation}

Note that, as mentioned above, because the kernel \(\kappa\) is
\(C^\infty\), the convolution \(f_\kappa\) is also \(C^\infty\) which we
can easily show by pulling the differentiation operator \(\partial_x\)
inside the integral so that we get \(f'_\kappa(x) = (f*\kappa')(x)\)
where the \(f'\) and \(\kappa'\) respectively denote the derivative with
respect to the variable \(x\). The convolution of a piecewise-linear
function with a \(C^\infty\) kernel, together with the kernel examples,
is shown in figure \ref{fig:convolution}.

We do not use convolution explicitly in our algorithms but we enforce a
minimum width for limit orders which has the same effect: if we convert
a \emph{buy-at-1,000} limit order to \emph{buy-from-990-to-1,001} then
this is akin to a convolution of the constraint function with a kernel
of about unity width.

\section{Convex Optimization}\label{convex-optimization}

\label{sec:convex}

The marginal price optimization algorithm for arbitrage and routing,
described in section \ref{sec:margpframework}, is the core of this
paper. It is closely related to the convex optimization algorithm
described in \cite{angeris21, angeris22, diamandis23} that we were
previously using. As described in section \ref{sec:cvxconverg}, when we
applied that algorithm for Bancor's ArbFastLane product \cite{fastlane}
to markets containing important segments with levered curves we ran into
unsurmountable convergence issues.

We will not go into the details of convex optimization here -- the
aforementioned references are excellent resources for that. However, a
description of the convex optimization setup is important for setting
the scene and context in which our core theorem of this paper, the
\emph{core equivalence theorem} \ref{thm:coreequiv}, establishing the
equivalence between our marginal price optimization algorithm and the
previously known convex optimization algorithm. This will happen in the
next subsection \ref{sec:cvxsetup}, and in the following section
\ref{sec:cvximpl} we will discuss implementation and results of the
convex optimization algorithm on unlevered curves where it works
extremely well. In the last subsection \ref{sec:cvxconverg} we will
finally discuss convergence issues, providing a specific example of a
problematic case.

\subsection{Setting up the problem}\label{setting-up-the-problem}

\label{sec:cvxsetup}

We will here briefly outline the methodology that we used as a basis for
our implementation. It closely follows \cite{angeris21} so interested
readers are encouraged to read both papers in conjunction.

For setting up a convex optimization problem, we need to define the
\emph{optimization variables}, the \emph{target function} and a set of
\emph{constraints}. The target function is the function that we want to
optimize, and the constraints are the conditions that the variables must
satisfy.

\textbf{Optimization Variables.} We are operating in the context of a
linked set of DEXes, and the natural variables are the token balances of
all those DEXes, as those fully define the state of the system that we
are interested in. For ease of notation and WLOG we only consider DEXes
that operate on token pairs, but it should be noted that conceptually,
multi-token exchanges work the same way, but the notation gets
significantly heavier.

We assume that we have \(M\) DEXes indexed by \(\nu=0\ldots M-1\). Those
DEXes have \(2M\) token balances \(x_0, \dots, x_{2M-1}\) where we
simply assume that adjacent balances belong to a given DEX, ie
\(x_{2\nu}\) and \(x_{2\nu+1}\) are the token balances of DEX \(\nu\).
We use \(\alpha=0\ldots2M-1\) as index for the balances.

\textbf{Positivity Constraints.} The basic constraint is that all token
balances must be non-negative, ie

\begin{equation}\label{eq:positivity}\begin{aligned}
\forall \alpha=0\ldots2M-1: \quad x_\alpha \geq 0
\end{aligned}\end{equation}

For unlevered curves, those constraints can often be omitted, because
they are not binding. However, they are relevant in the case of levered
AMMs where those balances are virtual balances according to equation
\ref{eq:invariantl}. Also, many solvers operate more efficiently if
positivity constraints are explicitly stated.

\textbf{Curve Constraints (unlevered).} The token balances satisfy the
following curve constraints or \emph{curve (in)equalities}

\begin{equation}\label{eq:curveconstr}\begin{aligned}
\forall \nu=0\ldots M-1: \quad x_{2\nu} \cdot x_{2\nu+1}\leq
\bar x_{2\nu} \cdot \bar x_{2\nu+1} \equiv \bar k_{\nu}
\end{aligned}\end{equation}

The barred quantities \(\bar x_\alpha\) are the initial values of the
token balances, so those are not optimization variables, but parameters
of the problem. The redundant terms \(\bar k_{\alpha}\) are only shown
to link back to the curve equation \ref{eq:invariant}.

\textbf{Curve Constraints (levered).} When using levered curves we
recreate the step from equation \ref{eq:invariant} to equation
\ref{eq:invariantl} and we rewrite equation \ref{eq:curveconstr} as

\begin{equation}\label{eq:curveconstrl}\begin{aligned}
\forall \nu: \quad
(x_{2\nu} + \bar x_{2\nu}^0) 
\cdot 
(x_{2\nu+1} + \bar x_{2\nu+1}^0)
\leq
(\bar x_{2\nu} + \bar x_{2\nu}^0) 
\cdot 
(\bar x_{2\nu+1} + \bar x_{2\nu+1}^0)
\end{aligned}\end{equation}

This equation states that the current state of the AMM is adjusted by
the virtual base balances \(\bar x_\alpha^0\), which are additional,
constant, parameters of the problem.

Technically, the constraints in \ref{eq:invariant} and
\ref{eq:invariantl} should be equalities, but as shown in
\cite{angeris21}, inequalities are required to make the problem convex.
However, the solution to the optimization problem will be found on the
boundary, so any solution will actually satisfy the equality constraints
as opposed to the inequality ones.

\textbf{Token Flows.} Up to here we have treated all token balances as
independent variables, which misses one very important piece of
information, notably what type of tokens they are (eg WETH, USDC etc).
This information is provided in the form of \emph{self-financing
constraints} which ensure that the sum of the balances of the same token
across all DEXes is constant, with the exception of the \emph{target
token} which is the token in which the profits are being extracted (see
below). In other words: those constraints ensure the we do not move
tokens other than the target token in or out of the system.

We assume here that we have \(N+1\) tokens indexed by \(i=0\ldots N\)
and that token \(i=0\) is the \emph{target token}. We want to define the
\emph{token flows} \(\phi_i\) as changes in token balances of token
\(i\) before and after the optimization. In order to do that, we define
the \emph{token matrix} \(\mathbf{T}\) as

\begin{equation}\label{eq:tmatrix}\begin{aligned}
\mathbf{T} = (T_{i \alpha}), \quad T_{i \alpha} = 
1\ \mathrm{if}\ (x_\alpha\ \mathrm{of\ type}\ i)  
\ \mathrm{else}\ 0
\end{aligned}\end{equation}

In other words, \(T_{i \alpha}\) is the indicator function for the token
associated with the DEX balance \(x_\alpha\) being of token type \(i\).

Using this token matrix, we can now express the \emph{token balance
function} \(\mathbf{B}: R^{2N} \rightarrow R^{N+1}\) as a function from
the state space into the ``balance space'', associating each state
\(x=(x_\alpha)\) a balance vector \(\mathbf{B}(x) = (B_i(x))\) where
\(B_i(x)\) is the sum of the token balances of token \(i\) across all
DEXes. Using the token matrix we can write this in matrix form as

\begin{equation}\label{eq:betam}\begin{aligned}
\mathbf{B}(x) = \mathbf{T} x
\end{aligned}\end{equation}

or, broken down into its components, as

\begin{equation}\label{eq:beta}\begin{aligned}
B_i(x) = \sum_{\alpha=0}^{2M-1} T_{i \alpha} x_\alpha
\end{aligned}\end{equation}

This finally allows us to define the aforementioned \emph{token flow
function} \(\phi: R^{2N} \rightarrow R^{N+1}\) as the difference between
the token balance function after optimizaton \(\phi(x)\) and the initial
token balance function \(B(\bar x)\) as

\begin{equation}\label{eq:phi}\begin{aligned}
\phi(x) = B(x) - B(\bar x)
\end{aligned}\end{equation}

The convention here is that outflows from the DEX system are negative,
and inflows to the DEX system are positive.

\textbf{Self Financing Constraints.} The self-financing constraints we
use for arbitrage calculation is that, other than for the target token
\(i=0\), the flow must be zero, yielding

\begin{equation}\label{eq:sfconstr}\begin{aligned}
\forall i=1\ldots N: \quad \phi_{i} = 0
\end{aligned}\end{equation}

Note that we start at \(i=1\) because we excluded the target token from
the token flow calculation. Whilst the optimal routing problem is out of
scope for this paper, we note that if we are interested in routing
instead of arbitrage, the above equation becomes

\begin{equation}\label{eq:sfconstrr}\begin{aligned}
\forall i=1\ldots N: \quad \phi_{i} = w_{i}
\end{aligned}\end{equation}

where \(\mathbf{w} = (0, w_1, \ldots, w_M)\) is the vector of desired
flows to route into or out of token 0.

\textbf{Target Function.} Last but not least we need to define our
target function (ie the function that we want the optimizer to minimize
or maximize). In our case, we want to maximize our profit, and our
profit is the outflow of target token \(\nu=0\) from the system, which
in our conventions is a negative number. Therefore our optimizer target
is

\begin{equation}\label{eq:target}\begin{aligned}
\mathrm{target} = \min \phi_0(x)
\end{aligned}\end{equation}

and the optimizer should minimize the negative number \(\phi_0(x)\)\footnote{This definition is WLOG in that we could optimize for any
other token but token 0, but this would unnecessarily complicated the
formulas.}.

We can tie all those definitions together in the following definition.
As mentioned above, the proof that this is a \emph{convex} optimization
problem is given in \cite{angeris21}.

\begin{definition}[Convex Optimization Formulation]\label{def:convexform}
The "\textbf{Convex Optimization Formulation} of the Arbitrage Problem" is the convex optimization problem with the target function from equation \ref{eq:target} which is subject to the positivity constraints in equation \ref{eq:positivity}, the self-financing constraints in \ref{eq:sfconstr}, and the curve constraints either in their unlevered form in \ref{eq:curveconstr} or their levered form in \ref{eq:curveconstrl}, or a combination thereof.
\end{definition}

\subsection{Implementation}\label{implementation}

\label{sec:cvximpl}

An example implementation of this algorithm is given in \cite{angeris21}
appendix B, and we refer the interested reader to that reference for a
worked example. Like in that example, when we started working on the
ArbFastLane arbitrage bot \cite{fastlane, fastlanerepo} we implemented
the algorithm in Python using the CVXPY library
\cite{cvxpy, boyd04, diamond16}.

See appendix \ref{app:tablescharts} for a detailed explanation of our
charts and tables.

\begin{figure}[htbp]
    \centering
    \includegraphics[width=\textwidth]{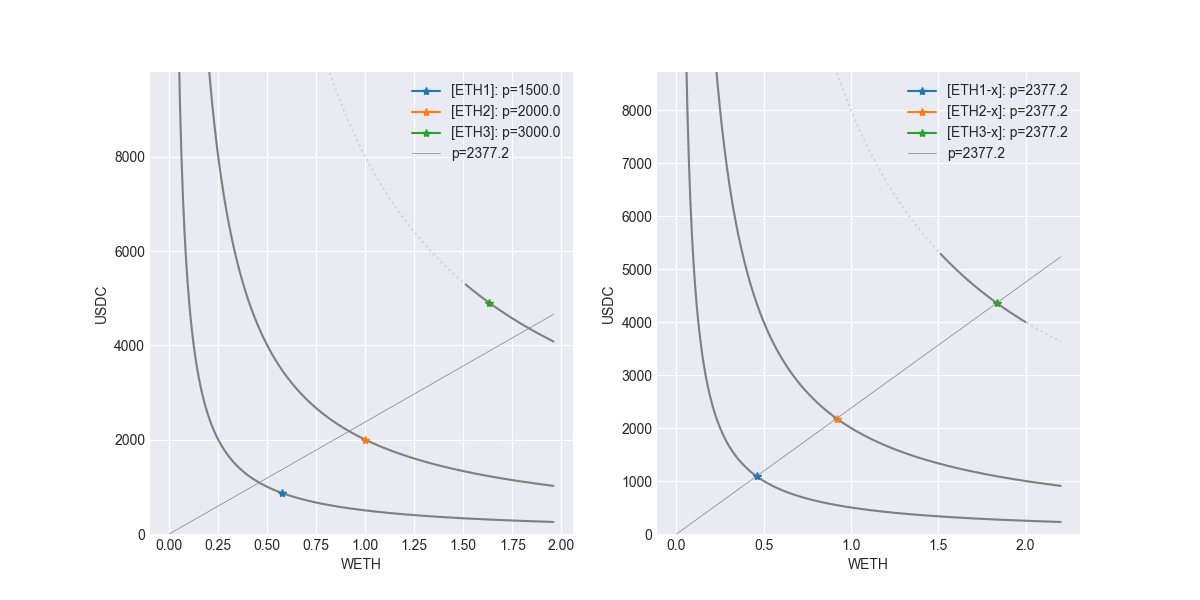}
    \caption[Single pair arbitrage with three curves]{\textit{Single pair arbitrage with three curves. }The left panel shows three curves of the same pair WETH/USDC, all at different prices (the stars are not aligned). The right panel shows the same curve after the arbitrage process, where all three stars are on the same line from the origin, at a price of 2377.}
    \label{fig:arbpair}
\end{figure}

\begin{table}[htbp]  
\centering
\small

\begin{tabular}{lrr}
\toprule
{} &     WETH &      USDC \\
\midrule
\textbf{PRICE    } &  2,377.2 &       1.0 \\
\textbf{ETH1     } &   -0.119 &   224.201 \\
\textbf{ETH2     } &   -0.083 &   180.452 \\
\textbf{ETH3     } &    0.201 &  -538.075 \\
\textbf{AMMIn    } &    0.201 &   404.653 \\
\textbf{AMMOut   } &   -0.201 &  -538.075 \\
\textbf{TOTAL NET} &   -0.000 &  -133.423 \\
\bottomrule
\end{tabular}

\caption{Trade instructions (single pair)}
\label{tbl:3_arb1x3instr}
\end{table}

In the pair arbitrage figure \ref{fig:arbpair} we show three curves, all
operating in the same pair WETH/USDC. As shown in the left panel,
initially they are at different prices. The result after running the
optimization algorithm is shown in the right panel, where all three
curves are now at the same price of 2377. Therefore, all three stars
indicating the current state of the curves are on the same straight line
through the origin. The associated trade instructions are in table
\ref{tbl:3_arb1x3instr}.

\begin{figure}[htbp]
    \centering
    \includegraphics[width=\textwidth]{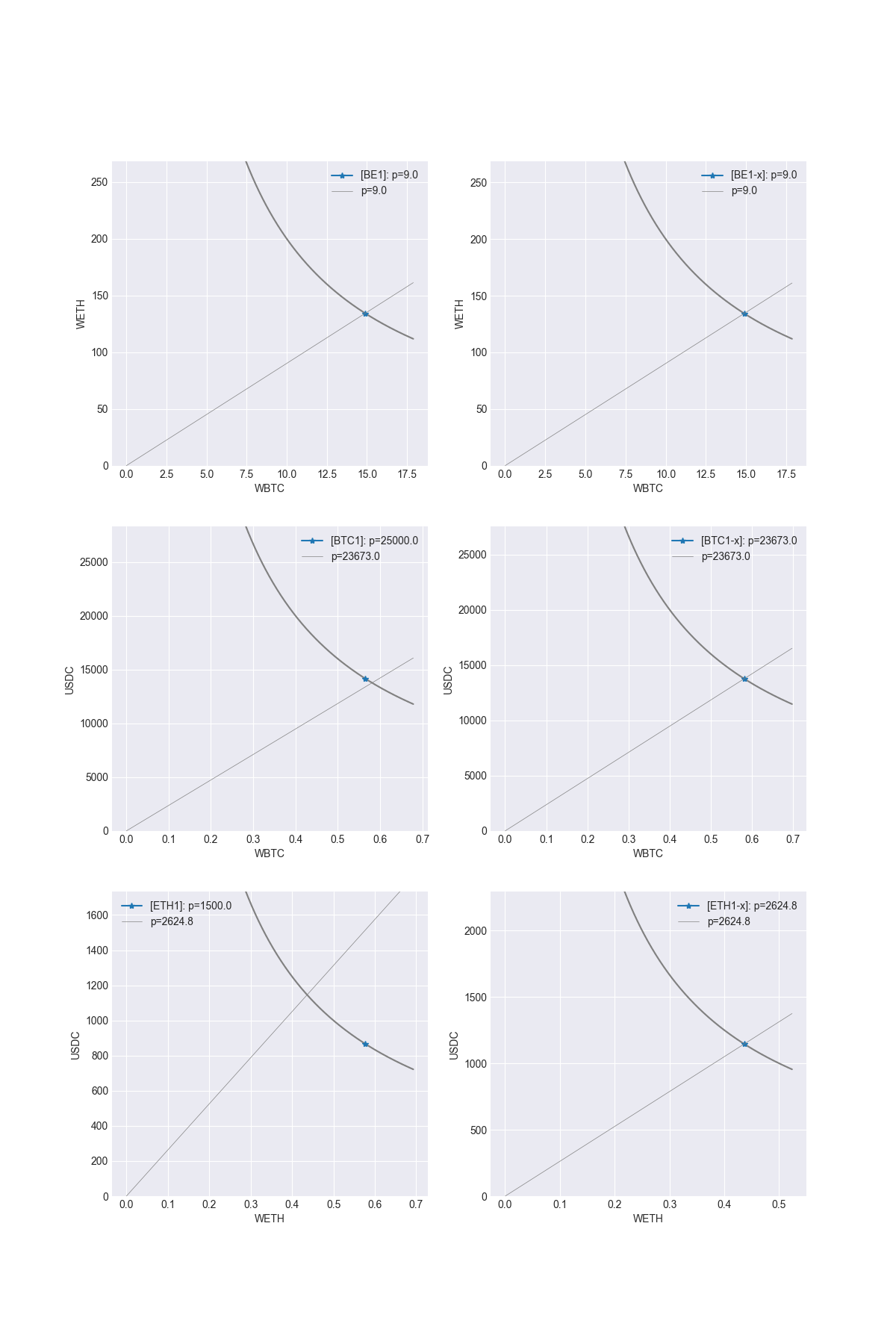}
    \caption[Triangle arbitrage with one curve each]{\textit{Triangle arbitrage with one curve each. }The three charts on the left represent curves in the triangle WETH/USDC/WBTC, at price points that allow a circular arbitrage. The right panel shows the same curves after the arbitrage process, where the stars are aligned so no circular arbitrage is possible.}
    \label{fig:arbtriangle1c}
\end{figure}

\begin{table}[htbp]  
\centering
\small

\begin{tabular}{lrrr}
\toprule
{} &     WETH &      WBTC &      USDC \\
\midrule
\textbf{PRICE    } &  2,624.8 &  23,673.0 &       1.0 \\
\textbf{ETH1     } &   -0.141 &        &   279.578 \\
\textbf{BTC1     } &       &     0.016 &  -380.457 \\
\textbf{BE1      } &    0.141 &    -0.016 &        \\
\textbf{AMMIn    } &    0.141 &     0.016 &   279.578 \\
\textbf{AMMOut   } &   -0.141 &    -0.016 &  -380.457 \\
\textbf{TOTAL NET} &   -0.000 &    -0.000 &  -100.879 \\
\bottomrule
\end{tabular}

\caption{Trade instructions (single triangle)}
\label{tbl:3_arb3x1instr}
\end{table}

We then look at a simple triangular arbitrage, where we add WBTC as a
third token, and where we have curves for each of the three constituent
pairs, as shown in the left column of charts in figure
\ref{fig:arbtriangle1c}. Note that, if we multiply the first two prices,
we do not get the third one, meaning that there is a circular arbitrage
opportunity. After the arbitrage process, the stars indicating the state
of the AMM changed their location, and the product of the first two
prices equals the last one, therefore no further arbitrage opportunities
are left. The associated trade instructions are in table
\ref{tbl:3_arb3x1instr}.

\begin{figure}[htbp]
    \centering
    \includegraphics[width=\textwidth]{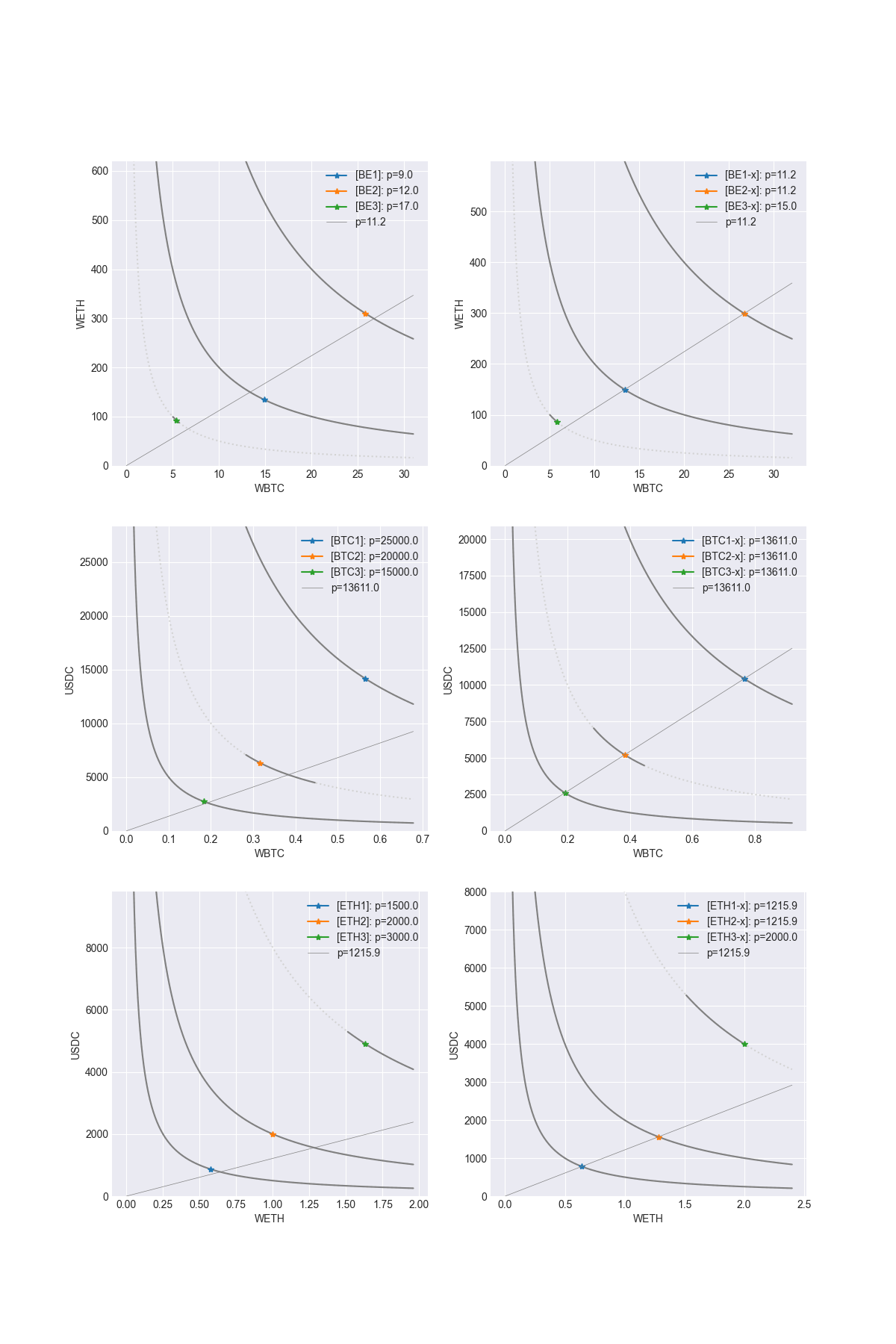}
    \caption[Combined triangle and pair arbitrage including levered curves]{\textit{Combined triangle and pair arbitrage including levered curves. }The curves on the left represent a market where both pair and triangular arbitrage is possible, the former being visible by the fact that the stars are not aligned. The right panel shows the same curves after the arbitrage process, where all stars are either aligned, or at the correct boundary in case of levered curves outside of their range.}
    \label{fig:arbtriangle3c}
\end{figure}

\begin{table}[htbp]  
\centering
\small

\begin{tabular}{lrrr}
\toprule
{} &     WETH &      WBTC &        USDC \\
\midrule
\textbf{PRICE    } &  1,215.9 &  13,611.0 &         1.0 \\
\textbf{ETH1     } &    0.064 &        &     -86.315 \\
\textbf{ETH2     } &    0.283 &        &    -440.579 \\
\textbf{ETH3     } &    0.367 &        &    -898.979 \\
\textbf{BTC1     } &       &     0.201 &  -3,707.206 \\
\textbf{BTC2     } &       &     0.067 &  -1,107.091 \\
\textbf{BTC3     } &       &     0.009 &    -129.880 \\
\textbf{BE1      } &   15.463 &    -1.541 &          \\
\textbf{BE2      } &  -10.584 &     0.913 &          \\
\textbf{BE3      } &   -5.593 &     0.350 &          \\
\textbf{AMMIn    } &   16.177 &     1.541 &       0.000 \\
\textbf{AMMOut   } &  -16.177 &    -1.541 &  -6,370.051 \\
\textbf{TOTAL NET} &   -0.000 &    -0.000 &  -6,370.051 \\
\bottomrule
\end{tabular}

\caption{Trade instruction (triangle and pairs)}
\label{tbl:3_arb3x3instr}
\end{table}

Finally, in figure \ref{fig:arbtriangle3c} in the left column, we look
at a triangular scenario with multiple curves per pair, presenting both
pairwise arbitrage opportunities (the stars are not on the same straight
line from the origin) and triangular ones (the products of the prices
once aligned do not match). In the right column of figure
\ref{fig:arbtriangle3c} we present the post arbitrage scenario, and the
associated instructions are in table \ref{tbl:3_arb3x3instr}. Note that
the stars indicating the current states of the AMMs are not all aligned:
those on the unlevered curves are, but some of the levered curves are
stuck at the boundary closest to the relevant price point. Again, the
circular price equation is satisfied but only for the aligned prices
corresponding to interior points on the curves.

\subsection{Convergence issues}\label{convergence-issues}

\label{sec:cvxconverg}

\begin{figure}[htbp]
    \centering
    \includegraphics[width=\textwidth]{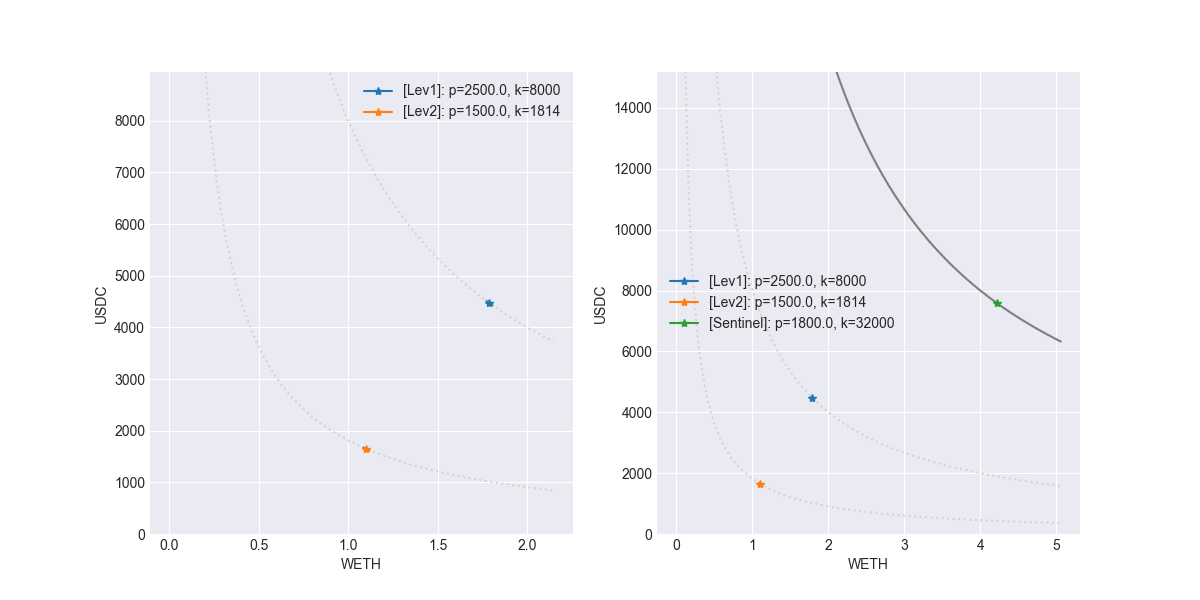}
    \caption[Example for curves where convergence fails]{\textit{Example for curves where convergence fails. }The left hand panel shows two levered curves that are in the money against each other and where the algorithm fails. Adding a sufficiently large sentinel curve (right panel) allows the algorithm to converge, with no trading on the sentinel curve.}
    \label{fig:problemcurves}
\end{figure}

In scenarios like the one above, that are dominated by unlevered or
sufficiently wide levered curves, we have found that the algorithm
converged well. It however ran into issues in scenarios dominated by
narrow levered curves. We have shown a simple example for those types of
curves in figure \ref{fig:problemcurves}. On the left hand panel we see
two levered curves that are in the money against each other: one curve
buys WETH at around 2,500, one sells it at around 1,500, with an overall
profit opportunity of around 17 USDC. None of our convex solvers would
converge with that problem. However, if we add a reasonably sized
unlevered \emph{sentinel curve} -- the unlevered curve added in the
right panel of figure \ref{fig:problemcurves} -- then convergence
succeeds, even though the sentinel curve does not in this case
participate in the arbitrage trade, proving that convergence on this
problem should be possible\footnote{Looking ahead, the marginal
price optimizer that is subject of this paper and that we describe in
section \ref{sec:margpframework}, however, does converge well in this
case; failure of the convex algorithm to converge on curves that we were
interested in, rather than the improvement in speed was the original
reason why we investigated the development of a new algorithm to replace
convex optimization} even without the sentinel curve. The associated
trade instructions are in table \ref{tbl:3_problemc_instr}.

\begin{table}[htbp]  
\centering
\small

\begin{tabular}{lrr}
\toprule
{} &     WETH &     USDC \\
\midrule
\textbf{PRICE    } &  1,800.2 &      1.0 \\
\textbf{Sentinel } &   -0.000 &    0.502 \\
\textbf{Lev1     } &    0.018 &  -44.947 \\
\textbf{Lev2     } &   -0.018 &   27.268 \\
\textbf{AMMIn    } &    0.018 &   27.769 \\
\textbf{AMMOut   } &   -0.018 &  -44.947 \\
\textbf{TOTAL NET} &   -0.000 &  -17.178 \\
\bottomrule
\end{tabular}

\caption{Trade instructions (levered curves with sentinel)}
\label{tbl:3_problemc_instr}
\end{table}

We have not formally shown this, but we suspect that the issues we see
are related to the phenomenon we have hinted at in a very simple case in
section \ref{sec:optimization}: there is a difference between how to
deal with smooth constraints and general linear constraints.
Specifically, in the smooth case, a gradient descent method can be used,
or some other numeric solver which allows us to solve for the condition
that \emph{the gradient of the constraint equals the gradient of the
target function}. For piecewise-linear constraints this no longer works,
and we need to use a method like the simplex method \cite{nelder65} that
jumps between the vertices of the feasible region. In practice, our
constraints are piecewise smooth, meaning sometimes we get interior
solutions like in the smooth case, and sometimes we get corner solutions
like in the linear case\footnote{See appendix \ref{app:radome} for a
description of the \emph{radome optimization problem} that is a very
similar geometric problem}. We have not found a convex solver that
that can be relied upon to consistently perform well\footnote{We
note that because we developed the marginal price optimization algorithm
before the publication of \cite{diamandis23} we never tried the
conjugate algorithm developed there} under those circumstances.

Financially, the smooth vs corner solution cases are easily understood:
an interior solution is a situation where -- in the region of interest
-- the curves provide liquidity in both directions, and no abrupt
changes in liquidity occur when moving prices. On the other hand, a
solution where levered curves trade, but end up at their boundary is a
corner solution. An example for an interior solution is any collection
of Uniswap v2 pools, or Uniswap v3 pools with sufficient liquidity in
and around the current tick that the trades do not fully empty the
current-tick pool in one direction. The archetypical example for a
corner solution are Carbon DeFi limit orders that are in the money
against each other (eg one curve selling 1 TKN at 100 USDC and another
curve buying it at 105 USDC). This is a highly non-smooth problem where
no gradient method will work. However, simplex methods usually work just
fine.

The specific issue for using a convex optimization algorithm in a
production environment is that it is very hard to predict which case we
are in before we have actually solved it. We go back to the \emph{``sell
TKN at 100 USDC, buy it at 105''} example. If both curves are
sufficiently narrow, the transaction will bring at least one of the two
curves to its boundary, meaning that after the transaction at least one
of the curves is either completely empty or completely full. However, if
the curves are wide enough, then this does not happen. Instead, the
transaction will stop at the point where the marginal prices on the two
curves coincide, and, therefore, where every additional dollar
transacted would have a negative marginal contribution to the
transaction profit.

\begin{figure}[htbp]
    \centering
    \includegraphics[width=\textwidth]{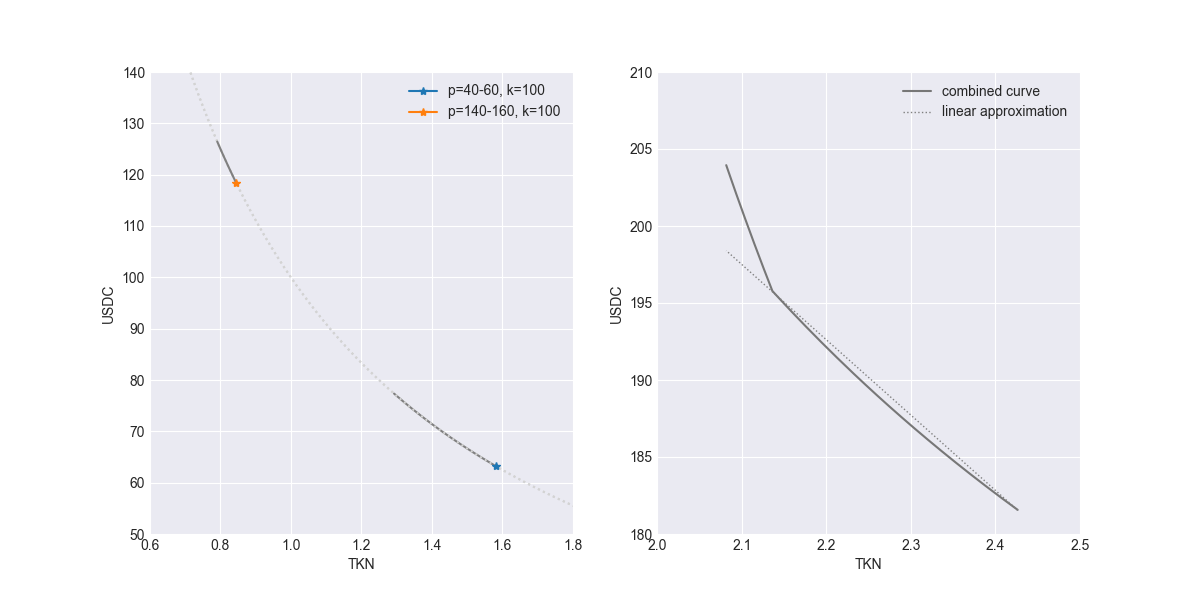}
    \caption[Combining two invariant curves]{\textit{Combining two invariant curves. }The left panel shows two equally sized, levered invariant curves covering disjoint price ranges. The right panel shows the combined invariant curve (the dotted straight line shows curvature of the combined curve).}
    \label{fig:curvescombined}
\end{figure}

To finish this section, we show some examples of problematic curves. In
principle we could have chosen figure \ref{fig:problemcurves} for this,
but for clarity of exposure we use here a different set of numbers. In
figure \ref{fig:curvescombined} we show two curves that cover
non-overlapping price ranges on the same pair. The ranges are relatively
wide, and therefore, the prices change substantially within the curve.
On the left hand panel we show the invariant function of the constituent
curves, and on the right hand one we show the combined invariant curve.
We have also drawn a straight line as a benchmark for the right segment
of the curve, showing that the actual combined curve is indeed not
straight\footnote{Compare the curvature of the solid black line
against the straight dotted line in the right curve segment}. What
will happen in this case is that for post-arbitrage market prices
between 40-60 and 140-160 the solution is an interior solution on one of
the curves (left) or curve segments (right), and for prices below 40,
between 60-140 and above 160 the solution is a corner solution where
first one and then the other curve flips its liquidity to the other
boundary. Depending on the regime we are in, a gradient or simplex
method is better suited to find the solution. See the appendix
\ref{app:radome} for a discussion of a closely related geometric problem
in higher dimensions, the ``radome optimization problem''.

\section{Marginal price optimization}\label{marginal-price-optimization}

\label{sec:margpframework}

We have seen in section \ref{sec:convex}, that the arbitrage and routing
problem is a convex optimization problem, and that we therefore should
be able to solve it using standard convex solvers. We have also seen in
section \ref{sec:cvxconverg} that there are potential pitfalls with
respect to levered curves, where convergence may be problematic. For us,
this was particularly problematic, because with the open source solvers
of the CVXPY package we used \cite{cvxpy}, we often were not able to
determine why the algorithm failed on a viable scenario, and what, if
anything, we could have done to make it converge. Ultimately, we were
not able to make them work reliably enough to work in the ArbFastLane
\cite{fastlane} product, which motivated the developepment of an
alternative method. This new method is the main focus of this paper: the
Marginal Price method for arbitrage finding (and routing). It is similar
to solving the conjugate convex problem proposed in \cite{diamandis23}
which was published around the time we put the finishing touches on our
method, and it is based on financial rather than pure mathematical
reasoning and somewhat further optimized for performance.

In this section, we describe the method starting from mathematical first
principles. For this, we first define the mathematical framework and fix
the notations for describing AMMs (section \ref{sec:margpfwamm}), and
then we do the same with respect to arbitrage transactions (section
\ref{sec:margpfwarb}). In section \ref{sec:margpfwcore}, finally, we
move on to our central result -- defining the \emph{Marginal Price
Formulation} of the arbitrage problem in definition \ref{def:margpform},
and showing the equivalence with its \emph{Convex Optimization
Formulation} from definition \ref{def:convexform}, in what we refer to
as the \emph{Core Equivalence Theorem}, theorem \ref{thm:coreequiv}.

\subsection{Definitions and results related to
AMMs}\label{definitions-and-results-related-to-amms}

\label{sec:margpfwamm}

\begin{definition}[Token baskets, Overlap, Larger/Smaller]\label{def:overlap}
If we have a market of tokens $i=1\ldots N$ we define a \textbf{token basket} $\mathbf{x}=(x_i),\ x_i\geq 0$ as a vector of non-negative numbers corresponding to the token holdings in that basket.  Two token baskets $\mathbf{x}$ and $\mathbf{y}$ are \textbf{non-overlapping} if $\forall i: x_i=0 \vee y_i=0$, otherwise the are \textbf{overlapping}.

A token basket $\mathbf{y}$ is said to be \textbf{larger than} a token basket $\mathbf{x}$ (denoted $\mathbf{y} > \mathbf{x}$) iff $\forall i: y_i \geq x_i$, and $\exists i^*: y_{i^*} > x_{i^*}$. The definitions of \textbf{smaller than} and \textbf{larger/smaller or equal} are analogous.
\end{definition}

It must be understood that the relation \(\mathbf{y} > \mathbf{x}\) only
introduces a \emph{partial order} on token baskets: for example there is
no relation between a basket that holds 1 WETH and one the holds 1,000
USDC.

\begin{definition}[Direction]\label{def:direction}
The \textbf{direction} of a token basket $\mathbf{x}$ is the equivalence class of token baskets that only differ from $\mathbf{x}$ by a scalar factor $\lambda$. In other words, if $\mathbf{x}=(x_i)$ is a token basket then the direction of $\mathbf{x}$ is the set $\{\lambda\cdot\mathbf{x} = (\lambda x_i)\ |\ \forall \lambda\neq 0\}$. We call the set where $\lambda>0$ the \textbf{positive direction} and the set where $\lambda<0$ the \textbf{negative direction}.
\end{definition}

\begin{definition}[Non-improving prices]\label{def:prices}
If we have an entity that exchanges a token amount $x>0$ into an amount $y>0$ of a different token then we define the \textbf{marginal price} at location $x$ in units of $y$ as the derivative $p_{\mathrm{marg}}(x) = \frac{dy}{dx}(x)$. Marginal prices are referred to as \textbf{non-improving} if whenever $x_2 > x_1$ then $p_{\mathrm{marg}}(x_2) \leq p_{\mathrm{marg}}(x_1)$.

This definition can be extended to token baskets in the obvious manner\footnote{We can reduce the multi-dimensional problem to a one-dimensional one by choosing two non-overlapping directions represented by $\mathbf{x}, \mathbf{y}$ where $\mathbf{x}$ represents the inputs and $\mathbf{y}$ the outputs. If we fix the multiplier $\lambda_x$ then the AMM will determine the multiplier $\lambda_y$ such that it considers the exchange of $\lambda_x\cdot\mathbf{x}$ against $\lambda_y\cdot\mathbf{y}$ as fair. Non-improving prices in this context means that the "marginal basket exchange rate" $\frac{d\lambda_y}{d\lambda_x}$ is decreasing in $\lambda_x$, ie whatever $\mathbf{x}, \mathbf{y}, \lambda_x$ we have $\frac{d^2\lambda_y}{d\lambda_x^2} < 0$}, and if the non-improving prices condition holds for all token baskets $\mathbf{x}, \mathbf{y}$ then we define this as \textbf{non-improving prices in every direction}.
\end{definition}

In other words, non-improving prices mean that the price for the
\emph{``second dollar sold''} cannot be better than the one for the
\emph{``first dollar sold''}, both for tokens and token baskets.
However, the prices can be the same.

\begin{definition}[AMM]\label{def:amm}
An Automated Market Maker, short \textbf{AMM}, operating within a market of tokens $i=1\ldots N$, is formally defined as an entity that exchanges a token basket $\mathbf{x}$ against a non-overlapping token basket $\mathbf{y}$ at non-improving prices in every direction. An \textbf{unlevered} AMM covers the full range of prices $(0,\infty)$ in every direction\footnote{Covering the full range of prices means that the function $\lambda_y(\lambda_x)$ is defined for all $\lambda_x\in(0,\infty)$ and bijective}. Any AMM that does not satisfy this condition is referred to as a \textbf{levered}.

AMMs can be \textbf{bi-directional} if they trade in both directions and traded tokens automatically move "to the other side of the curve", or they can be \textbf{directed} if tokens can only be traded once. Bi-directional AMMs can impose \textbf{trading fees} or not\footnote{Note that financially there is no difference between fees and a bid ask spread, and for directed AMMs the notion of fees -- the difference between the buying and selling price of the AMM -- does not make sense}.
\end{definition}

\begin{definition}[Bonding curves and invariant functions]\label{def:bonding}
An AMM is said to be driven by a bonding curve if, ignoring fees and gas, there exists a function $f$ (the so-called "\textbf{bonding curve}" or "\textbf{invariant function})" of all token baskets into the real numbers where the AMM is indifferent between all states that have $f(\mathbf{x}) = k = \mathrm{const}$.
\end{definition}

In other words, if the AMM is currently in the state represented by
\(\mathbf{x}\), and the basket \(\mathbf{y}\) represents a state such
that \(f(\mathbf{y})=f(\mathbf{x})=k\) then it will accept the basket of
all tokens for which \(y_i - x_i > 0\) in exchange for the basket of all
tokens where \(y_i - x_i < 0\).

\begin{proposition}[Convexity]\label{prop:convex}
An AMM driven by a bonding curve $f$ exhibits non-improving prices in all directions according to definition \ref{def:prices} iff the function $f(\mathbf{x})$ is convex in all directions corresponding to the tokens the AMM covers, and constant in the others.
\end{proposition}

\textbf{Proof.} Firstly, the fact that \(f\) must be constant (and is
therefore just on the border of being convex) in the directions of
tokens the AMM does not cover is trivial. The remainder is a well known
result (eg referred to in \cite{angeris21}). To sketch a full proof, in
the one dimensional case it is easy to see from the definition of the
price \(p\) which corresponds to the negative of the first derivative
\(p=-f'\), and the fact that for a convex function we have the second
derivative \(f''>0\). In the higher dimensional case we note that this
must hold in every token direction and can therefore be reformulated
with the \(\lambda_y(\lambda_x)\) functions defined above.
\(\blacksquare\)

Those definitions coincide with the AMM examples provided in section
\ref{sec:AMM} where equation \ref{eq:bancor} is an example for a the
bonding curve of a multi-asset AMM operating on token balances vectors
\(\mathbf{x}, \mathbf{y}\), and both equations \ref{eq:invariant} and
\ref{eq:invariantl} are examples for bonding curves of single-asset AMMs
operating on single token balances \(x,y\). We note that all those
bonding curves are convex. For levered AMMs, the values of \(x,y\) have
to be restricted to the range covered by the AMM.

\subsection{Definitions and results related to
arbitrage}\label{definitions-and-results-related-to-arbitrage}

\label{sec:margpfwarb}

\begin{definition}[Arbitrage-free]\label{def:arbfree}
A market consisting of tokens $i=1\ldots N$ is \textbf{arbitrage free} iff there is no sequence of possible trades that starts with a token basket of $\mathbf{x}$ and ends with a token basket of $\mathbf{y}$ where $\mathbf{y}>\mathbf{x}$.
\end{definition}

Note that in the presence of unlimited and cost free \textbf{flash
loans} (ie loans that can be taken out in unlimited size for all tokens
and that have to be repaid at the end of the transaction) the above
condition can be simplified to the requirement that it is not possible
to start with a balance of \(\mathbf{x}=\mathbf{0}\) and end with a
balance of \(\mathbf{y}>\mathbf{0}\) where the last inequality is to be
interpreted in line with definition \ref{def:overlap}.

\begin{definition}[Circularity]\label{def:circ}
A set of prices $p(i,j)$\footnote{The price convention is that $p(i,j)$ is the price of token $i$ expressed in units of token $j$} between tokens $i=1\ldots N$ is said to satisfy the \textbf{circularity condition} iff for all closed loops $(i_1, i_2, \ldots, i_l, i_1)$ of any length $l$ we have

$$
p(i_1,i_2)
\cdot
p(i_2,i_3)
\cdot\cdots\cdot
p(i_{l-1},i_{l})
\cdot
p(i_{l},i_{1})
= 1
$$
\end{definition}

In other words, the circularity condition states that if we exchange
infinitesimal amounts\footnote{Infinitesimal amounts because
curve-based AMMs will continuously adjust prices to the worse so this
condition will only ever hold in the limit; we sometimes in a slight
abuse of notation refer to infinitesimal transaction amounts as ``first
dollar traded''} of tokens along any closed loop, then we end up with
the same token amount we started with.

\begin{proposition}[Arbitrage-free unlevered AMMs]\label{prop:arbulamm}
A set of bi-directional unlevered AMMs that charge no fees and that are operating on a token market $i=1\ldots N$ is \textbf{arbitrage free} iff (1) for each token pair $(i,j)$ and for all AMMs that allow trading that pair they do so at the same marginal price $p_{\mathrm{marg}}(i,j)$ and (2) the marginal prices satisfy the circularity condition of definition \ref{def:circ}.
\end{proposition}

\textbf{Proof.} Before we move on to prove this we point out that this
is a well known result in finance, and we refer to \cite{hull93} as one
of many examples. However, as the proof is directly linked to the main
topic of this paper we present it here anyway, in a very condensed form.
Starting with (1), we note that bi-directional unlevered AMMs that do
not charge fees buy and sell at exactly the same price. Therefore, if we
have two AMMs that operate on the same token pair at two different
prices \(p_1\neq p_2\) then buying low, selling high would be an
arbitrage transaction. Note that here the convexity condition is
important, ie the first dollar traded must always be at the best price,
otherwise we may be able to get additional arbitrages by trading a
larger amount. Similarly in (2), we first note that we can move along
the loop in either direction, and resulting price in one direction
\(p_+\) will be the inverse of the price in the other direction, ie
\(p^+=1/p^-\). Unless the circular product is unity we can always choose
a direction in which the circular product in definition \ref{def:circ}
is strictly greater than one, and thus we can make an arbitrage profit
by trading along the loop in that direction. This concludes the proof in
both directions. \(\blacksquare\)

\begin{proposition}[Numeraire]\label{prop:numeraire}
A set of tokens $i=1\ldots N$ with marginal prices $p_{\mathrm{marg}}(i,j)$ satisfies the circularity condition iff there exists a sequence of positive numbers $(p_i)$, $i=1\ldots N$, typically referred to as "prices", such that for all token pairs $i,j=1\ldots N$ we have $p_{\mathrm{marg}}(i,j) = p_i/p_j$.
\end{proposition}

\textbf{Proof.} Proving that circularity follows from the existence of
the \(p_i\) is straightforward, replacing \(p_{\mathrm{marg}}(i,j)\)
with \(p_i/p_j\) in definition \ref{def:circ} and observing that all
terms cancel out. Going the other way, we define
\(p_i \equiv p_{\mathrm{marg}}(i,1)\) as the price of token \(i\) in
terms of the \emph{``numeraire token''} \(1\). For every token pair
\(i,j\) we can look at the loop \((1, i, j, 1)\). Because it is a closed
loop, the product of marginal prices is unity, and therefore we have

\begin{equation}\begin{aligned}
p_{\mathrm{marg}}(i,j) 
= \frac{p_{\mathrm{marg}}(i,1)}{p_{\mathrm{marg}}(j,1)}
\equiv p_i/p_j
\end{aligned}\end{equation}

which concludes the proof. \(\blacksquare\)

We note that the numeraire \(p_i\) is only unique up to a multiplicative
constant \(\lambda\), meaning that if \(p_i\) is a numeraire then
\(\bar p_i = \lambda p_i\) is also a numeraire that yields the same
prices. If for a given token \(i^*\) we chose the constant \(\lambda\)
such that \(p_{i^*}=1\) then we refer to this as \emph{``token \(i^*\)
being the numeraire''}. Given the above proposition we define the
concept below that allows us to abstract prices from a specific
numeraire:

\begin{definition}[Price vector]\label{def:pvec}
Given a set of tokens $i=1\ldots N$ we define the \textbf{price vector} $\pi=(\pi_1,\ldots,\pi_N)$ as the vector of prices in a specific numeraire. We also define the associated \textbf{price function}

$$
\pi_{ij} = \pi_i/\pi_j
$$

We define an equivalence relation between price vectors that we denote "$=$" where two price vectors $\pi^a$ and $\pi^b$ are equivalent iff their price functions coincide, ie

$$
\pi^a = \pi^b : \Longleftrightarrow \forall i,j: \pi^a_{ij} = \pi^b_{ij}
$$

meaning the ratios $\pi^a_i/\pi^a_j=\pi^b_i/\pi^b_j$ are the same .
\end{definition}

The purpose of the above definition is to abstract the price information
in a fully arbitraged market. We note that the price vector \(\pi\)
itself should never be used directly because it is only defined up to a
multiplicative constant. Instead, all usage of \(\pi\) should respect
the associated equivalence relation which can be assured by always using
the associated price function \(\pi_{ij}\) instead of the components
\(pi_i\).

However, the price vector \(\pi\) is a valid mathematical object that
resides in the reduced dimensional space where all prices are positive
and where the numeraire token is fixed at unity (ie
\(\pi\in R_{>0}^{N-1}\times\{1\})\).

At this stage, we are ready to move on to prove a more general
proposition that covers levered and unlevered AMMs:

\begin{theorem}[Arbitrage-free AMMs]\label{thm:arbamm}
A set of  levered or unlevered AMMs $\nu=1\ldots M$ operating on a token set $i=1\ldots N$ is \textbf{arbitrage free} iff there exists a vector of prices $p_i$ so that for every AMM trading a token pair $i,j$ the marginal price satisfies $p^\nu_{\mathrm{marg}}(i,j) \stackrel{\rightarrow}{=} p_i/p_j$ where the symbol $\stackrel{\rightarrow}{=}$ indicates that the marginal price on the left is the closest approximation to $p_i/p_j$ within the set of attainable prices of AMM $\nu$.
\end{theorem}

\textbf{Proof.} First we note that, in case of unlevered AMMs, this
proposition reduces to proposition \ref{prop:arbulamm} because (a) the
closest attainable price will simply be \(p_i/p_j\), so (b) all AMMs
exchanging tokens \(i,j\) will be set at the same price \(p_i/p_j\) that
(c) satisfies the circularity condition in definition \ref{def:circ}
because of proposition \ref{prop:numeraire}. For a levered AMM, we
firstly note that if the current marginal price is not at a boundary,
for small trades the levered AMM behaves like an unlevered AMM,
therefore the same reasoning applies and its marginal price must be at
\(p_i/p_j\) as in the unlevered case. If the price of the AMM was at the
boundary away from \(p_i/p_j\) then someone trading against the AMM
could buy low at \(p_i/p_j\) from another curve and sell high into the
AMM stuck at the far boundary, thereby moving the price closer towards
\(p_i/p_j\). The only point where this trade is not possible is at the
boundary closest to \(p_i/p_j\) because at this point the AMM will no
longer buy. \(\blacksquare\)

We next we define the \emph{price response function} (``\textbf{PRF}'')
that indicates how an AMM -- or a set of AMMs -- responds to a change in
price(s):

\begin{definition}[Price response function]\label{def:prf}
Given a set of AMMs $\nu=1\ldots M$, a set of tokens $i=1\ldots N$, the \textbf{individual PRF} $\mathbf{\rho}_\nu$ of AMM $\nu$ is an equivalence-respecting\footnote{In line with definition \ref{def:pvec}} function that maps a price vector $\pi$ to a set of token changes

$$
\rho_\nu(\pi) = (\Delta x_{\nu 1}, \ldots, \Delta x_{\nu N})
$$

The \textbf{aggregate PRF} of a set of AMMs $\mathbf{\rho}$ is the sum of the individual PRFs

$$
\rho(\pi) = \sum_{\nu=1}^{M} \rho_\nu(\pi)
$$

By convention, outflows from the AMM are negative, and inflows are positive, therefore the pre-trade balances $\mathbf{x}^{\mathrm{pre}}$ and post trade balances $\mathbf{x}^{\mathrm{post}}$ satisfy

$$
x^{\mathrm{post}}_{\nu i} = x^{\mathrm{pre}}_{\nu i} + \Delta x_{\nu i}
$$
\end{definition}

Arguably, the PRFs are the most important financial objects that we are
dealing with. They are equivalent to, but more financially relevant
than, the usual invariant functions in equations \ref{eq:bancor},
\ref{eq:invariant}, and \ref{eq:invariantl}. For a traditional AMM
without fees they are path independent, meaning that aggregating the PRF
results over any price path \(\pi^{{1}}, \pi^{{2}}, \ldots \pi^{{N}}\)
is the same as going directly to \(\pi^{{N}}\), the end point\footnote{This follows directly from the derivation of the PRF from
an invariant curve}. However, in the presence of fees, things change.
Firstly, longer paths lead to a higher fee bleed. Moreever, if fees
accumulate on the curve, the invariant curve changes and therefore, so
does the PRF. Finally, in a directed AMM like Carbon DeFi, curves do not
automatically \emph{reload}\footnote{Except via the associated
counter curve in the Carbon DeFi ``Strategy'', if so desired}, so in
this case the PRF is usually highly path-dependent.

\begin{definition}[Trade instructions]\label{def:ti}
A \textbf{trade instruction matrix} ("\textbf{TIM}" or simply "\textbf{trade instructions}") for a set of AMMs $\nu=1\ldots M$  operating in the token set $i=1\ldots N$ is a matrix $(\Delta x_{\nu i})$ that describes the flows of token $i$ into (positive) and out of (negative) the AMM $\nu$. A TIM is called "respecting a set of self financing constraints" if its aggregate over all $\nu$ fulfils the constraints from equation \ref{eq:sfconstrr}. It is called an "\textbf{arbitrage TIM}" if it fulfils the arbitrage SFC in equation \ref{eq:sfconstr}. In those two cases we refer to  $\sum_\nu \Delta x_{\nu i}$ as the \textbf{result}\footnote{This terminology is driven by the usage of the term "result" within an convex optimization context} of the arbitrage finding or routing process. In case of a pure abitrage, the negative result (a positive number) is also referred to as the \textbf{arbitrage profit}.
\end{definition}

\subsection{The Core Equivalence
Theorem}\label{the-core-equivalence-theorem}

\label{sec:margpfwcore}

We are now ready to present the mathematical core of this paper, the
claim that the marginal price optimization problem described in this
paper which forms the basis of operations for the FastLane Arbitrage bot
\cite{fastlane} is equivalent to the convex optimization problem
described in section \ref{sec:convex} based on \cite{angeris21}. To do
this, we first formally define the ``Marginal Price Formulation'' of the
problem:

\begin{definition}[Marginal Price Formulation]\label{def:margpform}
The "\textbf{Marginal Price Formulation} of the Arbitrage Problem" is the root finding problem to identify the price vector $\pi$ that satisfies the \textbf{arbitrage condition} 

$$
\rho(\pi)=-\lambda e_0
$$ 

where $e_0$ is the unit vector of the zeroth vector component\footnote{Again like in the footnote to equation \ref{eq:target} the choice of token 0 is WLOG and for simplicity of presentation only; this condition states that this is a pure arbitrage transaction where the profits, if any, are taken in token 0} and the "result" (in the sense of definition \ref{def:ti}) $\lambda\geq 0$ is a scalar determined by the algorithm via the trade instruction matrix that we will show to be non-negative below in proposition \ref{def:exuniq}. 

Note that whilst the optimal routing problem is out of scope in this paper we still want to record the fact that for routing the above equation will become

$$
\rho(\pi)=-\lambda e_0 + \mathbf{x}
$$ 

where the vector $\mathbf{x}$ are the desired in and outputs in tokens other than $0$ like in equation \ref{eq:sfconstrr}.

\end{definition}

We now show that the above \emph{Marginal Price Formulation} is
equivalent to the \emph{Convex Optimization Formulation} that was the
subject of section \ref{sec:convex}, and specifically defined in
definition \ref{def:convexform}. En passant we note that this is
essentially the well-known result from convex optimization theory
linking a problem and its conjugate as used in \cite{diamandis23}, but
as our focus is on finance as opposed to pure mathematics we want to
provide a self-contained proof more financial proof.

\begin{theorem}[Core Equivalence]\label{thm:coreequiv}
The problem of finding arbitrages (or route optimally) in a set of AMMs in the Marginal Price Formulation as described in definition \ref{def:margpform} is equivalent to the Convex Optimization Formulation as described in definition \ref{def:convexform}. Specifically, the trade instruction matrix (and therefore the arbitrage profit) obtained by both formulations will be the same.
\end{theorem}

\textbf{Proof.} To prove the above, we start with creating a list of
items where the two formulations coincide, and where they differ. The
formulations coincide in the following items

\begin{enumerate}
\def\labelenumi{\arabic{enumi}.}
\item
  They are solving the same arbitrage finding problem as presented in
  definition \ref{def:arbproblem}\footnote{For ease of presentation
  we here focus on arbitrage alone and leave the simple extension of the
  proof to routing application to the reader}.
\item
  They start with a set of AMM curves satisfying invariant equations
  along the lines of definition \ref{def:bonding}, and holding the
  associated amounts of tokens pre-arbitrage.
\item
  They seek a set of post-arbitrage token holdings, or equivalently a
  trade instruction matrix along the lines of definition \ref{def:ti},
  that implements the arbitrage transaction, and that satisfies the self
  financing constraints according to equation \ref{eq:sfconstr}.
\end{enumerate}

The convex optimization formulation (``COF'') and marginal price
formulation (``MPF'') differ in the following way:

\begin{enumerate}
\def\labelenumi{\arabic{enumi}.}
\item
  The COF seeks to minimize a target function in line with equation
  \ref{eq:target} whilst the MPF seeks to find the root according to
  definition \ref{def:margpform}.
\item
  The COF algorithm operates directly on the AMM token holdings
  \(x_{\nu i}\) whilst the MPF algorithm operates on a marginal price
  vector \(\pi_i\) according to definition \ref{def:pvec}.
\end{enumerate}

To prove equivalence, we have to either (a) show that they are the same,
or that (b) the MPF solution satisfies the COF conditions and vice
versa. We start with a non-rigorous argument for (a) by pointing out
that both are solving the financial real-world arbitrage problem
\ref{def:arbproblem} that, in a non-path-dependent environment, has a
unique solution\footnote{Unique in the sense that if there were
multiple solutions one could always trade from one to the other so any
algorithm that gets stuck on a sub-optimal solution does not in fact
solve the arbitrage problem; there can, however, be cases where more
than one set of trade instructions yields the same arbitrage profit in
which case both algorithms may find either of them depending on starting
conditions and algorithm details}. However, we have pointed out before
that Carbon DeFi positions introduce path dependence, and we can only
accept this as a proof when no directional curves are in the curve set.

For proving the theorem along the lines of (b) we first show that the
optimal solution in the COF framework satisfies the MPF conditions. For
this we point out that if marginal prices were not to satisfy the price
conditions in proposition \ref{prop:arbulamm} then additional profits
could be generated by ``buying low selling high'' and therefore the
convex optimization process had not worked as advertised\footnote{We
had some concern in cases where \emph{profit repatriation} was an issue,
but as we have shown in detail in appendix \ref{app:repatriation} that
is is not in fact the case}.

Now we go the other way and show that a solution of the MPF satisfies
the constraints of the COF and minimizes the target function. By design,
the MPF solution satisfies the positivity constraints from equation
\ref{eq:positivity}, the self-financing constraints from equation
\ref{eq:sfconstr} and curve constraints from either equation
\ref{eq:curveconstr} or \ref{eq:curveconstrl}. To show that this also
minimizes the (negative) target function we note that by construction
the state of the market is arbitrage free, and the existence of a state
with a bigger outflow under the same self-financing constraints would
imply that there were arbitrages available in the initial state.
\(\blacksquare\)

\begin{proposition}[Existence and uniqueness]\label{def:exuniq}
A solution to the routing problem based on equation \ref{eq:sfconstrr} may or may not exist, depending on the state of the AMMs and the routing constraints $\mathbf{x}$. A solution to the arbitrage problem based on equation \ref{eq:sfconstr} (ie, $\mathbf{x}=\mathbf{0}$) always exists and the result parameter $\lambda$ from definition \ref{def:margpform} will be positive, or zero if there is no arbitrage opportunity. If a solution exists, it will be unique.
\end{proposition}

\textbf{Proof.} For the \emph{may or may not exist} part, nothing needs
to be proved. All other properties follow directly from the Core
Equivalence Theorem \ref{thm:coreequiv}, and the fact that the empty
solution (zero everywhere) will either dominate the convex solution, or
be the solution. \(\blacksquare\)

We want to briefly elaborate on the \emph{may or may not exist} part,
using financial arguments. Firstly, unlevered curves can take up any
number of tokens, so any routing constraint pushing tokens into the
system will not usually be a problem. Provided there is a route from all
inputs to the desired output token, the routing problem will always have
a solution. Constraints taking tokens out of the system are limited by
the number of tokens available in the system however, so if this number
is being increased, at one point there will no longer be a solution.
Tokens that only live on levered curves will also have a maximum amount
that can be pushed in.

\section{Implementation and
convergence}\label{implementation-and-convergence}

\label{sec:margpimpl}

Whilst the marginal price algorithm is in principle the same on token
pairs and on token sets containing more than two tokens, there are
important numerical differences. Most importantly, on token pairs the
problem is a one-dimensional root finding problem, and according to the
intermediate value theorem \cite{abbott15} we are guaranteed to find a
root\footnote{Or rather, a root location, if we also consider step
functions that may not technically have a root, a case that is of
practical importance for us; see appendix \ref{app:bisection} for
details} if we can bracket it. Convergence will be in logarithmic time
-- every step will increase the precision by a factor of two.

As we discuss in appendix \ref{sec:hidim}, in higher dimensions there is
no equivalent to the the intermediate value theorem, and therefore no
equivalent to the bisection method that is guaranteed to converge. In
fact, a priori it is not clear that a solution even exists: we have a
function \(f:R^n \to R^n\) and we are looking for a point \(\mathbf{x}\)
such that \(f(\mathbf{x}) = \mathbf{0}\). In general, such a point does
not always exist, and if it exists it is not clear that it is unique.

Having said this -- our problem is more benign than the general
mathematical framework may suggest. After all, we are solving a real
world problem in finance, and as shown in \cite{angeris21}, the problem
is convex and therefore has a unique solution. Specifically, the
arbitrage problem is always dominated by the \emph{null solution}, so
either either a proper arbitrage solution exists, or \emph{do nothing}
is the formal solution to the arbitrage problem. The issue is therefore
less the question of existence and uniqueness\footnote{Again,
financially it is clear that trade instructions cannot be unique in the
general case; for example, consider two zero-slippage curves covering
the same pair where any routing of the required amount through the two
curves will be a solution}, but rather the question of how to find the
solution within a reasonable amount of time.

We have split the discussion into multiple parts. First, we discuss the
implementation in the case of a single token pair using bisection in
section \ref{sec:margpimplpair}. We then discuss the generic case using
Newton-Raphson / gradient in section \ref{sec:margpimplgen}, and finally
we deal with the topic of convergence in section \ref{sec:margpconverg}.

\subsection{Marginal price optimization on token
pairs}\label{marginal-price-optimization-on-token-pairs}

\label{sec:margpimplpair}

On token pairs, the arbitrage problem boils down to finding a single
price where the net flow of all other tokens than the target token is
zero for arbitrage, or a specific number for optimal routing\footnote{The marginal price routing algorithm is the one used when
trading on Carbon DeFi via the canonical user interface}.

\begin{figure}[htbp]
    \centering
    \includegraphics[width=\textwidth]{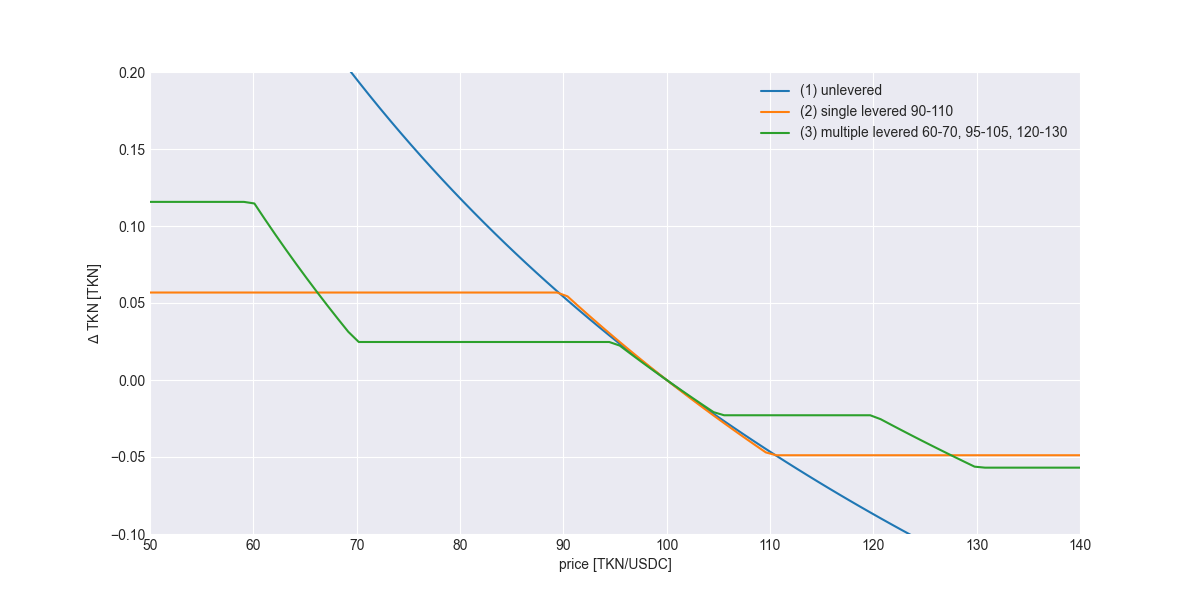}
    \caption[Price response function with no fees]{\textit{Price response function with no fees. }Price response function, defined as change in token amounts against price for (1) a single unlevered curve, (2) a single levered curve, and (3) multiple levered curves, all curves with no fees and all at current price \(p=100\).}
    \label{fig:dtknfromp}
\end{figure}

This means that in two dimensions, our general root finding problem
without fees looks generally like the different graphs depicted in
figure \ref{fig:dtknfromp}: an unlevered curve is a simple convex line
(blue, 1), a single levered curve is a convex segment in between two
flat areas (orange, 2), and multiple levered curves at different prices
correspond to a series of convex segments separated by flat areas
(green, 3).

\begin{figure}[htbp]
    \centering
    \includegraphics[width=\textwidth]{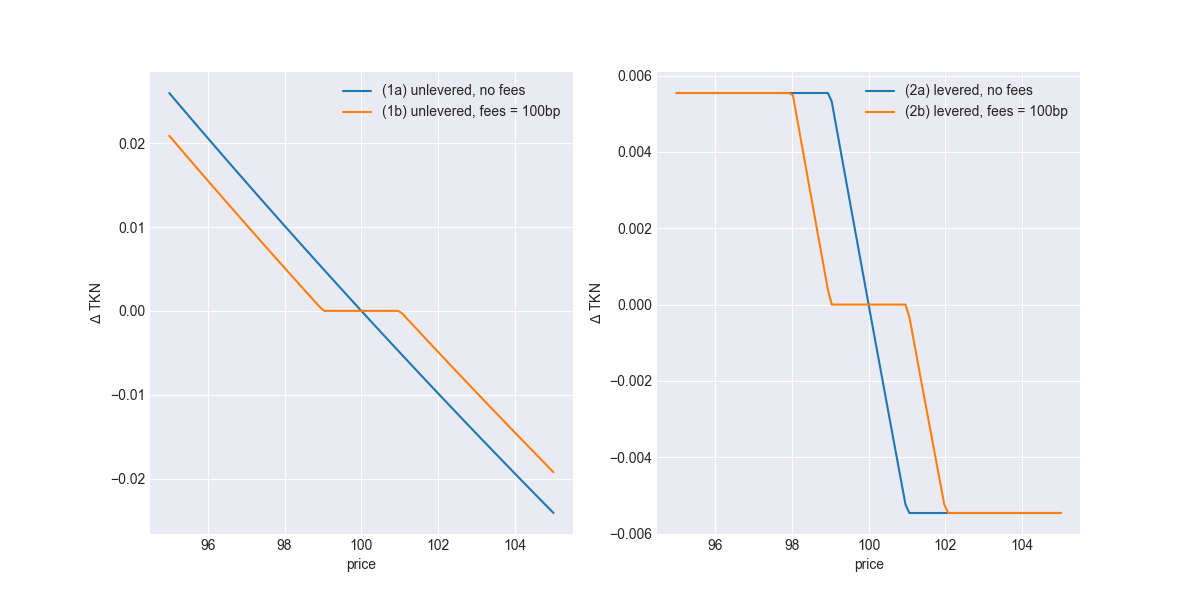}
    \caption[Comparison of price response function with and without fees]{\textit{Comparison of price response function with and without fees. }Price response function, defined as change in token amounts against price for (1) unlevered curves, (2) levered curves, (a) without and (b) with fees, all at current price \(p=100\).}
    \label{fig:dtknfrompfee}
\end{figure}

Note that if we include fees the picture changes considerably in that we
get a flat area inserted at the current price point which depicts the
boundary between buying and selling. The width of the flat area is the
current price multiplied with the percentage fee charged. This is shown
in figure \ref{fig:dtknfrompfee} for an unlevered curve in the left hand
panel and a levered curve in the right hand panel.

For one dimensional root finding problems, there are fundamentally two
methods

\begin{itemize}
\tightlist
\item
  the bisection or bracketing method described in appendix
  \ref{app:bisection}, and
\item
  the Newton-Raphson or gradient descent method described in appendix
  \ref{app:nr}.
\end{itemize}

As discussed in appendix \ref{sec:bsconv}, the bisection method is
extremely robust in that it will either fail right from the start
(because the bracketing did not yield two points with an opposite sign),
or it is guaranteed to converge to the location of either a root in case
the function is continuous (cf left hand panel in figure
\ref{fig:functions}), or a \emph{root location} where the target
function changes its sign in case the function is not continuous and
jumps across the x-axis (cf right hand panel in figure
\ref{fig:functions}).

Depending on the shape of the function, the Newton-Raphson method can be
much faster than the bisection method. As we discuss in appendix
\ref{app:nrconv}, if the function is linear, convergence is in one step.
Generally, on convex functions convergence is fast regardless of the
starting point (see figures \ref{fig:newton} and \ref{fig:newton2} for a
few examples). However, for functions that change convexity, or that are
not continuously differentiable, a number of bad things can happen,
notably the algorithm can go into an infinite cycle (see figure
\ref{fig:newton2} bottom right panel), or the new sampling point can be
catapulted to ``infinity'' when the function is very flat at the current
sampling point, and depending on the shape of the function it can be
impossible to recover from that.

For example, consider the green function (3) in figure
\ref{fig:dtknfromp} in the price area arounf \(p\simeq 80\). The
function is flat and therefore any gradient descent will fail. We could
regularize the function somewhat (eg by adding a small slope to ensure
that the gradient is never zero). However, even this regularization
would lead the algorithm very far out, into another flat area, from
where the regularization would lead it back. In this case we can have
either of two eventual outcomes: the algorithm may eventually end up in
the price area of the curve that contains the root\footnote{This is
the equivalent of ending up by chance an the correct face of the radome
in the example of appendix \ref{app:radome}}, or it may end up in an
infinite cycle. In practice, we will most likely see a large number of
iterations and we will run into our \emph{maximum iteration} boundary.
In any case -- those kind of scenarios are extremely inefficient for a
Newton Raphson algorithm: either it takes a long time to converge, or it
takes even longer until we decide that it does not converge in the
alloted time.

Generally in the pair case, the improved speed of the Newton-Raphson
method in our experience does not outweigh the risk of non-convergence,
so we exclusively use a bisection method when we deal with arbitrage or
routing within a single token pair.

We do need to qualify the robustness claim made above though because,
for non-continuous functions, it only applies in \emph{price-space}.
However we seek to specify transactions that live in
\emph{token-amount-space}, so convergence in price-space is not enough.
This is not an issue with the bisection algorithm specifically, but
rather with converting a token space problem into a price space problem
as is the foundation of the marginal price method. It is particularly
annoying, however, in case of the bisection method, because it somewhat
torpedoes the robustness properties of this algorithm.

The functions that pose problems are those that are discontinuous at
their root location (ie that jump from positive to negative or vice
versa). In practical applications this also includes functions that are
\emph{numerically discontinuous}, with which we mean functions that have
such a steep gradient that, within the resolution of the algorithm, they
appear to be discontinuous. Those functions are easily identified: they
are what in Carbon DeFi \cite{carbon22} we refer to as \emph{``limit
orders''} (ie orders where the parameter \(A=0\) and where the start and
end price are the same). In practice, very narrow ranges with
\(A \gtrapprox 0\) also pose a problem, because they are what we
referred to as ``numerically discontinuous'' above\footnote{Note
that technically \(A\) has the units of the square root of a price, so
it is not a scale free number, and the correct condition for a range
being \emph{``small''} is along the lines of
\(A/B < \sqrt{\varepsilon_0}\) where \(\varepsilon_0\) is some small
number}.

In the introductory section \ref{sec:optimization} on optimization, we
have introduced, in equation \ref{eq:convolution}, the \emph{convolution
method} to regularize a function. We could apply this method here, but
in practice we find it easier to enforce a minimum width for limit
orders. Specifically, we enforce values of \(A,B\) so that they satisfy
the condition above with \(\varepsilon_0 \simeq 10^{-6}\), and we adjust
both \(A\) and \(B\) if this is not the case, ensuring that the
adjustment is such that the effective price of the order, when fully
executed, remains the same.

\subsection{General marginal price
optimization}\label{general-marginal-price-optimization}

\label{sec:margpimplgen}

We now move on to the higher dimensional case, ie everything where more
than two tokens are involved. As discussed in appendix \ref{sec:hidim},
there is no equivalent of the bisection method in dimensions higher than
one, so we are forced to use a multi-dimensional Newton-Raphson method
as described in appendix \ref{app:nrhighdim}. If we have \(N+1\) tokens
\(i=0\ldots N\), this algorithm works as follows:

\begin{enumerate}
\def\labelenumi{\arabic{enumi}.}
\item
  Compute the \(N\times N\) Jacobian matrix \(J_{ij}\), consisting of
  the derivatives of all \(N\) function values -- the self-financing
  constraints of all but the target token according to equation
  \ref{eq:sfconstr} or \ref{eq:sfconstrr} -- to all \(N\) prices in the
  price vector defined in definition \ref{def:pvec}, according to
  equation \ref{eq:jacobian}.
\item
  Invert the Jacobian to solve the linear approximation of the function,
  as described in equations \ref{eq:jacobianapprox} and \ref{eq:update}.
\item
  Update the price vector according to equation \ref{eq:updateeta},
  taking into account the learning rate \(\eta\) if need be.
\item
  Repeat until convergence or failure (maximum iterations or away from
  reasonable domain).
\end{enumerate}

This algorithm looks deceptively easy, but for it to work in a
production setting, a number of points need to be considered, most
importantly the following:

\textbf{Log Prices.} We initially implemented the algorithm using actual
prices. This ran into a number of issues, most importantly that
sometimes the algorithm ended up in a situation where the price was
negative. Also, because price levels in crypto can range from below
\(10^{-6}\) to the dollar to almost \(100,000\), the numerical
conditioning is not ideal, in particular for calculating the derivatives
of the price response function (definition \ref{def:prf}) for the
Jacobian. We therefore switched to log prices; we define
\(\mathbf{x}=\log{\mathbf{\pi}}\) and we perform all calculations for
the algorithm in the log space \(\mathbf{x}\), not in the price space
\(\mathbf{\pi}\).

\textbf{Calculating Derivatives.} We calculate the Jacobian using
perturbations. Because we operate in log space, those are effectively
percentage-perturbations, which means we don't need to worry much about
the size of the calculation stencil. Although, we need to be somewhat
careful if we are at the price boundary of a levered position to ensure
that (a) the algorithm does return a result there, and (b) the result is
sensible\footnote{We currently calculate derivatives along the blue
(sum first) path we will define in diagram \ref{eq:diagramperf} in the
next section, section \ref{sec:perfcomp}. Please refer to the discussion
in that section for the implications thereof}.

\textbf{Curves with no closed form solutions.} Because we calculate the
derivative of the price response function (definition \ref{def:prf})
numerically via perturbation, and this calculation is at the core of the
algorithm, it is important that this calculation is fast and consistent\footnote{Please refer to the next section, specifically
\ref{sec:perfcompnt}, for a discussion how the calculation of the
Jacobian is the main numerical effort of the current algorithm}.
Calculating derivatives from something that itself relies on numerical
approximations is often slow and error prone -- so in cases where we do
not have a closed form solution for the PRF of a specific AMM, we
approximate the curve with sufficient number of levered constant product
AMMs placed next to each other and use those for the calculations.

\textbf{Singular Jacobian.} The Jacobian can become singular, in which
case the naïve algorithm fails. We have implemented a fallback algorithm
that inverts the Jacobian only on its image, and does not attempt
inversion on its null space. This improves performance in higher
dimensional cases, because there is at least a chance that the algorithm
either ends up in a non-singular place, or that the singularity
corresponds to prices that ultimately do not matter for the arbitrage
problem at hand. Note that singularities typically occur if at a certain
price level there are no curves that allow trading the pair
corresponding to that price. Because of the market structure on crypto
markets, where most AMM curves are against one of (W)ETH, WBTC or a USD
stable coin, this is particularly pertinent if an unusual target token
is chosen, which in turn suggests that it may be better to choose one of
the aforementioned tokens as target token.

\begin{figure}[htbp]
    \centering
    \includegraphics[width=\textwidth]{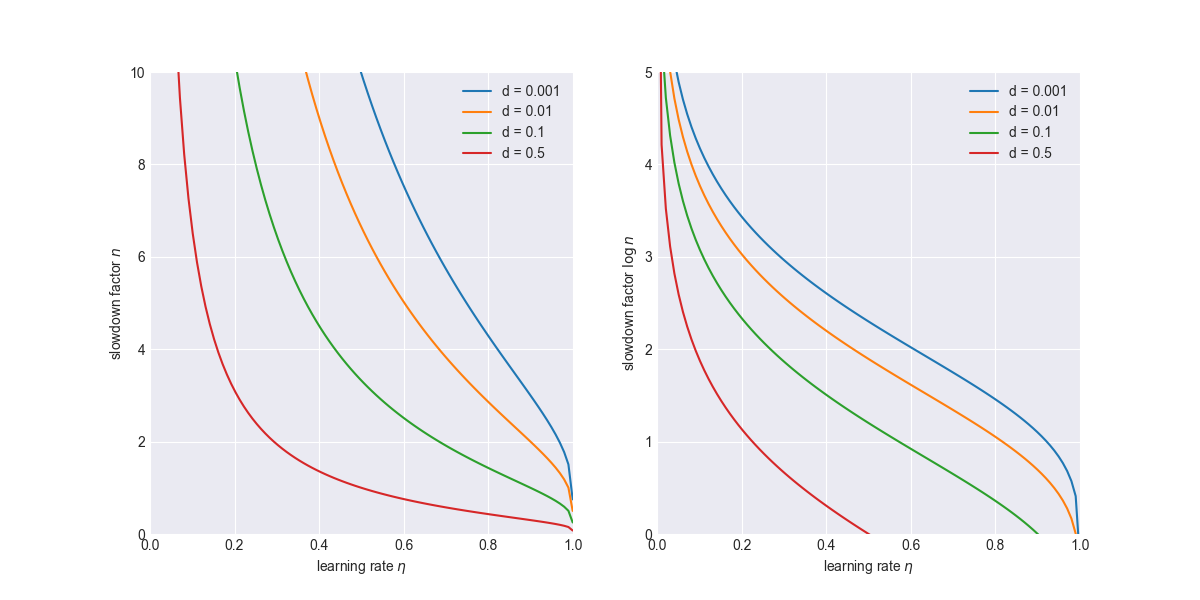}
    \caption[Slowdown in convergence as a function of the learning rate]{\textit{Slowdown in convergence as a function of the learning rate. }Slowdown factor (ie number of iterations required) \(n\) to achieve various convergence levels \(d\) as a function of the learning rate \(\eta\), regular (left) and log scale (right).}
    \label{fig:slowdown}
\end{figure}

\textbf{Learning rate.} We have experimented with changing the learning
rate \(\eta\) to improve convergence, but we found that it did not make
a significant difference. In reality, the biggest issue is that, because
of lack of liquidity in a certain price region, prices are catapulted
towards infinity, and catapulting them to towards ``\(\eta\) times
infinity'' does not seem to lead a significant improvement. On the other
hand, choosing \(\eta<1\) is costly. In figure \ref{fig:eta} in the
appendix we have illustrated the impact of the learning rate \(\eta\) on
the Newton Raphson algorithm. Abstracting the findings from this further
we have plotted in figure \ref{fig:slowdown} the \emph{slowdown factor}
\(n\) for a given \emph{convergence level} \(d\) as a function of the
learning rate \(\eta\). The convergence level \(d \in (0,1)\) describes
the residual distance to the real solution as a function of the original
distance (eg \(d=0.01\) means that the distance has been reduced by a
factor of \(100\)). The slowdown factor \(n\) is the number of
iterations required to reach this level of convergence. We note that
that slowdown factor is a real cost: running the algorithm is
computationally expensive, and the slowdown factor \(\eta\) goes
directly to the running cost bottom line, and it can also affect latency
which matters on fast chains.

\textbf{Convergence criteria.} We have initially used a \emph{relative}
convergence criteria (ie that the algorithm does not change prices any
longer). In some ways this is a good criterion, because we operate in
log space, so the criterion is effectively the \emph{average percentage
change}. The issue is that the algorithm in this case relatively often
indicated it had converged, even though it had simply ended up in a
region with no liquidity at all and, therefore, where the Jacobian was
flat zero. We could have tested for this, but we ultimately chose an
\emph{absolute} criterion along the lines of \emph{``average violation
of the self-financing constraints \textless{} 1 USD''}, which is more
financially meaningful. The downside in this case was that we always
needed to provide USD prices to the algorithm, even if no USD curves
were involved in the arbitrage problem at hand. Also, divergence can
take longer if the algorithm is stuck close to infinity -- relative
convergence would break right away, falsely claiming convergence.
Absolute convergence will eventually rightly report non-convergence, but
it will take longer to do so because it will run empty cycles.

\textbf{Convergence issues.} Regardless of the criteria used, the
algorithm sometimes diverges, even though a solution exists. We have
already discussed the issue in relation to the convex optimization
algorithm in section \ref{sec:cvxconverg}, and whilst the marginal price
algorithm converges better, it still has a substantial failure rate when
very thin curves are involved.

\textbf{Decomposition and linearization.} We will discuss the
linearization and routing problem in our upcoming paper on transaction
decomposition linearization \cite{loesch24b}, so here we present just a
brief introduction to the topic. As we already pointed out in the
example discussed in the intro to section \ref{sec:margpimpl}, the
algorithm generally creates a set of trade instruction that touches
many, and possibly all, curves. Having a transaction with so many curves
has two drawbacks. Most importantly, the more curves are involved, the
more likely the whole transaction is to fail because the blockchain
state has changed by the time the transaction is included. Additionally,
if there are limited flashloan opportunities and we have only limited
token amounts, linearization\footnote{The term ``linearization''
refers to the process of creating a transaction (based on own token
holdings or flashloans) that can be executed in a linear manner without
any intermediate token balances ever falling below zero} of the
transaction can be complex. The latter problem can be dealt with, but
the former remains: we usually want to decompose the transaction into
smaller ones, and prioritize their submission based on profitability and
complexity.

\subsection{Convergence}\label{convergence}

\label{sec:margpconverg}

Convergence of the marginal price algorithm is significantly better than
that of the convex optimization algorithm from \cite{angeris21}. For
example, the example discussed in section \ref{sec:cvxconverg} converges
well to the same result is in table \ref{tbl:3_problemc_instr} without
need for a sentinel curve like in figure \ref{fig:problemcurves}.
However, even the marginal price alorithm can run into convergence
issues. We have thus far identified two scenarios that do lead to
divergence:

\begin{itemize}
\tightlist
\item
  \textbf{escape scenario}: the gradient catapults the algorithm into a
  region where no curves are located, at which point it is either
  blocked, or jumps around erratically
\item
  \textbf{loop scenario}: the algorithm enters an infinite loop without
  ever converging to a solution, or at least not within the maximum
  iterations allowed
\end{itemize}

\begin{figure}[htbp]
    \centering
    \includegraphics[width=\textwidth]{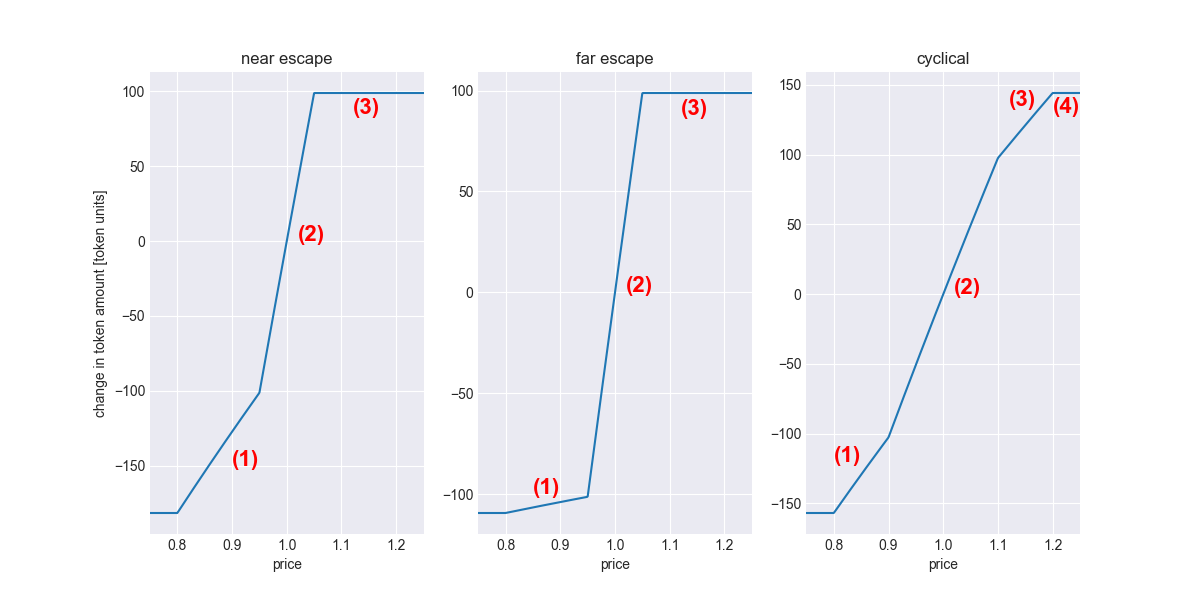}
    \caption[Divergent scenarios for the marginal price optimizer]{\textit{Divergent scenarios for the marginal price optimizer. }The panels above show price response functions where the result is in segment (2) and they diverge when starting in segment (1). The two left panels escape to segment (3), the left one closely, the middle one far away. The right panel enters into an infinite loop (1), (3), (1), etc.}
    \label{fig:margpdivergence}
\end{figure}

Both scenarios are shown in figure \ref{fig:margpdivergence}. All panels
depict price response functions of different scenarios, and in all case
the solution \(dx(p)=0\) is found in segment (2) and convergence is fast
provided the algorithm starts in, or ever reaches, segment (2). In each
of the cases however, whenever the algorithm starts in region (1) it
will diverge.

The left and middle panels show escape scenarios. In the left panel, the
curve segment (1) is relatively steep so the next point will be in
segment (3), close to segment (2). In the middle panel, the curve is
very flat and the escape is towards infinity. In those particular cases
this does not make a big difference -- both algorithms will detect a
zero gradient, at which point they will fail. What this shows however is
that a learning rate \(\eta\) (as discussed in section
\ref{sec:margpimplgen}) may or may not be helpful: Whilst in the left
panel, an \(\eta\simeq 0.8\) would bring us into segment (2) from where
we would converge, in the middle panel we would need a really small
\(\eta\) to not end up in the empty region (3). However, this would be
too detrimental for the overall speed of convergence of the algorithm,
as discussed in appendix \ref{app:nreta}.

The right panel shows a loop scenario similar to the one shown in the
bottom right panel of figure \ref{fig:newton2}. Here, we have three
curve segments, so (4) instead of (3) indicates the segment without
curves. Here, if we start in segment (1), the algorithm will bring us to
(3). From there, we will return to (1) and the cycle restarts. Depending
on the exact starting point and shape of the curve, the algorithm may
eventually escape to segment (4) or its counterpart on the left, may hit
segment (2) and will converge, or it may be stuck in an infinite loop.
In reality this does not usually matter because we cannot afford to run
the algorithm until we find out, so each of those three cases will hit
the maximum-iterations boundary.

Whilst neither of the scenarios converges, in the above two dimensional
case there is an important difference: the escape scenario diverges very
quickly -- as soon as the algorithm enters the empty space it will
detect a zero gradient, at which point it will know it has failed and
will terminate. The loop scenario is harder to detect, and in fact we
only detect it via hitting the maximum iterations threshold. This of
course is expensive, and careful management of this boundary is
important because it will significantly impact the performance of the
algorithm. We note that the bona fide escape scenario is relatively
rare: whenever there is an unlevered curve present in the analysis, the
gradient will never be fully zero, in which case we will end up in a
variation of the loop scenario.

In higher dimensions the issues are different, except that one can have
a convergent and all types of divergent scenarios present at the same
time, depending on the direction. We currently only terminate if the
Jacobian is zero. If it is singular -- meaning that we are in the empty
space in at least one of the directions -- we instead invert the
Jacobian on the invertible sub space only, on the hope that this returns
us to a region where the entire Jacobian is invertible again. We are not
currently certain whether or not this is the right choice because if we
do not quickly return to the core region of convergence we simply waste
more iterations on the problem before we diverge anyway. Currently we
are leaving it in because we value convergence over speed.

\section{Performance comparison}\label{performance-comparison}

\label{sec:perfcomp}

In this section, we report the results of our performance comparison
between the different algorithms. The software and hardware details are
in table \ref{tbl:6_system}. The performance of this machine is
comparable to the cloud servers we use in the production environment, so
the numbers, which range from milliseconds to seconds, are indicative of
what we can expect there.

\begin{table}[htbp]  
\centering
\small

\begin{tabular}{lr}
\toprule
{} &      version \\
\textbf{system    } &              \\
\midrule
\textbf{MacBook   } &  Air M3 2024 \\
\textbf{MacOS     } &         15.0 \\
\textbf{RAM       } &        16 GB \\
\textbf{python    } &       3.12.4 \\
\textbf{cvxpy     } &        1.5.2 \\
\textbf{numpy     } &       1.26.4 \\
\textbf{pandas    } &        1.5.3 \\
\textbf{matplotlib} &        3.9.1 \\
\textbf{networkx  } &          3.3 \\
\bottomrule
\end{tabular}

\caption{System configuraton for analysis}
\label{tbl:6_system}
\end{table}

Times are measured using simple wall clock time from start to end of the
calculation, including the profiling of the code in \ref{sec:perfcompnt}
which used instruments introduced into the code that recorded wall clock
readings along important way points of the execution and then aggregated
the results into an overall \emph{time-spent-per-stage} reading. We are
well aware of the theoretical limitations of this approach, and we have
taken measures to address those. For example, for measuring short
running processes we have repeated the measurements up to 10-100x. The
ultimate size of the effect observed in the results suggests that those
measures were sufficient for our purposes: our core result shows a 20x
to 100x+ improvement for the marginal price algorithm over the convex
ones from \cite{angeris21}\footnote{As it was published around the
same time as we put the finishing touches on our algorithm we have never
implemented the conjugate algorithm described in \cite{diamandis23} so
we cannot directly compare the two, but the performance improvement we
achieve against \cite{angeris21} that we are significantly faster,
albeit not by the same margin}, which is beyond any noise or bias that
could possibly be introduced by our measurement protocol.

The contestants in our line up are the following algorithms:

\begin{enumerate}
\def\labelenumi{\arabic{enumi}.}
\tightlist
\item
  Marginal Price optimization using a bisection algorithm as described
  in appendix \ref{app:bisection}, which we call the \textbf{pair mode}
  in our implementation because it only works if all curves are
  operating on the same pair of tokens
\item
  Marginal Price optimization using Newton-Raphson / gradient descent as
  described in appendix \ref{app:nr}, which is what we call the
  \textbf{full mode} because there is no restriction on the number of
  tokens
\item
  Convex optimization with CVXPY \cite{boyd04, diamond16, cvxpy} using
  the new default solver \textbf{Clarabel}
\item
  Ditto using the older solvers \textbf{ECOS} and \textbf{SCS} which ex
  post we group together because their performance here is very similar
\end{enumerate}

\subsection{Performance on token
pairs}\label{performance-on-token-pairs}

We start the analysis on token pairs, the only arena where all
contestants, including our marginal price pair mode, can compete. For
this, we run the algorithms on two tokens, and we vary the number of
curves from 2 to 2,000. We note that whilst using traditional AMMs even
the most crowded pairs will not usually present more than a dozen of
curves, we have developed the marginal price algorithm for the use
together with the CarbonDeFi protocol
\cite{carbon22, carbon22l, carbon}, where every single trading position
is a curve. For busy markets, a few thousand positions is not a
particularly high number.

\begin{figure}[htbp]
    \centering
    \includegraphics[width=\textwidth]{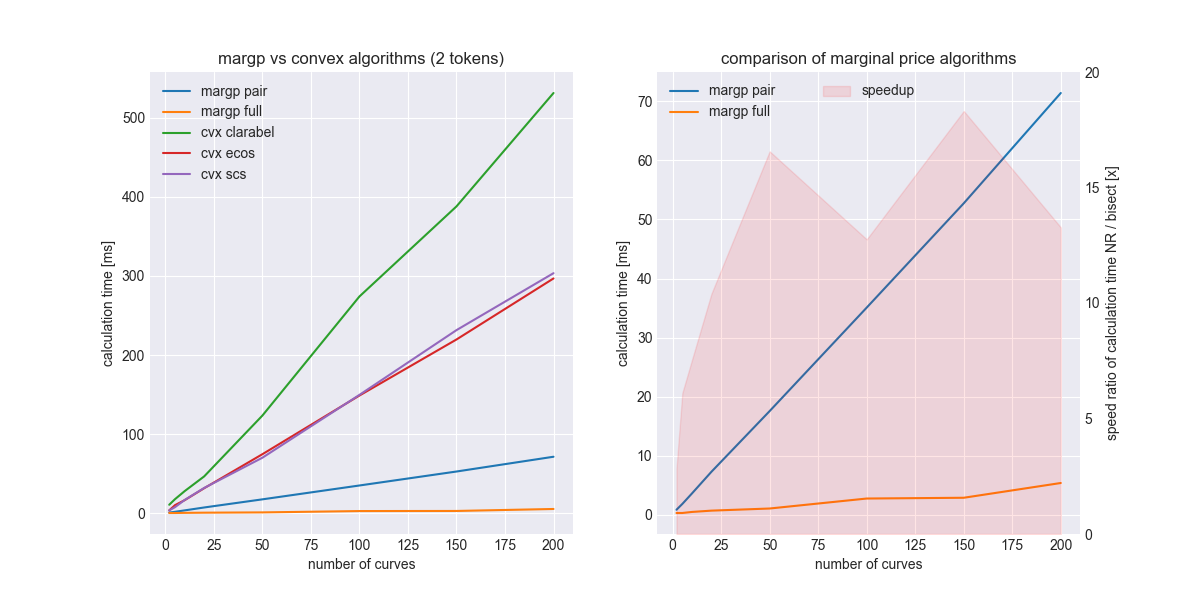}
    \caption[Calculation time versus number of curves (token pairs)]{\textit{Calculation time versus number of curves (token pairs). }The left hand panel shows the calculation time in ms for the different algorithms (orange and blue marginal price; the others convex) on token pairs. The the right hand panel shows the same data for the marginal price algorithms only, plus the speedup ratio between the two (red surface).}
    \label{fig:algocompare}
\end{figure}

The key results we have obtained for token pairs are summarized in
figure \ref{fig:algocompare}. Note that we limited our charts to 200
curves in this case -- our analysis ran further (up to 2,000) but
nothing meaningfully different happened there, so we chose the tighter
range for a more effective presentation. The left panel of figure
\ref{fig:algocompare} shows the performance of all algorithms on a
linear scale, and the right panel zooms into the two marginal price
algorithms only. We make the following observations, most of which are
representative for more general cases which we will discuss in the
following sections.

\begin{observation}[Performance comparison]\label{obs:perfpairs}
When comparing the performance of the different algorithms for token pairs in \ref{fig:algocompare}, we find a number of key differences:

\begin{enumerate}
\def\labelenumi{\arabic{enumi}.}
\tightlist
\item
    The marginal price modes outperform the convex modes by a massive margin, to the point that the curve of the full marginal price mode appears flat in the chart.

\item
    The pair mode is substantially slower than the full mode. The ratio is somewhat volatile but as the right panel shows, the speedup of the full mode versus the pair mode is at least 5-10x, except for the very smallest number of curves where the fixed costs dominate the algorithm. 

\item
    The convex algorithms group into ECOS and SCS on one hand, and Clarabel, the new algorithm on the other one. None of the algorithms are competitive to even the pair mode, let alone the full mode. Clarabel is performing significantly worse than the older ECOS/SCS.

\item
    None of the convex algorithms display a performance that would be useable for real life arbitrage purposes across a large number of curves -- the calculation time required scales approximately linearly in the number of curves, at about 1.5 seconds per 1,000 curves for ECOS/SCS and almost 3 seconds for Clarabel. For comparison, pair mode is at 0.4 seconds, and full mode at about 0.04 seconds per 1,000 curves.
\end{enumerate}
\end{observation}

\subsection{Scaling the number of
curves}\label{scaling-the-number-of-curves}

\label{sec:perfcompnc}

We now look more closely at what happens when we increase the number of
curves whilst holding the number of tokens constant. We have run the
analysis for different numbers of tokens between 2 and 20. The results
we show here are for 10 tokens, and they are representative for what he
have seen in the other cases.

\begin{figure}[htbp]
    \centering
    \includegraphics[width=\textwidth]{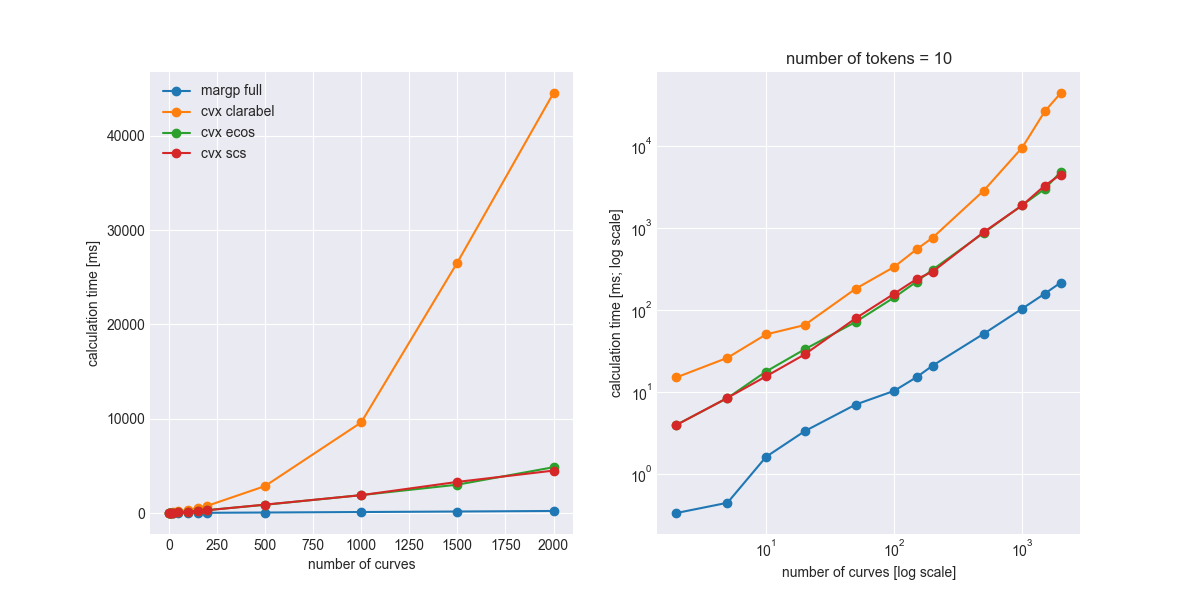}
    \caption[Calculation time versus number of curves (10 tokens)]{\textit{Calculation time versus number of curves (10 tokens). }Both panels show the calculation time for the different algorithms in a market with 10 tokens as a function of the number of curves. Blue is marginal price full mode, orange is Clarabel, and green/red are ECOS/SCS. The left hand panel is on a linear scale, the right hand panel is on a log/log scale.}
    \label{fig:bync2c10}
\end{figure}

The results are presented in figure \ref{fig:bync2c10} where the left
panel is the same chart as in the left panel of figure
\ref{fig:algocompare} except that the x-axis now goes up to 2,000
curves. Also, as the number of tokens is above two, there is no marginal
price \emph{pair mode}. Fundamentally, the results are the same as in
observation \ref{obs:perfpairs}, except that Clarabel seems to perform
even worse: whilst ECOS/SCS deteriorate linearly, the curve for Clarabel
looks quadratic. At more than 40 seconds to run a single analysis on
2,000 curves it is beyond any usefulness for us in practical settings.
It also dominates the chart in the left hand panel to the extent that
the performance figures for the other algorithms are hard to read.

We therefore redraw the same chart in the right hand panel of figure
figure \ref{fig:bync2c10} on a log/log scale. The marginal price
algorithm is the clear winner, with a consistent performance of about
10ms per 1,000 curves. The ECOS/SCS group is consistently behind by what
looks like an order of magnitude along the entire curve, and Clarabel is
another half order of magnitude behind that at the start, deteriorating
to a full order of magnitude at the large-number-of-curves end of the
spectrum.

\begin{table}[htbp]  
\centering
\small

\begin{tabular}{lrrr}
\toprule
\textbf{speedup versus:} & clarabel &   ecos &    scs \\
\textbf{n curves} &          &        &        \\
\midrule
\textbf{10      } &    31.3x &  11.0x &   9.6x \\
\textbf{100     } &    32.6x &  13.8x &  15.4x \\
\textbf{500     } &    56.0x &  17.1x &  17.4x \\
\textbf{1000    } &    92.7x &  18.3x &  18.3x \\
\textbf{2000    } &   206.6x &  22.5x &  20.9x \\
\bottomrule
\end{tabular}

\caption{Speedup marginal price versus convex (10 tokens)}
\label{tbl:6_speedup}
\end{table}

We we present the associated speedup numbers of the marginal price
algorithm over the convex algorithms in table table \ref{tbl:6_speedup}.
We see that ECOS/SCS are behind the marginal algorithms by a factor of
10-20x, with the distance getting larger at larger-number-of-curves end.
The speedup compared to Clarabel starts at 30x for 10 curves and reaches
200x+ for 2,000 curves, and is most likely getting even worse beyond
that. This is making CVXPY's new default solver definitely not a good
choice for this task, even when compared to ECOS/SCS.

\subsection{Mathematical interlude}\label{mathematical-interlude}

\label{sec:interlude}

The results of the following section \ref{sec:perfcompnt} have surprised
us at first. Before we continue, we need to provide some additional
context. As we discuss in a forthcoming paper \cite{loesch24b}, we only
ever use the optimization algorithm for two or three tokens at a time,
and we loop over the reasonable token combinations. Therefore,
performance of our implementation of the algorithm is not relevant for
us for large token numbers, but we do care about the performance for
large numbers of curves. This means that in practice we operate in the
region covered by section \ref{sec:perfcompnc}.

However, when running numbers for this paper, we found that our
implementation degrades significantly when the number of tokens gets
large, to the point that as shown in figure \ref{fig:byntll} we can go
beyond ECOS/SCS. When we looked into this, we concluded that this was an
artefact of the way we implemented the algorithm rather than a
fundamental limitation of the algorithm itself, for reasons that will
become clear in a moment. We at on stage will improve the algorithm
implementation, and we will report on the results in a future paper and
update this one accordingly. Technically, the changes are not trivial,
and we need to be careful making changes to our research system to
ensure that it stays reasonably close to our production system.

The fundamental issue is that we have the derivative of a sum of
functions, and the derivative and sum operations commute, leading to the
commutative diagram in equation \ref{eq:diagramperf}. This diagram needs
some explanation. Firstly, all vector quantities are denoted by bold
face, and all scalar quantities by regular face. We have a set of
vector-valued function \(\mathbf{f}_\nu(\mathbf{x})\) of a vector
\(\mathbf{x}\), and the aggregate function \(\mathbf{F}\) which is the
sum over all the constituent functions
\(\mathbf{F}=\sum_\nu\mathbf{f}_\nu\). We also have a derivate operator,
the \emph{Jacobian operator} \(\mathbf{J}\), which has a matrix-valued
result, where the element \(\mathbf{Jf}_{ij}\) of \(\mathbf{J}\) applied
to \(\mathbf{f}\) is the partial derivative of the \(i\)-th component
\(f^i\) with respect to the \(j\)-the element \(x_j\), as seen in the
top right corner in the diagram \ref{eq:diagramperf}. We also have a sum
operator \(\sum\) that operates either at the vector level if it
aggregates function values, or at the matrix level if it aggregates the
Jacobian values. The diagram shows two paths to calculate the Jacobian
of the aggregate function \(\mathbf{F}\), one in red and one in blue,
and the diagram commutes.

\begin{equation}\label{eq:diagramperf}\begin{aligned}
\begin{tikzcd}
    \mathbf{f}_\nu(\mathbf{x}) 
    \arrow[r, color=red, "J"] \arrow[d, color=blue, "\sum"'] 
    & \mathbf{J\,f}_\nu(\mathbf{x})_{ij} 
    = \frac{\partial f_\nu(\mathbf{x})^i}{\partial x_j} 
    \arrow[d, color=red, "\sum"] \\
    \mathbf{F}_\nu(\mathbf{x}) = \sum_\nu \mathbf{f}_\nu(\mathbf{x}) \arrow[r, color=blue, "J"'] 
    & 
    \mathbf{J\,F}_\nu(\mathbf{x})_{ij} = \sum_\nu \mathbf{J\,f}_\nu(\mathbf{x})_{ij}
\end{tikzcd}
\end{aligned}\end{equation}

The blue path -- sum first, then derivative -- is easier to implement
than the right one. For the blue path, all we have to do is to dispatch
the relevant components of the vector \(\mathbf{x}\) that
\(\mathbf{f}_\nu\) needs into the function, and aggregate the result
correctly on a coordinate level in the sum function \(\mathbf{F}\). We
can for example use dictionaries to create those sparse vector
structures which is easy to implement and produces very little overhead.
The function \(\mathbf{F}\) then has the same interface as the component
functions \(\mathbf{f}_\nu\), and an unmodified Jacobian algorithm can
be fed with the aggregate function \(\mathbf{F}\).

The red path -- derivative first, then sum -- is somewhat harder to
implement because the aggregation for \(\mathbf{J}\) happens at the
matrix level as opposed to at the vector level, so we need to use
different aggregation algorithms for \(\mathbf{F}\) and
\(\mathbf{J} \mathbf{f}\).

However, the red path has one important advantage that massively reduces
the computational effort: all our functions only depend on two variables
that for presentational purposed here we call \(y,z\). Also, they only
return two values. Therefore, the Jacobian of \(\mathbf{F}\) when
embedded in the larger matrix looks like in equation \ref{eq:jmatrix},
ie it is mostly zero.

\begin{equation}\label{eq:jmatrix}\begin{aligned}
\begin{pmatrix}
  \cdot & \cdot & \cdot & \cdot & \cdot & \cdot & \cdot & \cdot  \\
  \cdot & (\partial_y f)_y & \cdot & (\partial_z f)_y & \cdot & \cdot & \cdot & \cdot  \\
  \cdot & \cdot & \cdot & \cdot & \cdot & \cdot & \cdot & \cdot  \\
  \cdot & (\partial_y f)_z & \cdot & (\partial_z f)_z & \cdot & \cdot & \cdot & \cdot  \\
  \cdot & \cdot & \cdot & \cdot & \cdot & \cdot & \cdot & \cdot  \\
  \cdot & \cdot & \cdot & \cdot & \cdot & \cdot & \cdot & \cdot  \\
  \cdot & \cdot & \cdot & \cdot & \cdot & \cdot & \cdot & \cdot  \\
  \cdot & \cdot & \cdot & \cdot & \cdot & \cdot & \cdot & \cdot
\end{pmatrix}
\end{aligned}\end{equation}

The consequences of this are important enough to warrant their own
proposition:

\begin{proposition}[Jacobian complexity]\label{prop:jaccompl}
We assume a system with $K$ tokens and $N$ curves operating on these tokens. Provided all curves only allow exchanging between a uniformly bounded number of tokens, the complexity of calculation their individual Jacobian is $O(1)$ with respect to $K$, and therefore the overall calculation complexity along the red path (Jacobian first) is $O(N)$. The complexity of calculating the Jacobian along the blue path (sum first) however is $O(N\cdot K^2)$.
\end{proposition}

\textbf{Proof}. If the number of tokens is uniformly bounded, typically
\(2\) or \(3\) but importantly not depending on \(K\), then the \(O(1)\)
dependence is obvious. The \(O(N)\) is then introduced by the sum, and
there is no dependence on \(K\) provided the zeroes in the smart vectors
are handled intelligently. On the other path however, each calculation
of \(\mathbf{F}\) is \(O(N)\) and it is executed \(O(K^2)\) times, thus
concluding the proof. \(\blacksquare\)

In other words, the blue path is a really expensive way of calculating
zeroes. In addition to that, the function \(f\)\footnote{A reminder
that the function \(F\) is the aggregate price response function as
defined in \ref{def:prf}} is actually a function of the ratio of its
variables only (ie \(f(y,z) = \bar f(y/z)\)). Using the chain rule we
obtain the following identities that allow us to cut the number of
calculations in half once more:

\begin{equation}\label{eq:chainid}\begin{aligned}
y\,\partial_y f 
= -\frac{z}{\partial_z f}
= \frac{y}{z}\ \bar f'(\frac{y}{z})
\end{aligned}\end{equation}

\subsection{Scaling the number of
tokens}\label{scaling-the-number-of-tokens}

\label{sec:perfcompnt}

\begin{figure}[htbp]
    \centering
    \includegraphics[width=\textwidth]{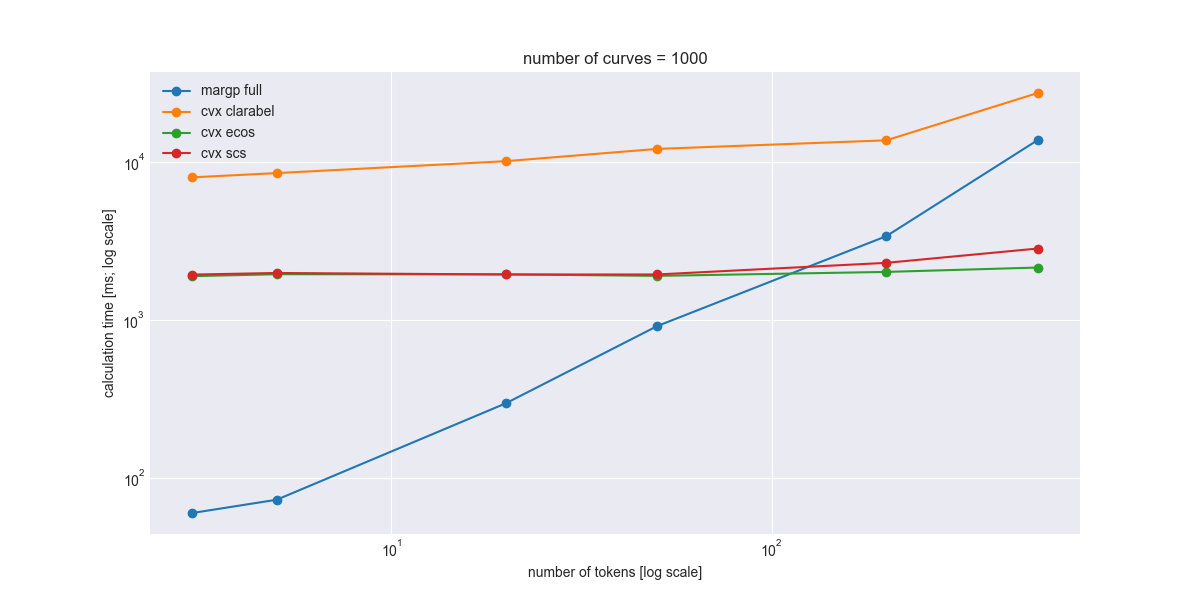}
    \caption[Calculation time versus number of tokens (1,000 curves)]{\textit{Calculation time versus number of tokens (1,000 curves). }This chart shows the calculation time for the marginal price algorithm (blue), and the convex Clarabel (orange) and ECOS/SCS (green/red) algorithms as a function of the number of tokens. The chart is on a log/log scale.}
    \label{fig:byntll}
\end{figure}

With the mathematics out of the way, we can now discuss the scaling of
our implemented algorithm -- which follows the blue path in diagram
\ref{eq:diagramperf} -- with respect to the number of tokens. For the
analysis in this section, we have kept the number of curves at a
constant 1,000, which brings us to the region of 1 second calculation
time for ECOS/SCS and 10 seconds for Clarabel. The results are presented
in figure \ref{fig:byntll} and we make the following observations:

\begin{itemize}
\tightlist
\item
  The marginal price algorithm starts out very well, at about 50ms.
\item
  The performance however deteriorates quite dramatically with the
  number of tokens as the log/log plot in figure \ref{fig:byntll} shows,
  and the crossover with ECOS/SCS is at about 100 tokens; our algorithm
  even seems even on its way to catch up with Clarabel.
\item
  The performance of the convex algorithms is almost flat with ECOS/SCS
  at about 1 second, and with Clarabel at about 10 seconds and
  increasing at the end.
\end{itemize}

This is what we would expect. Additional numbers of tokens for the
convex solvers means adding more constraints, but those constraints are
covering fewer variables per constraint because the overall number of
variables is fixed by the number of curves. ECOS/SCS seem to be mostly
oblivious to an increase in token numbers, and even for Clarabel the
impact seems muted. However, as shown in proposition
\ref{prop:jaccompl}, the blue path algorithm we are using for the
marginal price method is \(O(K^2)\) with respect to the number of
tokens, and this is what we see in the chart \ref{fig:byntll}.

\begin{figure}[htbp]
    \centering
    \includegraphics[width=\textwidth]{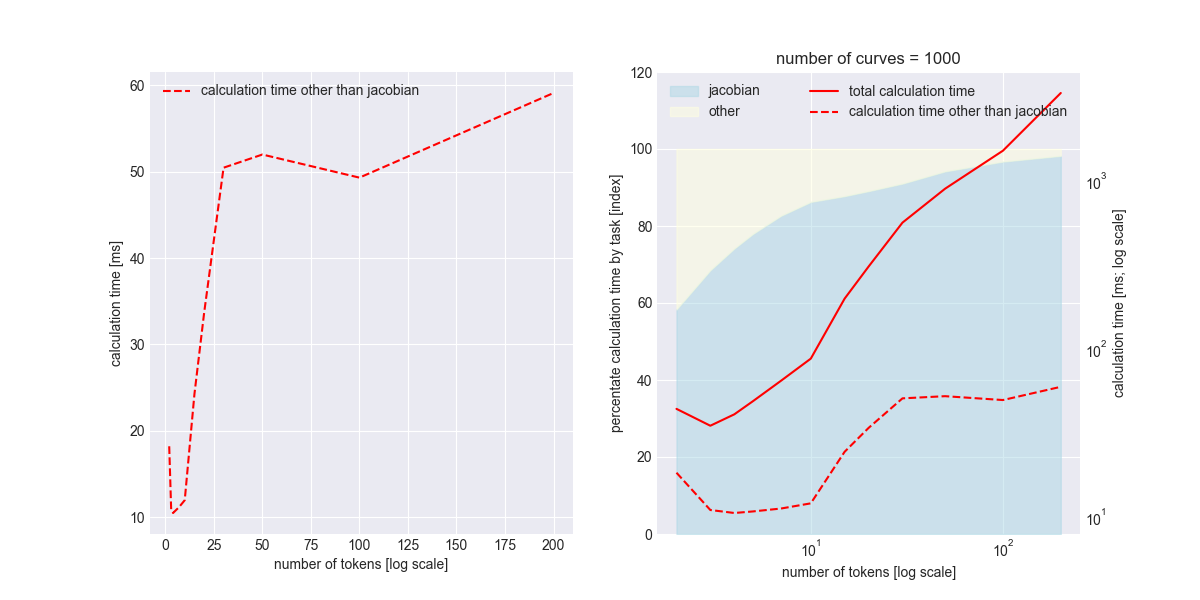}
    \caption[Impact of calculating the Jacobian on the overall performance]{\textit{Impact of calculating the Jacobian on the overall performance. }The left hand panel shows the calculation time as function of number of tokens (with 1000 curves) for everything but the calculation of the Jacobian. The right hand panel redraws this on a log/log scale, and adds the calculation time for the Jacobian (solid red). The surface chart in the background shows the percentage of time spent on the Jacobian calculation (linear scale).}
    \label{fig:byntcombo}
\end{figure}

We have not at this stage been able to prove this by implementing the
red path algorithm, for the reasons discussed above at the beginning of
section \ref{sec:interlude}. However, we did profile our current
algorithm and measured the cost of the Jacobian calculation as well as
that of the other operations, the results of which are shown in figure
\ref{fig:byntcombo}. We see that the Jacobian calculation is already
starting at about 50 percent of the total processing time for only two
tokens, and this increases dramatically with the number of tokens. At
100 tokens, virtually all processing time is consumed by the the
calculation of the Jacobian. We can see this by looking at the surface
plot in the left panel, and also by the fact that the time requirements
for the non-Jacobian calculations displayed in the left hand panel
become quickly flat at about 50ms over all 1,000 curves, regardless of
the number of tokens.

\section{Conclusion}\label{conclusion}

\label{sec:conclusion}

This concludes the first paper describing the mathematics and algorithms
underlying our FastLane Arbitrage bot \cite{fastlane}. This bot is
monitoring various chains, and specifically the DEXes and Carbon DeFi
deployments thereon, and is looking for arbitrage opportunities there
(ie it is looking for trades that allow it to make a profit without
taking any risk). This first paper is focussing on the newly developed
\emph{marginal price framework} for identifying all arbitrage
opportunities in a specific market or submarket.

The core of this paper is section \ref{sec:margpframework} where we have
described the the \emph{marginal price optimization algorithm}\footnote{We refer to it as ``optimization'' for historical reasons
because of its relationship to convex optimization; technically however
it is a root finding algorithm } for arbitrage and optimal routing and
where we have shown in theorem \ref{thm:coreequiv} that it is
outcome-equivalent to the convex optimization problem developed in
\cite{angeris21, angeris22, diamandis23} and described in section
\ref{sec:convex}. This algorithm dramatically simplifies the
optimization problem by only searching on the optimal surface described
by the marginal prices between the tokens in the system. This reduces
the number of variables from \emph{two per curve} to \emph{one per
token}\footnote{In fact, number of tokens minus one}, and converts
the optimization problem into an often better conditioned root finding
problem.

We discussed in section \ref{sec:perfcomp} that our algorithm
outperforms convex optimization\footnote{Specifically, the convex
optimization algorithm in token space proposed first in
\cite{angeris21}; we believe our algorithm still outperforms the
conjugate algorithm proposed in \cite{diamandis23} but as its
publication of this algorithm coincided with the finalization of our
arbitrage bot we never tried to implement it} by a factor of 75x on
ECOS/SCS solvers, and a factor of 200x for CVXPY's new standard solver
Clarabel. Also, whilst our algorithm still experiences convergence
issues in markets that are dominated by certain configurations of
levered curves, it is significantly more robust in this respect, and the
reasons for divergence are well understood.

We have presented two different implementations of the marginal price
algorithm, the \emph{pair optimizer} based on the bisection algorithm,
described in section \ref{sec:margpimplpair}, and the \emph{full
optimizer} that is based on the Newton-Raphson / gradient descent
algorithm described in section \ref{sec:margpimplgen}. As the name
implies, the pair optimizer only work on pairs, and we have found that
whilst it is slower than the full optimizer by a factor of up to 10x, it
is more robust -- in fact it always converges, regardless of market
conditions. Therefore, the decision which optimizer to use for pairs can
be hard\footnote{One may also consider combined applications that
start with the full optimizer on pairs, and escalate cases of
non-convergence to the pair optimzier}, depending whether one values
resource use and latency or robustness more.

This paper only covered part of the technology underlying the FastLane
Arbitrage bot. Notably what is missing is

\begin{itemize}
\item
  \emph{transaction decomposition} where we limit the number of curves
  involved in an arbitrage transaction to maximise the chances that an
  arbitrage transaction will go through, and
\item
  \emph{transaction linearization} where we ensure that arbitrages can
  be executed in a linear manner in situations where flashloans are
  available only for certain tokens, and where the arbitrageurs only as
  access to limited token amounts.token holdings are limited.
\end{itemize}

We also developed another algorithm -- the \_Graph Mode\_\_ -- that uses
a completely different approach to identifying arbitrage opportunities,
and that does not suffer from the same convergence issues as the
marginal price algorithm. It however displays other issues, notably
around scaling, that we are still working through. Ultimately we will
probably setle on a \emph{horse-for-courses} approach where both
algorithms are used in their respective sphere.

All of the above are subject to forthcoming papers that we currently
have in preparation, and that we will publish in due course.

\pagebreak
\addcontentsline{toc}{section}{References}
\bibliographystyle{plain}
\bibliography{references}

\pagebreak
\section*{Appendix}
\appendix

\section{The Radome Optimization
Problem}\label{the-radome-optimization-problem}

\label{app:radome}

\begin{figure}[htbp]
    \centering
    \includegraphics[width=\textwidth]{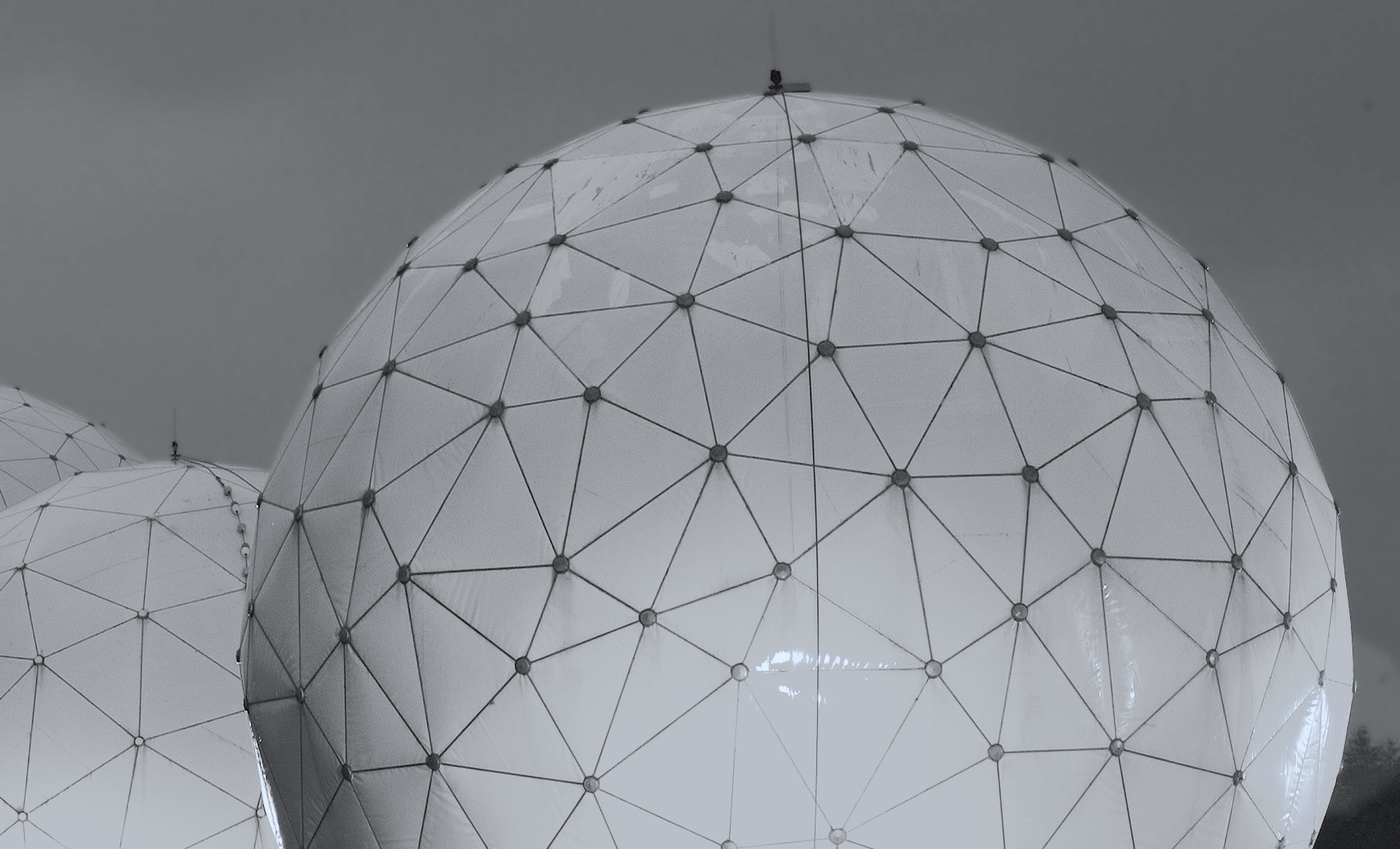}
    \caption[Example for a radome]{\textit{Example for a radome. }Example for a radar dome (``radome''), a structure that consists of flat or near flat segments touching one another along non-smooth connector lines (image credit Wikipedia)}
    \label{fig:radome}
\end{figure}

A ``radome'' \cite{radome} is a structure that protects a radar antenna
from the elements, and typically has the geometry shown in figure
\ref{fig:radome}. It also provides an excellent example for a class of
convex optimization problems, referred to herein as the \emph{``radome
optimization problems''}.

\begin{definitionnn}[Radome Optimization]\label{def:radomeproblem}
The "\textbf{Radome Optimization Problems}" are a class of convex optimization problems that arise when searching for an optimal point on a radome. They differ in two aspects, firstly the exact nature of the shape of the radome, and secondly the objective function that is to be optimized.

In terms of \textbf{radome shapes} we consider the following types (a) the \textbf{balloon type}, a smooth surface that is convex in all directions (b) the \textbf{planar type}, a collection of flat surfaces that is convex in all directions, like shown in figure \ref{fig:radome} and (c) the \textbf{mixed type}, a "mostly smooth" surface that can be thought of as replacing the flat surfaces in the planar type with convex surfaces, in a manner that retains the vertices and edges of the planar type and overall convexity.

In terms of \textbf{objective functions} we consider (1) a \textbf{linear} function like "height" or "distance to the sun", or (2) a \textbf{non-linear} one like "distance to a nearby point".
\end{definitionnn}

The problems A1 and A2 are traditional smooth optimization problems that
can be solved with Lagrange multipliers \cite{goldstein02}: at the
optimal point, the gradient of the constraint must be equal to the
gradient of the objective function, the latter being constant
\((0,0,1)\) in case of the \emph{``height''} function in (1) and a
vector pointing in the direction of the sun in the general linear case.
In case (2) the gradient is orthogonal to the usual equidistance
surfaces around the target point.

If we assume the radome is the unit sphere then the solution of A1 in
the height case is the north pole, and more generally
\(\mathbf{x}/||\mathbf{x}||\) where \(\mathbf{x}\) is a vector pointing
in the direction of the sun.

The planar type radome -- the usual form as depicted in figure
\ref{fig:radome} -- is an example for a set of linear constraints: for
each of the faces \(\nu=1\ldots M\) we have a linear condition on the
points \(\mathbf{x}\) inside that radome that reads

\begin{equation}\begin{aligned}
\mathbf{n}_\nu \cdot \mathbf{x} 
\leq 
\mathbf{n}_\nu \cdot \mathbf{x_\nu}
\end{aligned}\end{equation}

where \(\mathbf{n}_\nu\) is the normal vector of the face \(\nu\), and
\(\mathbf{x_\nu}\) is any point on that face. We note that the normal
vector \(\mathbf{n}_\nu\) is the equivalent of the gradient of the
constraint in the smooth case. Therefore the Lagrange condition above
cannot usually be strictly satisfied\footnote{To formalize this
statement we could introduce a measure on the space of gradients that is
uniform on the unit sphere and when the sphere is deformed into a planar
structure we find that all mass will be in the vertices in the sense
that faces and edges are of measure zero}. In case we have linear
constraint here we arrive at the well-known \emph{linear-programming}
problem as described in many text books (eg \cite{dantzig97}). In this
case the solution is almost always\footnote{In the sense of the
measure introduced in the previous footnote} at a vertex of the
radome, which it can be found, for example, using a \emph{simplex
algorithms}.

The interesting cases -- and most relevant for us -- are B2 and C1; B2
because it is the most intuitive and C1 because it most closely relates
to the actual problem we are solving, which has a linear objective
function on a mixed-type-constraint hyper-surface.

Starting with B2, the solution has two steps: we need to find the face
that is closest to the target point, and on the face we need to find the
actual closest point. The algorithm for finding the optimal face is
similar to the simplex algorithm in linear programming, except that we
have a duality where faces now play the roles of vertices and vice
versa. Once we have identified the face we can then use a gradient
algorithm to find the optimal point on that face\footnote{In case of
a simple metric, we could of course do this part of the calculation
analytically, but for didactical reasons we assume that we do not have
this shortcut available to us}.

In case of C1 and C2, the constraint itself is neither linear nor
entirely smooth. Depending on the convexity of the faces, there is a now
a finite chance\footnote{The term ``chance'' is again to be
understood in terms of the measure on the gradients introduced above;
here, because the faces have a non-zero convexity, there is a non-zero
mass of the distribution of gradients on the faces, so not all of the
mass is concentrated in the vertices} of the solution being an
\emph{``interior solution''} (on one of the faces) and a finite chance
of it being a \emph{``corner solution''} (on a vertex or edge). Again we
may use a simplex-type algorithm to identify the optimal face, and on
the face we may use a gradient descent to lead us to the optimal point.
Importantly, we cannot simply start with a gradient descent before we
have identified the correct face because due to the non-smoothness of
the constraint function the local geometry is not a good guide to the
global geometry of the problem.

\section{The Repatriation Problem}\label{the-repatriation-problem}

\label{app:repatriation}

Here we discuss the \emph{repatriation problem} referred to in section
\ref{sec:margpfwcore} in more detail. High level, the issue is when an
arbitrage opportunity exists, but the profits cannot be extracted
because there does not exist sufficient capacity to trade into the
chosen target token. For example, consider the following set of curves
providing the arbitrage:

\begin{itemize}
\tightlist
\item
  Buy/sell USDC for USDT at 0.99 USDT per USDC (1m USDT capacity)
\item
  Sell/buy USDT for USDT at 1.01 USDT per USDC (significantly larger
  capacity)
\end{itemize}

Assuming those curves are infinitesimally thin, one possible trade here
is to start with \(1m\) USDT, sell them for \(~1.01m\) USDC on the first
curve, and to sell those against USDC for \(~1.02m\) USDT, yielding a
profit of \(~20k\) USDT. Alternatively one can start with \(~0.99m\)
USDC, sell them for \(1m\) USDT on the second curve, and sell those for
\(1.01m\) USDC. Again, the profit is \(~20k\) USDC. Using thicker curves\footnote{The MPF algorithms will not be able to deal with
infinitesimally thin curves to accurately determine trading volumes, so
in practice we will need to regularize the curves by imposing a minimum
width; the SOF algorithm operates on amounts rather than prices so in
principle would be able to handle it; however we have found in practice
that the SOF algorithms available to us would perform worse than our MPF
algorithm} will somewhat blunt that opportunity and for the purpose of
the argument we assume that we can make USDx \(15k\) in profit in either
direction.

In either the convex optimization formulation (``COF'') according to
definition \ref{def:convexform} or the marginal price formulation
according to definition \ref{def:margpform} we need to define a target
token in which to extract the profits. If it is either USDC or USDT both
will find the same solution -- a target token of USDT will yield the
first of the solutions above, and a target token of USDC will yield the
second one.

We now assume that the target token is WETH, and we will consider a
number of different cases.

\textbf{High capacity curve.} The baseline positive case is that there
is a high capacity curve linking USDx to WETH. In this case both
algorithms will converge nicely, the MPF one by choosing the price on
the WETH curve that absorbs the 15k profit, and the COF that works on
quantities rather than prices by pushing those 15k through the WETH
curve at whatever price that yields.

\textbf{Low capacity unlevered curve.} We now consider any unlevered
curve linking USDx to WETH. The magic of unlevered curves is that they
can absorb any token amounts -- the only question is at which price.
Fundamentally this is not different from the \emph{high capacity curve}
case above: both algorithms will converge, one by pushing the \(15k\)
USDx into the curve, the other one by finding a price point at which the
\(15k\) USDx can be absorbed. The difference to the high capacity case
is only that the curve is restricted by its WETH token balance in what
it can release -- if it holds 1 WETH then max output will be (below but
possibly close to) \(~1\) WETH; if it holds \(0.1\) WETH it will be
\(~0.1\) WETH and so on. In other words: the algorithms will converge\footnote{Unless the MPF algorithm has a price cutoff that is being
hit}, albeit possibly to a solution that is not particularly
advantageous.

\textbf{Limited capacity levered curve.} We now consider a levered curve
that is linking USDx into WETH. The key difference between this and the
\emph{low capacity unlevered curve} case is that because prices are
bounded there is a maximum amount of USDx that this curve can absorb.
Once it has dispensed all its WETH, and reached the, from the trader's
perspective, worst possible price in the range covered, it will simply
stop trading. The COF algorithm should converge: provided the solver can
deal with the boundary condition it will understand the maximum amount
that can be repatriated and will adjust the trade amounts through the
USDx curves accordingly. The MPF algorithm will do the same, provided
that (a) it can adjust the prices on the USDx curves sufficiently finely
to ensure the profit matches the capacity on the curve, and (b) the
price adjustment of the USDx/WETH curve is done in a way that either it
does not overshoot the boundary, or the algorithm can deal with
overshooting into the no-man's land beyond the curve without failing.

\textbf{No curve.} Finally we assume that there are no curves linking
either of the USDx tokens to WETH, directly or indirectly. How will this
manifest itself in the two cases? In the SOF it will depend on the
solver. Ultimately, because there is no path into WETH none of the
operations the optimization algorithm performs on the USDx curves will
impact that target function. Therefore, ultimately it will fail because
whatever it does, the target function will remain unaffected. In the MPF
case we can see this even more clearly. The Jacobian defined in equation
\ref{eq:jacobian} will be singular because all derivatives between USDx
and WETH will be zero\footnote{The ``no indirect connections''
condition here is important; if there are trade opportunities between
USDx and WETH that go via other tokens then the Jacobian will not be
singular}. This in turn means that the update rule in equation
\ref{eq:updateeta} will fail\footnote{Note that it will fail despite
the modified update rule that ignore the null space when trying to
invert the Jacobian, the issue being that the null space is the one
connecting USDx and WETH} and the algorithm will run into the max
iteration limit without being able to fulfil the solver condition of
\(\Delta \mathrm{USDC}=0\) and \(\Delta \mathrm{USDT}=0\).

In summary -- both algorithms, if sufficiently well done, converge to
the same result under the repatriation problem. There will be no
convergence in case no repatriation is possible because there are no
curves, and if repatriation is limited then they will restrict the
amount of USDx arbitrage performed.

\section{The Multiple Solutions
Problem}\label{the-multiple-solutions-problem}

\label{app:multiple}

Here we discuss the \emph{multiple solutions problem} referred to in
section \ref{sec:margpfwcore} in more detail. High level the issue is
that there could be multiple solutions to the optimization problem that
perform equally well in terms of arbitrage profits, but they result in
different trading instructions. The archetypical example is
buy-low-sell-high where there are multiple options to trade into at the
same conditions and where the volume is limited on the other side.
Consider the following set of three curves, all with a capacity of 1m
USDT:

\begin{itemize}
\tightlist
\item
  Curve A: Buy/sell USDC for USDT at 0.99 USDT per USDC
\item
  Curves B1, B2: Buy/sell USDC for USDT at 1.01 USDT per USDC\footnote{The curves B1, B2 could be composite curves trading
  through one or multiple other tokens, and there could be more than two
  curves, none of which would substantially change the argument}
\end{itemize}

The trade is buying \(~1m\) USDC on curve A at 0.99 USDT and selling it
into B1 or B2 or any combination thereof at 1.01 USDT, for a profit of
\(~20k\) USDT. The multiple solutions come from the fact that, provided
there is no slippage, there is no difference selling into B1, B2 or into
any combination thereof that has the right capacity.

The point about no slippage is important here. If there were slippage on
the curves, there would be a unique solution to the optimization
problem: partitioning the trade so that the post-slippage marginal
prices are the same across all curves. This issue only arises because
the marginal prices do not move with volume.

The algorithm under the convex optimization framework (``COF'') operates
on quantities, and moving quantities between B1 and B2 does not impact
the target function. Any well designed algorithm will not stumble over
this, and the chosen partition will depend on details of the algorithm
and starting conditions. The algorithm under the marginal price
framework (``MPF'') however operates on prices instead of volumes -- the
latter are implied. Therefore the MPF algo cannot operate on no-slippage
curves, and in our implementation of the algorithm we regularize the
curves by enforcing a certain minimum width. Therefore the MPF algorithm
will always converge to a unique post-regularization solution,
equalizing the marginal prices on all curves and distributing the trade
volume accordingly.

\pagebreak

\section{Numerical Methods}\label{numerical-methods}

\label{app:numerics}

In this appendix, we discuss the bisection and Newton-Raphson methods,
and how they can be used for finding minima and maxima and roots.

\subsection{Bisection Method}\label{bisection-method}

\label{app:bisection}

The bisection method is a root-finding method for one-dimensional
functions that has very interesting properties. Notably, it is very
robust in that it is guaranteed to converge (depending on the
convergence criteria chosen) on a very wide range of functions,
including those where roots do not really exist.

\subsubsection{Finding roots}\label{finding-roots}

The algorithm is very simple: given function \(f(x)\) and a
\emph{bracketing} interval \([a,b]\) such that \(f(a)\) and \(f(b)\) are
of different sign, we can find a root of any continuous function by
repeatedly bisecting the interval, checking the sign of the mid point,
and moving the boundary of the interval that has the same sign as the
mid point to the mid point.

This algorithm yields a sequence \((a_i, b_i)\) of intervals such that
\(f(a_i)\) and \(f(b_i)\) have different signs, and the length of the
interval \(|b_i - a_i|\) halves at every step, and therefore converges
to zero. Because of the intermediate value proposition, we know that if
\(f\) is continuous, there is (at least) one root in the interval
\([a_i, b_i]\), and therefore the location of the root can be
approximated to arbitrary precision.

\begin{figure}[htbp]
    \centering
    \includegraphics[width=\textwidth]{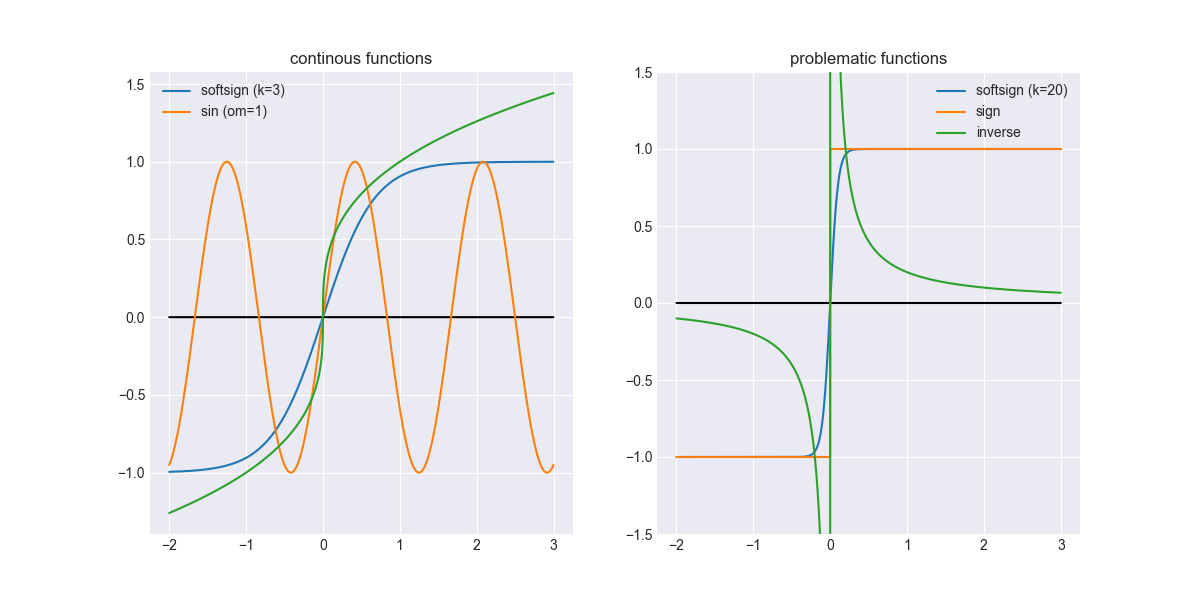}
    \caption[Example functions for root search]{\textit{Example functions for root search. }Both panels show functions discussed in the text for the performance of the root search algorithm on them. The left hand panel has benign functions that have one or multiple regular roots. The right hand panel shows functions that have a root location but either very badly conditioned root (blue), or not roots at all, with a finite jump (orange) or an infinite one (green).}
    \label{fig:functions}
\end{figure}

In the figure \ref{fig:functions} we are providing a number of example
of functions and we briefly discuss how the bisection method can be used
to find the roots. We start with the left hand panel where we have
continuous functions, and we assume that the starting interval \([a,b]\)
is such that \(f(a)\) and \(f(b)\) have different signs, but not
necessarily symmetric around the origin. For the highly regular
\emph{``softsign''} function, we can see that the bisection method will
always converges to the root at \(x=0\). Moreover, because the
derivative is uniformaly bounded in the sense of equation
\ref{eq:bounded}, we can propagate the error on the x axis to the y
axis, meaning that we not only know that position of the root with a
certain precision \(\varepsilon \simeq b_i-a_i\), but we also know that
the error on the y axis is approximately bounded by \(c \varepsilon\).

\begin{equation}\label{eq:bounded}\begin{aligned}
\exists c \, \forall x \in [a,b]:\ |f'(x)| \leq c
\end{aligned}\end{equation}

With the trigonometric \emph{``sine''} function, we can see that the
bisection method will converge to a root as well, provided we start with
an interval that has opposing signs. However, the function has multiple
roots, and to which of them we converge will depend on the choice of the
starting interval.

Finally we have the \emph{``root''} function \(f(x)=\sqrt[3]{x}\). This
function has a root at \(x=0\), but at this point the derivative is
unbounded. Therefore the error on the y axis can be quite big, even if
the convergence on the x axis is advanced.

In the right hand panel of the figure \ref{fig:functions} we have some
more pathological, specifically discontinuous functions. Firstly, to
make the obvious point: two out of the three functions do not have
roots, in the sense that \(f(x) = 0\). However, they have \emph{``root
locations''} \(x_0\) where they change sign:

\begin{equation}\label{eq:rootloc}\begin{aligned}
\exists x_0, \varepsilon, s\ \forall x^- \in (x_0-\varepsilon, x_0)\
\forall x^+ \in (x_0, x_0+\varepsilon): 
s f(x^+) \geq 0, s f(x^-) \leq 0
\end{aligned}\end{equation}

Note that not all functions have this property. For example, consider
\(f(x) = \sin(1/x)\) with \(f(0)=0\) around \(x=0\). This function has
an infinite number of roots that are dense around \(x=0\), so there is
no \(\varepsilon\) that could separate them. However, if the functions
we consider are continuous except for a finite number of points in every
finite interval, that above equation will apply and distinct roots
locations can be identified and found with the bisection method. The
value on that location of course may be undefined, in the sense that the
left and right limits do not coincide (and neither of them may be zero)
in any case (the \emph{``sign''} function), or those limits do not even
exist (the \emph{``inverse''} function \(f(x)=1/x\)). However,
especially in the ``finite jump'' case we note that we can always
regularize the function (eg by convolution), leading to something akin
to the \emph{``softsign''} function, which as \(C^\infty\) with bounded
derivative (for every fixed value of the regularization parameter), and
which matches finite-size functions reasonably well. As previously
mentioned, this case is particularly important to us, because limit
orders correspond to discontinuous functions. To make them continuous,
even differentiable, we regularize these orders by converting them into
very narrow range orders.

\subsubsection{Finding minima and
maxima}\label{finding-minima-and-maxima}

The bisection method can also be used to find (interior) minima and
maxima of functions by using the fact that the derivative of a function
is zero and changes sign at those points. In other words -- if we are
looking for minima or maxima we can use the bisection method to find the
roots of the derivative of the function. The derivative can be
calculated either analytically, or numerically for example using the
\emph{finite difference formula} \ref{eq:finitediff} for some small
value of \(h\):

\begin{equation}\label{eq:finitediff}\begin{aligned}
f'(x) \simeq \frac{f(x+h)-f(x-h)}{2h}
\end{aligned}\end{equation}

Again, this algorithm is very robust. Specifically, we again consider
the case of a non-differentiable function like the one in figure
\ref{fig:targetfuncs}. As we can see in the right hand panel, the
derivative is a step function that is constant between the steps (ie it
locally looks like the sign function in figure \ref{fig:functions}). We
have seen that those functions can be regularized (eg using a
convolution method), yielding functions similar to the softsign function
in the same figure. This function is differentiable and therefore the
bisection method will converge to the correct root of the regularized
function. However, we actually do not have to regularize the function
for the algorithm to converge. As discussed above it will converge, on
the x axis, to the root location as defined above. The error is more
benign than in the root finding case. As we are looking for a maximum or
minimum, by design the area around the root is somewhat flat, therefore
in practical applications the error on the y axis is typically small.
However, it must be stressed that the term ``in practical applications''
is an important caveat here. In particular we can imagine a very spiky
function -- think Dirac Delta. For example, if we look at a Gaussian
kernel function as defined in equation \ref{eq:kernels} and we take a
very big value of \(\lambda\) -- which is a popular choice for a
Dirac-Delta-like \(C^\infty\) function -- then the bisection method will
find the location of the spike, but it may not be that good at
estimating its exact height.

\subsubsection{Convergence}\label{convergence}

\label{sec:bsconv}

To build on our previous discussions we now briefly discuss the
convergence properties of the algorithm. We only discuss the root
finding case here, but the discussion can be easily extended to the
minima and maxima case by replacing the function with its derivative.

\begin{figure}[htbp]
    \centering
    \includegraphics[width=\textwidth]{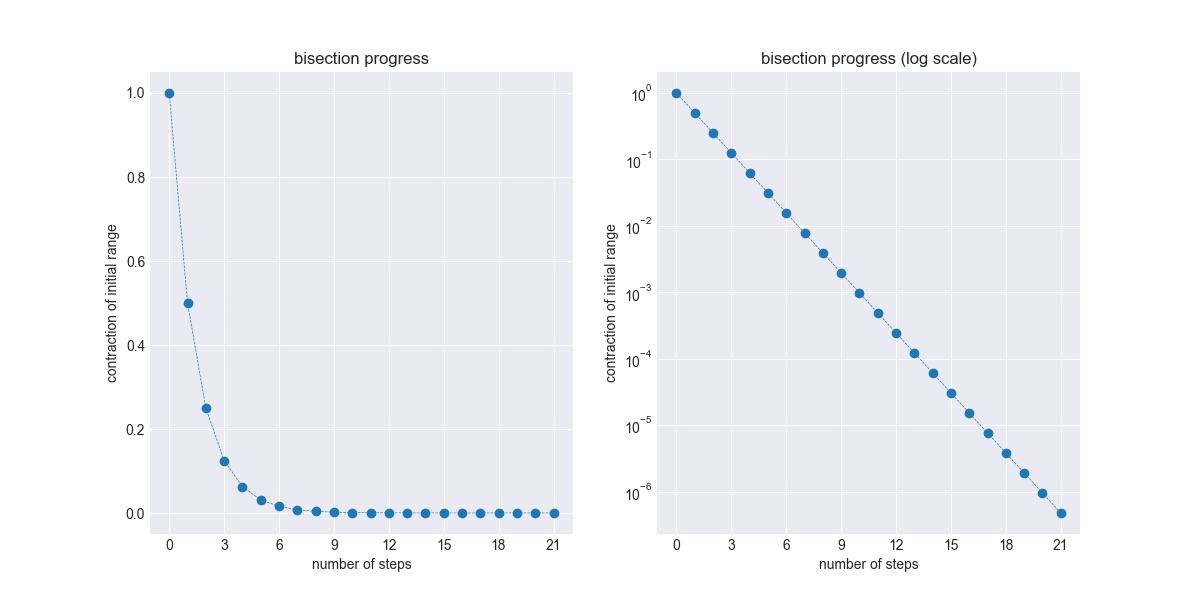}
    \caption[Bisection progress over time]{\textit{Bisection progress over time. }Both panels show how the size of the bisection bracket contracts with number of steps, compared to the initial size of the bracket. The left hand panel is on a linear scale, the right hand panel is on a log scale.}
    \label{fig:progress}
\end{figure}

Firstly, we know that, provided the start conditions are fulfilled, the
algorithm always converges \emph{``exponentially''} on the x-axis, and
we know exactly the speed of convergence: after \(n\) steps, the size of
the interval will be \(2^{-n}\) of the original interval size. See
figure \ref{fig:progress}.

Convergence on the y-axis depends on the type of function we are looking
at.

\textbf{Differentiable with bounded derivative everywhere}. If the
function has a bounded derivative in the initial interval (or any
interval subsequently chosen by the algorithm) then convergence on the
y-axis is equally exponential in nature. This is the case we will mostly
encounter in our problem set, and the bisection method is an excellent
and reliable method to find roots in this case.

An important sub case of this is where the function is
\textbf{differentiable twice everywhere}, which implies that in any
compact interval -- and we only care about those for numerical
applications because we need to choose our initial \([a,b]\) -- the
derivative is bounded, which means the above condition applies.

For the next condition we introduce the notion of separated points a
variation of which we have already seen in equation \ref{eq:rootloc}: a
set of points \(\{x_i\}\) is separated if there exists an
\(\varepsilon\) so that for all \(i\neq j\) we have
\(|x_i-x_j|>\varepsilon\).

\textbf{Continously differentiable everywhere except for separated
points}. The next case we consider is the case where the function is
differentiable and the derivative is continuous everywhere in the
initial interval, except for a set of separated points. The interesting
point here of course is if the root is amongst those points as,
otherwise, we will eventually find an interval where we are essentially
back in the first case.

So, if the root location is one of those points, the algorithm will
converge to it, as defined in equation \ref{eq:rootloc}, but we can't
make any predictions about the error on the y-axis. For example, look at
the right hand panel of figure \ref{fig:functions} for examples of such
functions. For example, if we use the sign function, unless we by chance
hit the root exactly, the value at the mid point will either be 1 or -1.
Even worse is the case of the inverse function \(1/x\) where the mid
point of the converged interval will be the larger in absolute value the
smaller the interval is, on top of the uncertainty around its sign.

In other words -- in this case, whilst we can be certain about
convergence on the x-axis, convergence on the y-axis is undetermined,
and in the worst case the longer we run the algorithm, the further the
value diverges from zero.

There are more pathological cases of functions that we can consider
here. For example, we can have a function with an infinite number of
roots in any finite interval containing a specific point (eg, the
aforementioned \(\sin(1/x)\) around zero). In this case, the algorithm
will still converge on the x-axis but we do not know where to. Whilst
this case is interesting from a mathematical point of view, it is not
relevant for our problem set as we will only encounter functions of the
first two types in our problem set.

\subsubsection{Higher dimensions}\label{higher-dimensions}

\label{sec:hidim}

\begin{figure}[htbp]
    \centering
    \includegraphics[width=\textwidth]{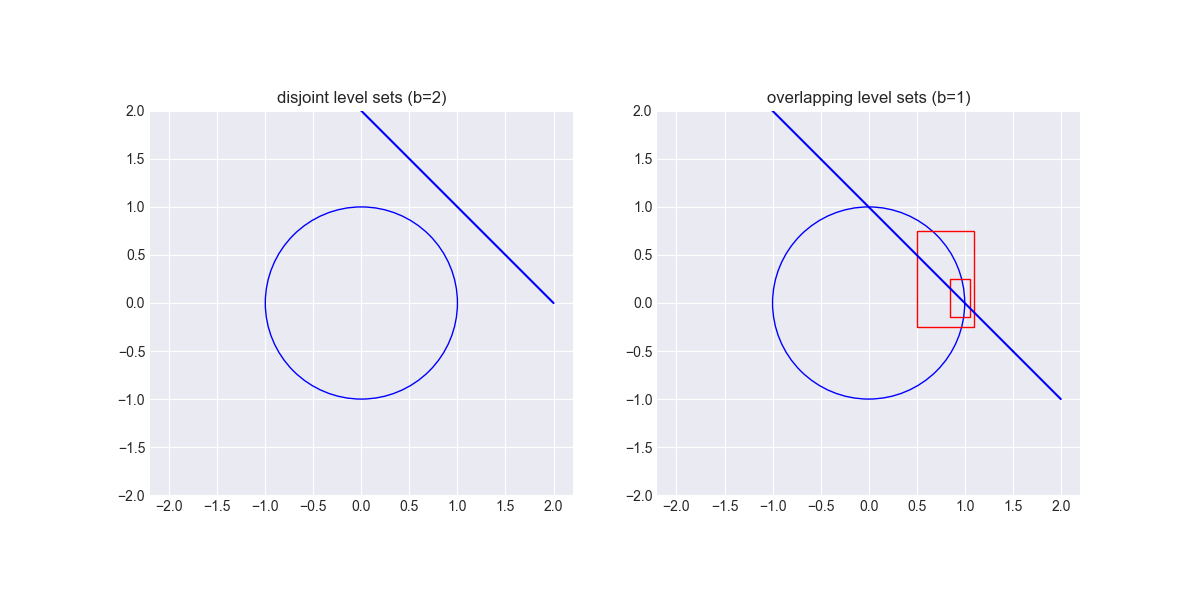}
    \caption[Attempting multi-dimensional bracketing]{\textit{Attempting multi-dimensional bracketing. }Both panels show the level sets \(f_i(\mathbf{x})=0\) (\(i=0,1)\) for the two components of vector valued function \(\mathbf{f}\). In the left panel, those level sets do not intersect, so the equation \(\mathbf{f}(\mathbf{x})=\mathbf{0}\) does not have a solution. On the right hand panel, the level sets intersect so there are solutions. The red rectangles represent an attempt at two dimensional bracketing discussed in the text.}
    \label{fig:multidim}
\end{figure}

The bisection method is hard to extend to higher dimension because the
geometry of the problem is more complex, and because the intermediate
value proposition no longer applies. To understand this, we want to look
at the following two dimensional problem

\begin{equation}\begin{aligned}
f(\mathbf{x}) = 
\begin{bmatrix}
f(x) \\
f(y)
\end{bmatrix}
=
\begin{bmatrix}
x+y-b \\
x^2 + y^2 - 1
\end{bmatrix}
=
\begin{bmatrix}
0 \\
0
\end{bmatrix}
\end{aligned}\end{equation}

that depends on a parameter \(b\). Note that of course in two dimension
we not only have two variables but also two functions, and we are
looking joint root \(f(\mathbf{x}) = \mathbf{0}\). We have drawn the
level sets for two different parameter values of \(b\) in the figure
\ref{fig:multidim}. In both panels we see the circle of radius one that
is the level set of the second function. In the left panel we also see
the line \(y=2-x\), whilst in the right panel we see the line \(y=1-x\).

Firstly we note that the left panel does not have a solution to the
problem as the two level sets do not intersect and therefore there is
not joint root. However, the right panel does have two solutions, at the
intersection points of the circle and the line.

We recall that for a bisection method we need a change of sign, and we
note that above (below) the line and outside (inside) the circle sign of
the respective function is positive (negative). We note that only on the
right panel we can identify a (hyper)rectangle where all possible
signatures are present: if we start from the bottom left and go
clockwise we have \((-,-)\), \((+,-)\), \((+,+)\), \((-,+)\). We also
note that there is a root in this space. This is not by chance. If we
can contract the rectangle to zero without changing the border
signatures (as indicated with the smaller rectangle) we will eventually
converge against a root location. However, in this process we encounter
a number of problems

\begin{itemize}
\item
  How do we know that there is a (joint) root? In high dimensions this
  problem is much harder to solve than in one dimension.
\item
  Relatedly, how do we identify the hyper-rectangle that satisfies the
  correct signature conditions at the boundary?
\item
  Finally, if we have such rectangle, how do contract it without
  violating the boundary conditions?
\end{itemize}

None of those problems are unsolvable. However, they are hard enough
that bisection in higher dimensions is not a popular method for generic
problems, and we will only consider it in one dimension.

\subsection{Newton-Raphson Method}\label{newton-raphson-method}

\label{app:nr}

\begin{figure}[htbp]
    \centering
    \includegraphics[width=\textwidth]{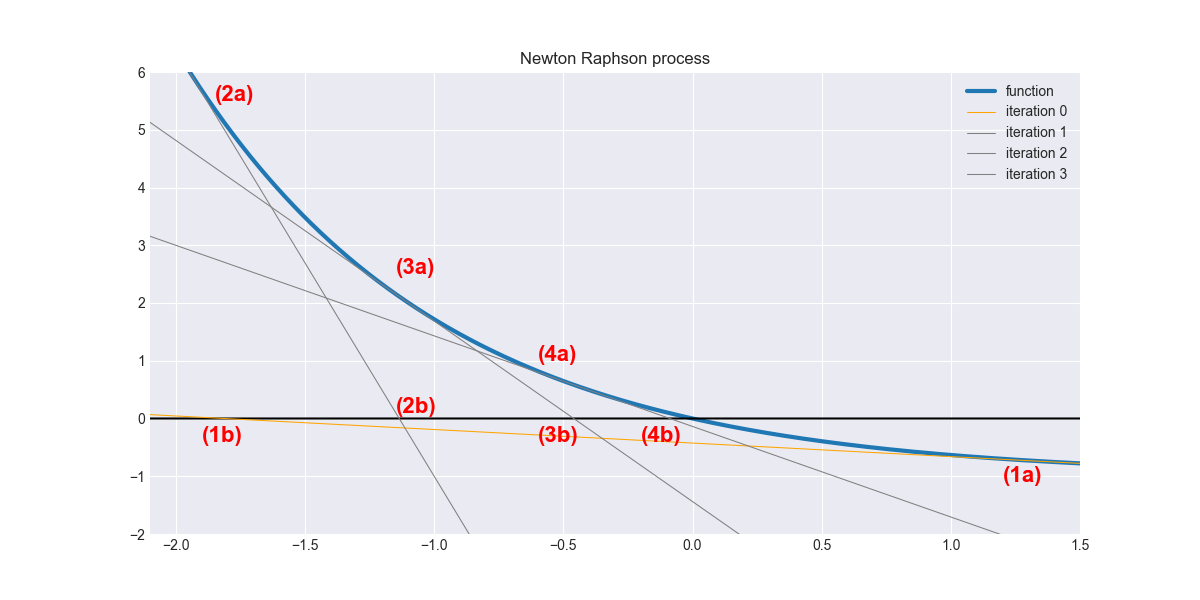}
    \caption[Newton-Raphson worked example]{\textit{Newton-Raphson worked example. }The above figure shows the Newton-Raphson method in action. The blue line represents the function whose root is to be determined. The algorithm starts at point (1a) and gets moved along the orange tanget to (1b). The process then repeats from (2a) which is the same x coordinate as (1b) via the black tangents up to the point (4b) that is in our example judged sufficiently close to being a root.}
    \label{fig:newton}
\end{figure}

The Newton-Raphson method (also named ``Gradient Descent'') uses the
first derivative (``gradient'', in higher dimensions) of a function to
find its roots. A worked example is provided in the figure
\ref{fig:newton}. Here the algorithm starts at a point \(x_0\simeq 1\)
(1a). This point is transported along the orange tangent line to the
point (1b) where that tangent line intersects the x axis. Point (2a) has
the same x coordinate as (1b), and the process is repeated. We see that
(4b) is already very close to the root, which is a general result of
gradient descent methods. On benign functions they converge very quickly
to the root -- and of malignant functions they may not converge at all,
but we well discuss this in more detail below.

\subsubsection{Convergence}\label{convergence}

\label{app:nrconv}

We now want to discuss the convergence properties of the Newton-Raphson
method. The first thing to note is that on a linear function it will
converge in a single step, by design. The issues are generally
introduced by convexity. A few observations:

\begin{itemize}
\item
  If the convexity is directed ``towards'' the x axis (ie what happens
  in figure \ref{fig:newton} in the area where \(f(x)>0\)), and the root
  exists, convergence is guaranteed and swift.
\item
  If the convexity is directed ``away'' from the x axis (ie what happens
  in figure \ref{fig:newton} in the area where \(f(x)<0\)), and the root
  exists, then the algorithm will move to the other side of the root.
\item
  If convexity has the same sign across the entire real axis, and the
  root exists, the algorithm will converge, because either the convexity
  is directed towards the x axis in the region of interest, or the first
  step transports the algorithm to the other side of the root into a
  region where the convexity is directed towards the x axis.
\end{itemize}

\begin{figure}[htbp]
    \centering
    \includegraphics[width=\textwidth]{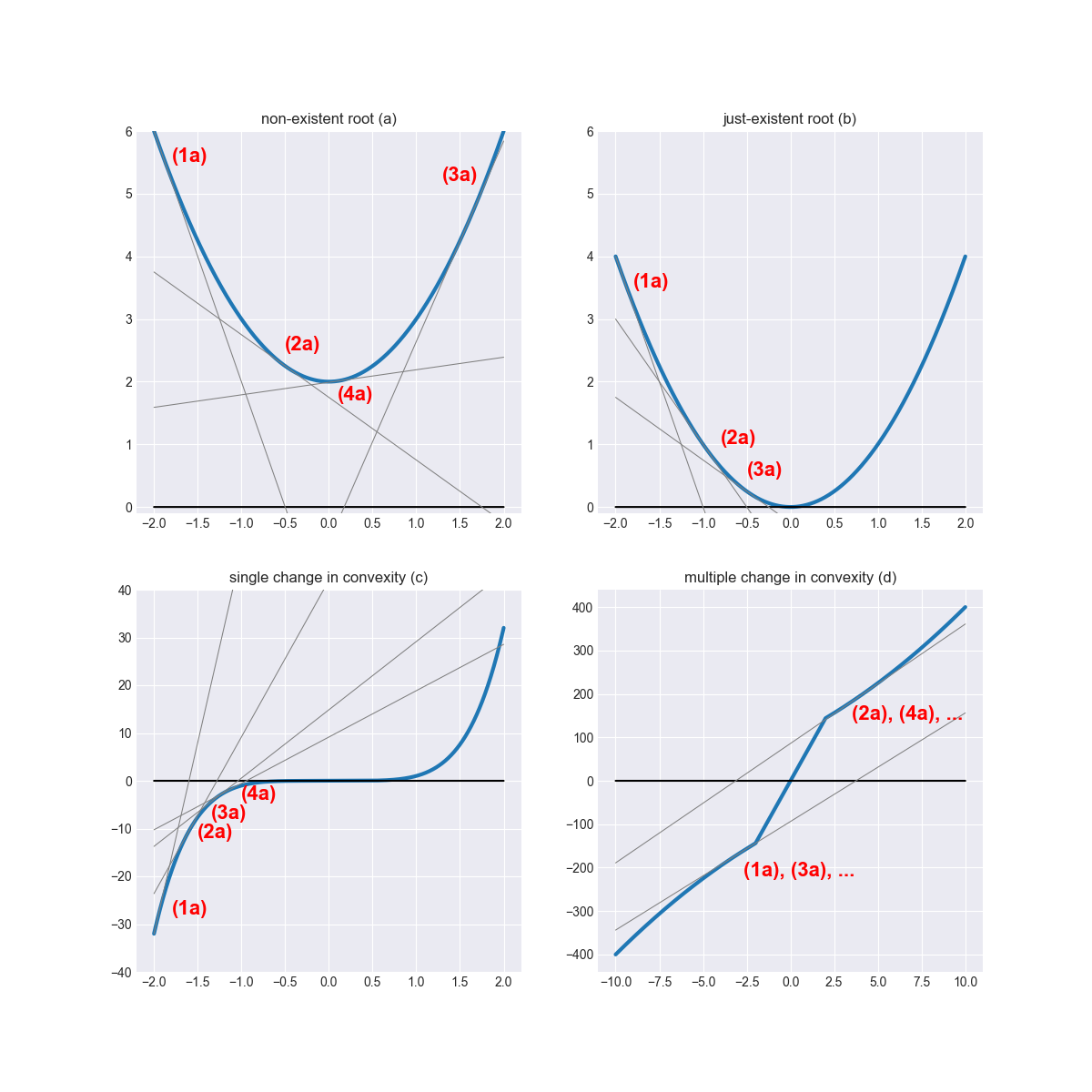}
    \caption[Examples of Newton-Raphson with potentially problematic convergence]{\textit{Examples of Newton-Raphson with potentially problematic convergence. }Those four panels show how the Newton Raphson algorithm performs on some problematic functions. The labels are like in figure \ref{fig:newton} except we omit the (b) labels. The top left panel has no root, and the algorithm enters an infinite cycle. The top right panel as a comparison shows the same function with a root. There convergence is fine. The bottom left panel shows a function with a single inflection point where convergence is fine. The bottom right panel shows a function with two inflection points that has a perfectly well conditioned root, but where the algorithm enters an infinite cycle because of the badly chosen starting point.}
    \label{fig:newton2}
\end{figure}

Now we look at a few things that can go wrong. Firstly, we look at what
happens if there is no root to be found. This case is in figure
\ref{fig:newton2} in the top-left panel (a). We start at point (1a) and
because there is no root, step (2a) overshoots to the other side of the
minimum. Step (3a) then brings is back very close to the minimum, and
because the function there is very flat, step (4a) brings us very far to
the left. In this case, the algorithm will enter an infinite cycle that
will only break if either we hit exactly the minimum where \(f'(x)=0\)
and we get a division by zero, or we leave the domain where the function
has been defined, or we hit a numerical limit. Note the top-right panel
which is the same function, except that it (just) has a root. There, the
convergence is just fine.

What we have just seen is that the algorithm does not converge if there
is no solution that can be found. This can be a nuisance sometimes, but
arguably it is not a major issue -- after all there is no root to be
found, and non-convergence is a good albeit possibly expensive indicator
of that. The other case we look at is if there is a change in convexity.
As the bottom-left panel (c) shows, this can be alright is there is
single inflection point. However, as the bottom-right panel (d) shows,
multiple inflection points, especially with very high or even infinite
convexity values, can lead the algorithm into an infinite cycle. This is
particularly vexing as the function is generally very well behaved, and
had we chosen a starting point in the inner region, convergence would
have been immediate.

This is the big downside of a gradient-based algorithm: because it
extrapolates local behavior across the entire curve it can massively
overshoot. If this goes into a region where the function is not defined
then it is hard to recover from that. Also there are configurations like
the one that we've seen above that lead to infinite cycles around the
root -- and unfortunately those are not that rare. It turns out that in
our specific problem, market scenarios with multiple levered curves
regularly lead to such situations.

\subsubsection{Introducing the learning rate
eta}\label{introducing-the-learning-rate-eta}

\label{app:nreta}

We have seen above that in many instances the Newton-Raphson method can
overshoot the root because it extrapolates local behavior across the
entire curve. This problem can be mitigated by introducing a learning
rate \(\eta < 1\) that reduces the step size of the update, allowing the
algorithm to adapt to change in the local conditions. This flexibility
however comes at a cost in that the convergence of the algorithm in the
quasi linear case is now slower.

To give an example, we consider the function \(f(x) = mx+b\) at the
point \(x=x^{(0)}\). Standard Newton-Raphson would converge in a single
step to the root \(x^{(1)}=-b/m\). If we introduce a learning rate
\(\eta\) the update rule becomes instead

\begin{equation}\begin{aligned}
x^{(s+1)} = x^{(s)} + \eta (x^{(s)} 
- x^{(\infty)})
\end{aligned}\end{equation}

where \(x^{(\infty)} = -b/m\) is the actual root of the function. In
other words, like Zeno's arrow, at every step we get closer to the root
by a constant percentage. However, unlike in Zeno's paradox, every step
is constant time so for \(\eta<1\) the algorithm will indeed never quite
reach the target.

\begin{figure}[htbp]
    \centering
    \includegraphics[width=\textwidth]{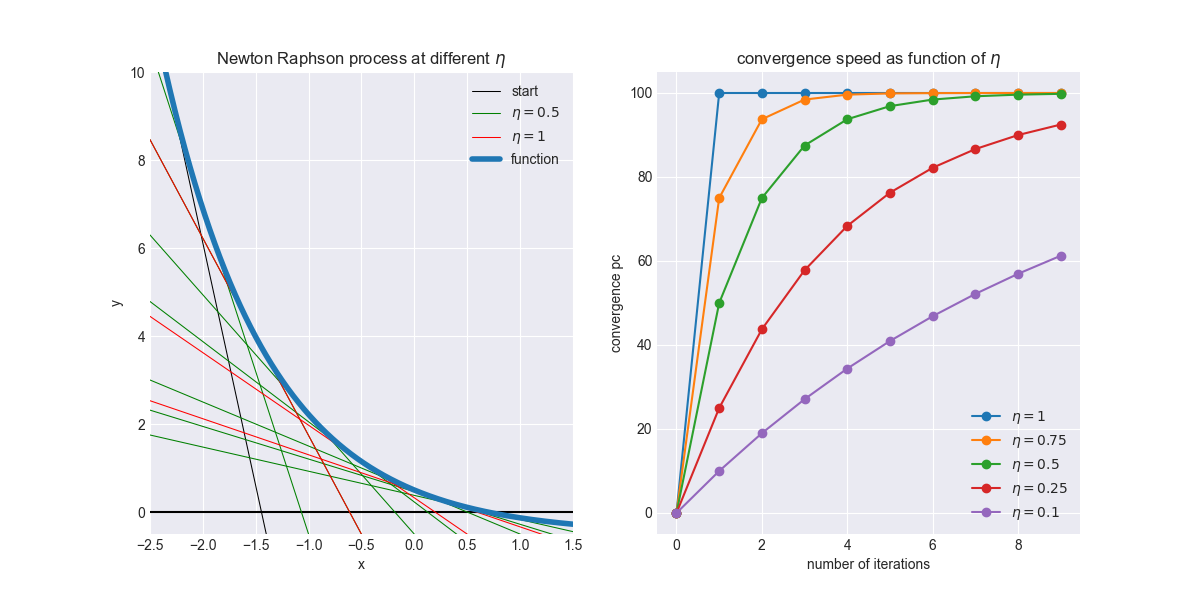}
    \caption[Impact of the learning rate on the Newton Raphson algorithm]{\textit{Impact of the learning rate on the Newton Raphson algorithm. }The left panel shows the mechanics of the Newton Raphson algorithm with learning rate \(\eta=1\) (red) and \(\eta=0.5\) (green), in line with the depiction in figure \ref{fig:newton} but with the labels (a), (b) omitted. The black line represents the first tangent. The red lines place the points (a) directly above the intersection and continue from there. The green lines only move the fraction \(\eta\) of the distance and place their new points (a) there. The right panel shows the convergence of the algorithm on a linear function. For \(\eta=1\) convergence is immediate, and for lower values of \(\eta\) it takes increasingly more steps to approach the target value which is never reached.}
    \label{fig:eta}
\end{figure}

In figure \ref{fig:eta} we provided some analysis and illustration on
the impact of the parameter \(\eta\). In the left hand panel we are
looking at the ``Zeno convergence'', indicating how fast a linear
function converges to the actual value. As we can see -- at \(\eta=1\)
convergence is perfect in step 1, but for any \(\eta<1\) convergence is
slower and always asymptotic. On the right hand panel we visualized how
a non-linear function converges according to different values of
\(\eta\): the red one is \(\eta=1\) and in this case convergence is very
quick. The green one is \(\eta=0.5\), and here it takes significantly
more steps to converge.

\subsubsection{Higher dimensions}\label{higher-dimensions}

\label{app:nrhighdim}

This algorithm easily generalizes to higher dimensions, where the
gradient is replaced by the Jacobian matrix, and the tangent line by the
tangent space. In the simple implementation shown above, effectively the
function is replaced by its best linear approximation, and the root of
that approximation is found. This ``linear root'' is then used to start
the next step of the process.

More formally, if we have a function \(f:\mathbb{R}^n\to\mathbb{R}^n\),
the Jacobian matrix \(J\) at the point
\(\mathbf{\hat x} = (\hat x_1,\ldots,\hat x_n)\) is defined as

\begin{equation}\label{eq:jacobian}\begin{aligned}
J_{ij}(\mathbf{\hat x}) = \frac{\partial f_i}{\partial x_j}(\mathbf{\hat x})
\end{aligned}\end{equation}

and the linear approximation of the function \(f\) around
\(\mathbf{\hat x}\) is

\begin{equation}\label{eq:jacobianapprox}\begin{aligned}
f(\mathbf{x}) \simeq f(\mathbf{\hat x}) + J(\mathbf{\hat x})\cdot (\mathbf{x}-\mathbf{\hat x})
\end{aligned}\end{equation}

The root of this linear approximation can be found by setting the right
hand side to \(\mathbf{0}\) and solving for \(\mathbf{x}\).
Specifically, if at step \(s\) we are at the point
\(\mathbf{\hat x}^{(s)}\) then, using the linear approximation for
\(\mathbf{\hat x}^{(s+1)}\), we get

\begin{equation}\label{eq:update}\begin{aligned}
\mathbf{\hat x}^{(s+1)} 
= 
\mathbf{\hat x}^{(s)} 
- 
J^{-1}(\mathbf{\hat x}^{(s)})
\cdot
f(\mathbf{\hat x}^{(s)})
\end{aligned}\end{equation}

where \(J^{-1}(\mathbf{\hat x}^{(s)})\) is the inverse of the Jacobian
matrix at the point \(\mathbf{\hat x}^{(s)}\).

Introducing the learning rate \(\eta\) we get the update rule\footnote{In practice we use an update rule that first tries to
invert the Jacobian, and if it is singular it uses another algorithm
that inverts it only on its range, and does not attempt inversion on its
null space}

\begin{equation}\label{eq:updateeta}\begin{aligned}
\mathbf{\hat x}^{(s+1)} 
= 
\mathbf{\hat x}^{(s)} 
+
\Delta \mathbf{\hat x}^{(s)}
\end{aligned}\end{equation} where \begin{equation}\begin{aligned}
\Delta \mathbf{\hat x}^{(s)}
=
- \eta\ J^{-1}(\mathbf{\hat x}^{(s)})\cdot f(\mathbf{\hat x}^{(s)})\end{aligned}\end{equation}

\pagebreak

\section{Explanation of key charts and
tables}\label{explanation-of-key-charts-and-tables}

\label{app:tablescharts}

\subsection{Curve charts}\label{curve-charts}

\begin{figure}[htbp]
    \centering
    \includegraphics[width=\textwidth]{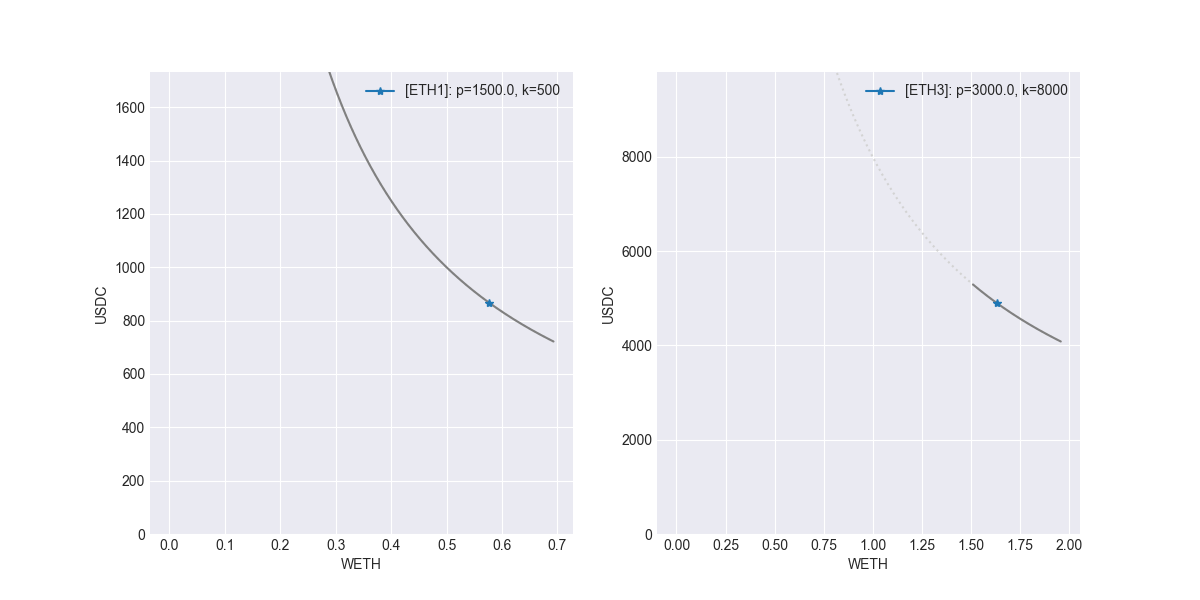}
    \caption[Representation of a single AMM curve with state]{\textit{Representation of a single AMM curve with state. }The left panel shows an unlevered invariance (bonding) curve one the WETH/USDC pair: the AMM is indifferent between all combinations of token holdings on this curve, and currently it is at the point indicated by the star. The right hand curve shows a levered curve where the interpretation is the same, except that the state cannot go beyond the solid part of the curve. The dotted part is the associated unlevered curve.}
    \label{fig:curves}
\end{figure}

Our key means of representing the state of the world are the \emph{curve
charts}. Fundamentally, they represent the invariance curves of the
respective AMMs, with the current position marked with a star. We
consider two types of curves, \emph{levered} and \emph{unlevered} ones.
In the former, for example described by the traditional AMM equation
\ref{eq:invariant}, liquidity is placed all along the curve that
therefore covers the entire price range between \(p=0\) and
\(p=\infty\). Such a curve is shown in the left hand panel of figure
\ref{fig:curves}. A levered curve on the other hand only places
liquidity inside a certain price range, and the mechanism operates with
virtual token balances as described in equation \ref{eq:invariantl}.
Such a curve is shown in the right hand panel of figure \ref{fig:curves}
where liquidity is only available for virtual ETH balances between
roughly \(1\ldots 2\) WETH and \(400\ldots 700\) USDC. In both cases,
the current state of the AMM on the curve is depicted by a star.
Ignoring fees, all trades must happen on the solid part of the
respective curve. The dotted part on the right hand panel is only for
information purposes -- this is where the AMM would trade if the curve
was unlevered with the same virtual token balances.

Here, both curves are trading WETH against USDC. The left, unlevered one
has a pool constant \(k=500\), and the current price is \(p=1500\) USDC
per WETH\footnote{The price is always quoted as \(dy\) per
\(dx\)}. The current state, as indicated by the blue star, is at
virtual balances of \(~0.58\) WETH and \(~866\) USDC, the ratio of the
two numbers corresponding to the price of \(1500\). This curve trades
over the entire price range. The right hand curve is a levered curve
that only trades in the small area where the curve is solid. The current
price, again indicated by the blue star, is \(3000\), and the range is
determined by the both ends of the solid curve at prices of \(2,000\)
and \(3500\) respectively.

\subsection{Trade instruction tables}\label{trade-instruction-tables}

\label{sec:titables}

In section \ref{sec:cvximpl} we have run the convex optimization
algorithm on a number of curves, notably figure \ref{fig:arbpair}
(pair), figure \ref{fig:arbtriangle1c} (triangle), figure
\ref{fig:arbtriangle3c} (triangle and pairs), and figure
\ref{fig:problemcurves} (levered pair).

In table \ref{tbl:3_arb1x3instr} (corresponding to figure
\ref{fig:arbpair}) we see the trade instruction table associated with
the pair trade. It contains the following lines:

\begin{itemize}
\item
  \textbf{Price line}: the post trade price of the respective token, in
  units of the \emph{``target token''} in which the profit is taken
  (which can in this line be identified by a value of \(1\)).
\item
  \textbf{Curve lines} (\emph{ETH1-3}): the trading lines corresponding
  to specific curves, as seen from the AMM. A negative number is an
  outflow of the respective token, and a positive number an inflow.
\item
  \textbf{Aggregate lines} (\emph{AMMIn, AMMOut}): aggregates all curves
  (``the AMM'', as opposed to ``the trader''), separating (positive)
  inflows and (negative) outflows.
\item
  \textbf{Net line}: the difference between aggregate in and outflows;
  in an arbitrage transaction all flows but one should be approximately
  zero, meaning that, in aggregate, no token flows happen in those
  tokens, and one flow (the one in the target token) should be negative,
  indicating the profit taken by the trader.
\end{itemize}

In table \ref{tbl:3_arb3x1instr} (corresponding to figure
\ref{fig:arbtriangle1c}) we are looking at a \emph{triangle arbitrage},
meaning that we have three tokens (WETH, WBTC and USDC) and one curve
each for the corresponding pairs. The structure of the table is the same
as in table \ref{tbl:3_arb1x3instr} except that we have one additional
token column. In table \ref{tbl:3_arb3x3instr} (corresponding to figure
\ref{fig:arbtriangle3c}) finally we have 3 tokens and 3 curves per
token, so a total of 9 curve lines.

\pagebreak

\section{Implementation example}\label{implementation-example}

\label{app:implapp}

In this appendix we describe in more detail the assumptions made and
data used in section \ref{sec:margpimpl}. We have 7 tokens, TKN0 to
TKN6, and their \emph{base prices} (ie the prices they would have
without deviations) are given in table \ref{tbl:4_prices}. For
simplicity they follow a geometric progression. For example, TKN4 has
twice the price of TKN3, and four times that of TKN2.

\begin{table}[htbp]  
\centering
\small

\begin{tabular}{lrrrrrrr}
\toprule
{} &  TKN0 &  TKN1 &  TKN2 &  TKN3 &  TKN4 &  TKN5 &  TKN6 \\
\midrule
\textbf{USD price} &     1 &     2 &     4 &     8 &    16 &    32 &    64 \\
\bottomrule
\end{tabular}

\caption{Token base prices}
\label{tbl:4_prices}
\end{table}

In the mostly-no-arbitrage case, corresponding to the curves in
\ref{tbl:4_curves}, we have 17 curves C00 to C16 that exhibit prices
close to their base price. There are also curves Ca0 to Ca2 which have a
higher capacity and will be the main route for taking the arbitrage that
we will consider by adding the curve in table \ref{tbl:4_curvesx1}.
However, those three curves on their own do not provide an arbitrage
opportunity, as the prices are very close to the base prices.

\begin{table}[htbp]  
\centering
\small

\begin{tabular}{lrrrrrr}
\toprule
{} &       Pair &    Price & Base Price & Deviation &          L &       L\_USD \\
\textbf{cid} &            &          &            &           &            &             \\
\midrule
\textbf{C00} &  TKN6/TKN3 &   8.1411 &          8 &      1.8\% &     46,341 &   1,048,576 \\
\textbf{C01} &  TKN6/TKN3 &   8.0320 &          8 &      0.4\% &     46,341 &   1,048,576 \\
\textbf{C02} &  TKN0/TKN2 &   0.2524 &        1/4 &      1.0\% &    524,288 &   1,048,576 \\
\textbf{C03} &  TKN4/TKN3 &   2.0448 &          2 &      2.2\% &     92,682 &   1,048,576 \\
\textbf{C04} &  TKN3/TKN2 &   2.0374 &          2 &      1.9\% &    185,364 &   1,048,576 \\
\textbf{C05} &  TKN3/TKN2 &   1.9805 &          2 &     -1.0\% &    185,364 &   1,048,576 \\
\textbf{C06} &  TKN4/TKN1 &   8.0760 &          8 &      1.0\% &    185,364 &   1,048,576 \\
\textbf{C07} &  TKN4/TKN1 &   7.9879 &          8 &     -0.2\% &    185,364 &   1,048,576 \\
\textbf{C08} &  TKN2/TKN1 &   1.9979 &          2 &     -0.1\% &    370,728 &   1,048,576 \\
\textbf{C09} &  TKN6/TKN0 &  64.2628 &         64 &      0.4\% &    131,072 &   1,048,576 \\
\textbf{C10} &  TKN4/TKN2 &   4.0058 &          4 &      0.1\% &    131,072 &   1,048,576 \\
\textbf{C11} &  TKN4/TKN5 &   0.5073 &        1/2 &      1.5\% &     46,341 &   1,048,576 \\
\textbf{C12} &  TKN6/TKN4 &   4.0304 &          4 &      0.8\% &     32,768 &   1,048,576 \\
\textbf{C13} &  TKN1/TKN2 &   0.5006 &        1/2 &      0.1\% &    370,728 &   1,048,576 \\
\textbf{C14} &  TKN0/TKN5 &   0.0314 &       1/32 &      0.4\% &    185,364 &   1,048,576 \\
\textbf{C15} &  TKN0/TKN5 &   0.0314 &       1/32 &      0.3\% &    185,364 &   1,048,576 \\
\textbf{C16} &  TKN2/TKN3 &   0.5075 &        1/2 &      1.5\% &    185,364 &   1,048,576 \\
\textbf{Ca0} &  TKN1/TKN2 &   0.5000 &        1/2 &     -0.0\% &  2,621,440 &   7,414,552 \\
\textbf{Ca1} &  TKN2/TKN3 &   0.5000 &        1/2 &      0.0\% &  3,707,276 &  20,971,520 \\
\textbf{Ca2} &  TKN3/TKN4 &   0.5000 &        1/2 &      0.0\% &  3,707,276 &  41,943,040 \\
\bottomrule
\end{tabular}

\caption{Curve set with limited arbitrage opportunities}
\label{tbl:4_curves}
\end{table}

\begin{table}[htbp]  
\centering
\small

\begin{tabular}{lrrrrrr}
\toprule
{} &       Pair &   Price & Base Price & Deviation &          L &       L\_USD \\
\textbf{cid} &            &         &            &           &            &             \\
\midrule
\textbf{CaX} &  TKN1/TKN4 &  0.1500 &        1/8 &     20.0\% &  1,853,638 &  10,485,760 \\
\bottomrule
\end{tabular}

\caption{Additional curve providing arbitrage opportunity}
\label{tbl:4_curvesx1}
\end{table}

We can confirm that there is only a very small arbitrage in the original
scenario corresponding to table \ref{tbl:4_curves} when we look at the
trade instructions in table \ref{tbl:4_instr} and \ref{tbl:4_instr_x2}.
The total arbitrage is about 400 USD, both when extracted via TKN0, and
via TKN2.

When we add the curve CaX from table \ref{tbl:4_curvesx1} as shown in
tables \ref{tbl:4_instr1} and \ref{tbl:4_instr1_x2}. However, that
arbitrage massively increases to an amount of about 40,000 USD. This is
unsurprising, given that the price of this curve is 20\% off the base
price, and the capacity the curve is 10 times bigger than that of the
standard curves, and commensurate with the curves Ca0 to Ca2 that close
the arbitrage\footnote{Curve CaX trades TKN1 and TKN4, and the three
other curves close this in a square via TKN2 and TKN3}. As expected,
most of the trading activity happens in those four curves and their
associated tokens TKN1 to TKN4, plus TKN0 if the profit is taken in that
token.

It is important to understand that the trade instructions form a
connected system that moves tokens around. Once those instructions are
created, individual curves cannot simply be removed, even if their
contribution to the arbitrage is minimal. Each curve plays a role, and
if one is removed, the associated flows must be rerouted through other
curves. This rerouting may or may not significantly increase costs,
depending on the importance of the curve to a specific trade.

There is no straightforward way for us to determine which curves are
important by simply examining the trade instruction table. The only
reliable method we know is to rerun the optimization algorithm without
certain curves and compare the results. For example in table
\ref{tbl:4_instr1_no00} we repeat the analysis from table
\ref{tbl:4_instr1}, but we remove the curve C00, which handled over
1,000 USD in trading volume. Despite this, the final profit is only
reduced by 2 USD, or about 0.5 bp of the total profit. However, removing
this curve causes significant rerouting through the other curves. The
details of this rerouting are not immediately apparent from examining
only the trade instructions tables (tables \ref{tbl:4_instr},
\ref{tbl:4_instr_x2}, \ref{tbl:4_instr1}, \ref{tbl:4_instr1_x2},
\ref{tbl:4_instr1_no00}, and the summary table \ref{tbl:4_instr1a}).

\begin{table}[htbp]  
\centering
\small

\begin{tabular}{lrrrrrrr}
\toprule
{} &    TKN0 &    TKN1 &    TKN2 &    TKN3 &    TKN4 &   TKN5 &   TKN6 \\
\midrule
\textbf{PRICE    } &    1.00 &    1.99 &    3.99 &    7.97 &   15.96 &  31.74 &  64.38 \\
\textbf{C00      } &      &      &      &    -536 &      &     &     66 \\
\textbf{C01      } &      &      &      &     353 &      &     &    -44 \\
\textbf{C02      } &   3,190 &      &    -803 &      &      &     &     \\
\textbf{C03      } &      &      &      &  -1,401 &     693 &     &     \\
\textbf{C04      } &      &      &  -2,405 &   1,191 &      &     &     \\
\textbf{C05      } &      &      &   1,315 &    -661 &      &     &     \\
\textbf{C06      } &      &  -2,105 &      &      &     262 &     &     \\
\textbf{C07      } &      &     777 &      &      &     -97 &     &     \\
\textbf{C08      } &      &     352 &    -176 &      &      &     &     \\
\textbf{C09      } &     979 &      &      &      &      &     &    -15 \\
\textbf{C10      } &      &      &     -39 &      &      10 &     &     \\
\textbf{C11      } &      &      &      &      &     284 &   -143 &     \\
\textbf{C12      } &      &      &      &      &      29 &     &     -7 \\
\textbf{C13      } &      &     400 &    -200 &      &      &     &     \\
\textbf{C14      } &  -1,990 &      &      &      &      &     63 &     \\
\textbf{C15      } &  -2,565 &      &      &      &      &     81 &     \\
\textbf{C16      } &      &      &   1,968 &    -991 &      &     &     \\
\textbf{Ca0      } &      &     575 &    -288 &      &      &     &     \\
\textbf{Ca1      } &      &      &     628 &    -314 &      &     &     \\
\textbf{Ca2      } &      &      &      &   2,360 &  -1,179 &     &     \\
\textbf{AMMIn    } &   4,169 &   2,105 &   3,911 &   3,904 &   1,276 &    143 &     66 \\
\textbf{AMMOut   } &  -4,555 &  -2,105 &  -3,911 &  -3,904 &  -1,276 &   -143 &    -66 \\
\textbf{TOTAL NET} &    -386 &       0 &       0 &       0 &       0 &      0 &      0 \\
\bottomrule
\end{tabular}

\caption{Trade instructions with little arbitrage}
\label{tbl:4_instr}
\end{table}

\begin{table}[htbp]  
\centering
\small

\begin{tabular}{lrrrrrrr}
\toprule
{} &    TKN0 &    TKN1 &    TKN2 &    TKN3 &    TKN4 &  TKN5 &   TKN6 \\
\midrule
\textbf{PRICE    } &    0.25 &    0.50 &    1.00 &    2.00 &    4.00 &  7.96 &  16.15 \\
\textbf{C00      } &      &      &      &    -541 &      &    &     67 \\
\textbf{C01      } &      &      &      &     348 &      &    &    -43 \\
\textbf{C02      } &   3,355 &      &    -844 &      &      &    &     \\
\textbf{C03      } &      &      &      &  -1,402 &     693 &    &     \\
\textbf{C04      } &      &      &  -2,407 &   1,192 &      &    &     \\
\textbf{C05      } &      &      &   1,313 &    -660 &      &    &     \\
\textbf{C06      } &      &  -2,109 &      &      &     262 &    &     \\
\textbf{C07      } &      &     773 &      &      &     -97 &    &     \\
\textbf{C08      } &      &     353 &    -177 &      &      &    &     \\
\textbf{C09      } &   1,097 &      &      &      &      &    &    -17 \\
\textbf{C10      } &      &      &     -42 &      &      11 &    &     \\
\textbf{C11      } &      &      &      &      &     277 &  -140 &     \\
\textbf{C12      } &      &      &      &      &      26 &    &     -7 \\
\textbf{C13      } &      &     401 &    -201 &      &      &    &     \\
\textbf{C14      } &  -1,939 &      &      &      &      &    61 &     \\
\textbf{C15      } &  -2,513 &      &      &      &      &    79 &     \\
\textbf{C16      } &      &      &   1,966 &    -990 &      &    &     \\
\textbf{Ca0      } &      &     582 &    -291 &      &      &    &     \\
\textbf{Ca1      } &      &      &     586 &    -293 &      &    &     \\
\textbf{Ca2      } &      &      &      &   2,346 &  -1,173 &    &     \\
\textbf{AMMIn    } &   4,452 &   2,109 &   3,865 &   3,886 &   1,269 &   140 &     67 \\
\textbf{AMMOut   } &  -4,452 &  -2,109 &  -3,962 &  -3,886 &  -1,269 &  -140 &    -67 \\
\textbf{TOTAL NET} &      -0 &      -0 &     -97 &      -0 &      -0 &    -0 &     -0 \\
\bottomrule
\end{tabular}

\caption{Trade instructions with little arbitrage, extracting via TKN2}
\label{tbl:4_instr_x2}
\end{table}

\begin{table}[htbp]  
\centering
\small

\begin{tabular}{lrrrrrrr}
\toprule
{} &     TKN0 &      TKN1 &     TKN2 &     TKN3 &     TKN4 &   TKN5 &   TKN6 \\
\midrule
\textbf{PRICE    } &     1.00 &      2.10 &     3.94 &     7.69 &    15.18 &  31.21 &  62.44 \\
\textbf{C00      } &       &        &       &     -165 &       &     &     20 \\
\textbf{C01      } &       &        &       &      724 &       &     &    -90 \\
\textbf{C02      } &   -2,985 &        &      756 &       &       &     &     \\
\textbf{C03      } &       &        &       &   -2,321 &    1,155 &     &     \\
\textbf{C04      } &       &        &   -5,581 &    2,798 &       &     &     \\
\textbf{C05      } &       &        &   -1,860 &      946 &       &     &     \\
\textbf{C06      } &       &   -28,135 &       &       &    3,680 &     &     \\
\textbf{C07      } &       &   -25,254 &       &       &    3,322 &     &     \\
\textbf{C08      } &       &   -15,990 &    8,255 &       &       &     &     \\
\textbf{C09      } &  -14,978 &        &       &       &       &     &    236 \\
\textbf{C10      } &       &        &   -5,034 &       &    1,281 &     &     \\
\textbf{C11      } &       &        &       &       &    1,393 &   -692 &     \\
\textbf{C12      } &       &        &       &       &      680 &     &   -167 \\
\textbf{C13      } &       &   -15,942 &    8,231 &       &       &     &     \\
\textbf{C14      } &  -10,626 &        &       &       &       &    337 &     \\
\textbf{C15      } &  -11,200 &        &       &       &       &    355 &     \\
\textbf{C16      } &       &        &   -1,208 &      616 &       &     &     \\
\textbf{Ca0      } &       &  -114,981 &   59,331 &       &       &     &     \\
\textbf{Ca1      } &       &        &  -62,890 &   31,827 &       &     &     \\
\textbf{Ca2      } &       &        &       &  -34,425 &   17,326 &     &     \\
\textbf{CaX      } &       &   200,302 &       &       &  -28,838 &     &     \\
\textbf{AMMIn    } &        0 &   200,302 &   76,573 &   36,911 &   28,838 &    692 &    257 \\
\textbf{AMMOut   } &  -39,789 &  -200,302 &  -76,573 &  -36,911 &  -28,838 &   -692 &   -257 \\
\textbf{TOTAL NET} &  -39,789 &        -0 &       -0 &       -0 &       -0 &     -0 &     -0 \\
\bottomrule
\end{tabular}

\caption{Trade instructions with arbitrage curve}
\label{tbl:4_instr1}
\end{table}

\begin{table}[htbp]  
\centering
\small

\begin{tabular}{lrrrrrrr}
\toprule
{} &     TKN0 &      TKN1 &     TKN2 &     TKN3 &     TKN4 &  TKN5 &   TKN6 \\
\midrule
\textbf{PRICE    } &     0.25 &      0.53 &     1.00 &     1.95 &     3.85 &  7.75 &  15.71 \\
\textbf{C00      } &       &        &       &     -692 &       &    &     85 \\
\textbf{C01      } &       &        &       &      197 &       &    &    -24 \\
\textbf{C02      } &   14,145 &        &   -3,523 &       &       &    &     \\
\textbf{C03      } &       &        &       &   -2,348 &    1,169 &    &     \\
\textbf{C04      } &       &        &   -5,763 &    2,891 &       &    &     \\
\textbf{C05      } &       &        &   -2,042 &    1,039 &       &    &     \\
\textbf{C06      } &       &   -28,331 &       &       &    3,707 &    &     \\
\textbf{C07      } &       &   -25,450 &       &       &    3,349 &    &     \\
\textbf{C08      } &       &   -15,731 &    8,117 &       &       &    &     \\
\textbf{C09      } &   -2,863 &        &       &       &       &    &     45 \\
\textbf{C10      } &       &        &   -5,266 &       &    1,342 &    &     \\
\textbf{C11      } &       &        &       &       &      709 &  -356 &     \\
\textbf{C12      } &       &        &       &       &      429 &    &   -106 \\
\textbf{C13      } &       &   -15,683 &    8,093 &       &       &    &     \\
\textbf{C14      } &   -5,354 &        &       &       &       &   169 &     \\
\textbf{C15      } &   -5,928 &        &       &       &       &   187 &     \\
\textbf{C16      } &       &        &   -1,389 &      709 &       &    &     \\
\textbf{Ca0      } &       &  -113,146 &   58,354 &       &       &    &     \\
\textbf{Ca1      } &       &        &  -66,520 &   33,687 &       &    &     \\
\textbf{Ca2      } &       &        &       &  -35,484 &   17,863 &    &     \\
\textbf{CaX      } &       &   198,340 &       &       &  -28,567 &    &     \\
\textbf{AMMIn    } &   14,145 &   198,340 &   74,564 &   38,524 &   28,567 &   356 &    130 \\
\textbf{AMMOut   } &  -14,145 &  -198,340 &  -84,503 &  -38,524 &  -28,567 &  -356 &   -130 \\
\textbf{TOTAL NET} &       -0 &        -0 &   -9,939 &       -0 &       -0 &    -0 &     -0 \\
\bottomrule
\end{tabular}

\caption{Trade instructions with arbitrage curve, extracting via TKN2}
\label{tbl:4_instr1_x2}
\end{table}

\begin{table}[htbp]  
\centering
\small

\begin{tabular}{lrrrrrrr}
\toprule
{} &     TKN0 &      TKN1 &     TKN2 &     TKN3 &     TKN4 &   TKN5 &   TKN6 \\
\midrule
\textbf{PRICE    } &     1.00 &      2.10 &     3.94 &     7.69 &    15.18 &  31.22 &  62.41 \\
\textbf{C01      } &       &        &       &      660 &       &     &    -82 \\
\textbf{C02      } &   -2,798 &        &      708 &       &       &     &     \\
\textbf{C03      } &       &        &       &   -2,323 &    1,156 &     &     \\
\textbf{C04      } &       &        &   -5,579 &    2,797 &       &     &     \\
\textbf{C05      } &       &        &   -1,858 &      945 &       &     &     \\
\textbf{C06      } &       &   -28,136 &       &       &    3,681 &     &     \\
\textbf{C07      } &       &   -25,255 &       &       &    3,322 &     &     \\
\textbf{C08      } &       &   -15,989 &    8,255 &       &       &     &     \\
\textbf{C09      } &  -15,282 &        &       &       &       &     &    241 \\
\textbf{C10      } &       &        &   -5,035 &       &    1,282 &     &     \\
\textbf{C11      } &       &        &       &       &    1,385 &   -688 &     \\
\textbf{C12      } &       &        &       &       &      649 &     &   -160 \\
\textbf{C13      } &       &   -15,941 &    8,231 &       &       &     &     \\
\textbf{C14      } &  -10,566 &        &       &       &       &    335 &     \\
\textbf{C15      } &  -11,141 &        &       &       &       &    353 &     \\
\textbf{C16      } &       &        &   -1,205 &      615 &       &     &     \\
\textbf{Ca0      } &       &  -114,972 &   59,326 &       &       &     &     \\
\textbf{Ca1      } &       &        &  -62,842 &   31,802 &       &     &     \\
\textbf{Ca2      } &       &        &       &  -34,496 &   17,362 &     &     \\
\textbf{CaX      } &       &   200,293 &       &       &  -28,837 &     &     \\
\textbf{AMMIn    } &        0 &   200,293 &   76,519 &   36,819 &   28,837 &    688 &    241 \\
\textbf{AMMOut   } &  -39,787 &  -200,293 &  -76,519 &  -36,819 &  -28,837 &   -688 &   -241 \\
\textbf{TOTAL NET} &  -39,787 &         0 &        0 &        0 &        0 &      0 &      0 \\
\bottomrule
\end{tabular}

\caption{Trade instructions with arbitrage curve, and removing C00}
\label{tbl:4_instr1_no00}
\end{table}

\begin{table}[htbp]  
\centering
\small

\begin{tabular}{lrrrr}
\toprule
{} &      TKN1 &     TKN2 &     TKN3 &     TKN4 \\
\midrule
\textbf{PRICE    } &      0.54 &     1.00 &     1.94 &     3.82 \\
\textbf{Ca0      } &  -141,173 &   73,381 &       &       \\
\textbf{Ca1      } &        &  -80,385 &   40,818 &       \\
\textbf{Ca2      } &        &       &  -40,818 &   20,569 \\
\textbf{CaX      } &   141,173 &       &       &  -20,569 \\
\textbf{AMMIn    } &   141,173 &   73,381 &   40,818 &   20,569 \\
\textbf{AMMOut   } &  -141,173 &  -80,385 &  -40,818 &  -20,569 \\
\textbf{TOTAL NET} &         0 &   -7,004 &        0 &        0 \\
\bottomrule
\end{tabular}

\caption{Trade instructions with arbitrage curves only}
\label{tbl:4_instr1a}
\end{table}

\end{document}